# Validation of non-equilibrium kinetics in $CO_2$-$N_2$ plasmas

C. Fromentin[1], T. Silva[1], T. C. Dias[1], E. Baratte[2], O. Guaitella[2] and V. Guerra[1]

[1] Instituto de Plasmas e Fusão Nuclear, Instituto Superior Técnico, Universidade de Lisboa, Portugal

[2] Laboratoire de Physique des Plasmas (UMR 7648), CNRS, Univ. Paris Saclay, Sorbonne Université, École Polytechnique, France

## Abstract

This work explores the effect of $N_2$ addition on $CO_2$ dissociation and on the vibrational kinetics of $CO_2$ and CO under various non-equilibrium plasma conditions. A self-consistent kinetic model, previously validated for pure $CO_2$ and $CO_2$-$O_2$ discharges, is further extended by adding the kinetics of $N_2$. The vibrational kinetics considered include levels up to $v = 10$ for CO, $v = 59$ for $N_2$ and up to $v_1 = 2$ and $v_2 = v_3 = 5$, respectively for the symmetric stretch, bending and asymmetric stretch modes of $CO_2$, and account for electron-impact excitation and de-excitation (e-V), vibration-to-translation (V-T) and vibration-to-vibration energy exchange (V-V) processes. The kinetic scheme is validated by comparing the model predictions with recent experimental data measured in a DC glow discharge operating in pure $CO_2$ and in $CO_2$-$N_2$ mixtures, at pressures in the range 0.6 - 4 Torr (80.00 – 533.33 Pa) and a current of 50 mA. The experimental results show a higher vibrational temperature of the different modes of $CO_2$ and CO and an increased dissociation fraction of $CO_2$, that can reach values as high as 70 %, when $N_2$ is added to the plasma. On the one hand, the simulations suggest that the former effect is the result of the $CO_2$-$N_2$ and CO-$N_2$ V-V transfers and the reduction of quenching due to the decrease of atomic oxygen concentration; on the other hand, the dilution of $CO_2$ and dissociation products, CO and $O_2$, reduces the importance of back reactions and contributes to the higher $CO_2$ dissociation fraction with increased $N_2$ content in the mixture, while the $N_2(B^3\Pi_g)$ electronically excited state further enhances the $CO_2$ dissociation.

## I. Introduction

To reach climate neutrality (balance between greenhouse gas emission sources and sinks [1]) and limit the irreversibility, severity and likelihood of climate hazards and their widespread impacts on human and natural ecosystems [2,3], by holding the temperature increase below 2°C in line with the Paris Agreement [4], breakthrough innovations are required. Some existing and under-development technologies for $CO_2$ valorisation in the world are briefly reviewed in [5]. The conversion of $CO_2$ into value-added chemicals and liquid fuels, via the $CO_2$ Capture and Utilization (CCU) approach is a potential way to not only decrease the $CO_2$ emissions, but also generate more economic value, reduce the dependence of fossil fuels and create the possibility to recycle $CO_2$ and close the carbon cycle [6]. The dissociation of $CO_2$ into CO is the critical step of any CCU process. Indeed, $CO_2$ is, thermodynamically, a very stable molecule so there are large energy barriers to overcome for its conversion. A very promising approach to achieve efficient dissociation of $CO_2$ is the use of non-thermal plasmas (NTPs) [7–18] benefiting from non-equilibrium conditions where high energy electrons coexist with cold ions and neutrals. The kinetics of nonequilibrium processes can be stimulated by vibrational excitation of molecules in the plasma [19]. Depending mainly on the electron temperature [20], NTPs can selectively transfer energy from the electrons (gaining energy from the electric fied) to the heavy particles in the plasma to initiate certain chemical processes like the dissociation of $CO_2$ with lower energy compared to conventional methods [21]. $CO_2$ dissociation has been the focus of many studies, exploring electron-impact [22–24], vibrationally driven [19,24–27] and thermal [28–30] dissociation routes.

In the last few years, a combination of computational simulations and experimental campaigns have contributed to a better understanding of $CO_2$-containing discharges [31–38] and to the development of the corresponding "reaction mechanisms", i.e., a set of reactions and rate coefficients validated against benchmark experiments. In particular, our modelling results were validated step-by-step against experimental data obtained using direct current (DC) glow discharges (plasmas sustained by high voltages inside a pair of electrodes) and advanced plasma diagnostics, namely Fourier Transform Infrared (FTIR) absorption spectroscopy [36] and optical emission spectroscopy (OES) [39]. By selecting specific working conditions in low-pressure glow discharges, the accuracy of the Electron Energy Distribution Function (EEDF) calculations, electron-impact vibrational excitation and de-excitation (e-V), vibration-to-vibration (V-V) and vibration-to-translation (V-T) energy transfer processes coefficients, and the rate coefficients of the main plasma processes could be tested in low $CO_2$ excitation regime. Our previous modelling research works were focused on:

(i) The electron-neutral scattering cross sections for $CO_2$ [40] and CO [41],
(ii) The validation of the electron-impact dissociation cross sections of $CO_2$ [22],
(iii) The study of the time-resolved evolution of the lower vibrationally excited $CO_2$ levels during the afterglow of $CO_2$ discharges, validating a set of V-V and V-T processes involving the lower ~ 70 states and the corresponding rate coefficients [32],
(iv) The extension of the model by including the e-V processes and studying the effect of electrons on the distribution of vibrationally excited $CO_2$ levels in pulsed and continuous glow discharges [33] to understand the transfer of electron energy toward $CO_2$ vibrations,
(v) The study of the dynamics of gas heating in the afterglow of pulsed $CO_2$ and $CO_2$–$N_2$ glow discharges, further validating the V-V and V-T mechanisms and rate coefficients [35],
(vi) The preliminary study of the $N_2$ influence on $CO_2$ the vibrational distribution functions in pulsed glow discharges using a simplified kinetic scheme [34],
(vii) The development of a reaction mechanism for 'vibrationally cold $CO_2$ plasmas' taking into account the $CO_2$ dissociation products in $CO_2$ discharges in conditions where the vibrational kinetics can modify the EEDF but has no direct influence on chemical reactions [42],
(viii) The elucidation of the role of the electronically excited metastable state $CO(a^3\Pi_r)$ in $CO_2$ dissociation and CO recombination [36,38,42,43],
(ix) The study of the $CO_2$ dissociation under Martian environment for oxygen production [31,44],
(x) The inclusion of a self-consistent description of the $CO_2$ and CO vibrational kinetics involving the dissociation products, namely CO, $O_2$, O validated in $CO_2$ and $CO_2$-$O_2$ discharges [38].

Herein we extend the study of the coupled electron, vibrational and chemical kinetics developed in [31–34,38,42], already tested and validated in $CO_2$-containing discharges and afterglows and for various operating conditions, with the addition and validation of the $N_2$ electron, vibration and chemical kinetics. For this purpose we include 59 vibrational levels of $N_2$, as was initiated in [31,34]. The V-T and V-V processes for these levels with $CO_2$ and $N_2$ molecules and the dissociation products are included, as well as the heavy-particle reactions involving nitrogen species. The model predictions are compared with measurements in DC discharges ignited in different $CO_2$–$N_2$ mixtures.

Several reasons motivate the study of $CO_2$-$N_2$ low temperature plasmas. $N_2$ represents a large fraction in most industrial flue gas [45] along with other impurities like oxygen and water, so the influence it may have on the kinetics and chemistry of the plasma must be investigated, as done previously for $O_2$ [38] and $H_2O$ [46,47]. Moreover, the combined presence of nitrogen and oxygen species is bound to lead to the formation of $N_xO_y$ by-products which are dangerous compounds [48] but could also be an intermediary for nitrogen fixation [49,50]. Furthermore, by varying the $N_2$ content in $CO_2$-$N_2$ mixtures we enlarge the parameter space and can have a thorough validation of the model and gain a deeper understanding of the kinetics of $CO_2$ plasmas. In the context of in situ resource utilization (ISRU) on Mars, studying the effect of $N_2$ on $CO_2$ dissociation is relevant given the presence of $N_2$ in the Martian atmosphere (96 % $CO_2$ – 2 % Ar – 2 % $N_2$). Decomposition of $CO_2$ under Martin environment would promote the local production of breathable $O_2$, fertilizers and transportation fuel on Mars [31,44,51–54], where the conditions are optimal for $CO_2$ dissociation by plasmas [31]. Moreover, once oxygen has been made available, extracted from $CO_2$ thanks to membranes for instance [52], $N_2$ from the Martian atmosphere can be used for the local production of $N_xO_y$ essential for the synthesis of fertilizers on site [54]. Finally, our model for the simulation of $CO_2$-$N_2$ DC discharges is a first step towards the simulation of $CO_2/N_2/H_2O$ mixtures as investigated in [55–57] where the chemistry induced by atmospheric pressure DC discharges above a water surface is studied. Since this mixture represents a model of prebiotic atmosphere of the Earth, it is interesting for the study of amino acids formation and for the verification of the theory of the origins of life. Some recent works have studied $CO_2$–$N_2$ mixtures, with a focus on the entry problems [58–62], the characterisation of electrical discharges in $CO_2/Ar/N_2$ mixtures [31,63–65] and the $CO_2$ conversion [63,66–71].

Current research on plasma-based $CO_2$ conversion in presence of $N_2$ is done mainly with dielectric barrier discharges (DBD) [67,69,72,73], microwave plasmas (MW) [68,74], ns pulse discharges [75–77] and DC glow discharges [34,66,78,79] where $CO_2$ dissociation pathways and energy efficiency are analysed. It has been shown that the admixture of $N_2$ has a beneficial impact on the $CO_2$ decomposition by plasmas [34,70]. Several reasons can be assigned to this effect: the enhancement of the reduced electric field, the effect of electronically and vibrationally excited states on $CO_2$ dissociation directly [66,80] or via changes of the EEDF [24,26,31,42], and the dilution with $N_2$ limiting the influence of back reaction mechanisms producing $CO_2$ from CO [34]. Favouring the excitation of the asymmetric vibrational mode of $CO_2$ by adding $N_2$ is well known in the context of $CO_2$–$N_2$ maser/laser systems extensively studied in the 60s [81–84]. The selective excitation of the asymmetric mode of $CO_2$ by nearly resonant vibrational energy transfers from $N_2$ was reported to enhance lasing power. Indeed, the first excited level of $N_2$ has an energy very close to the first asymmetric level of $CO_2$ ($\Delta E = 18$ cm$^{-1}$ = $2.23 \cdot 10^{-3}$ eV), which is smaller than the typical average kinetic energy kT [85], allowing nitrogen to easily exchange vibration quanta with the asymmetric mode of $CO_2$. This near-resonant transfer can be beneficial for vibrational dissociation of $CO_2$ as vibrationally excited $CO_2$ can undergo molecular dissociation through the so-called ladder climbing mechanism or by electron-impact stepwise processes. The anharmonicity of the molecule creates a preferential excitation of molecules in higher vibrational levels compared to those in lower levels, in V-V collisions, which leads to an overpopulating of the higher vibrational levels [86,87] and possible dissociation of highly vibrating $CO_2$ molecules without a significant amount of kinetic energy required. This purely vibrational mechanism requires ∼ 5.5 eV and the products are obtained in the ground state, whereas the $CO_2$ dissociation by direct electron-impact excitation requires more than 7 eV and leads to the formation of an O atom in an electronically excited state [21]. The asymmetric mode $v_3$ is of major interest here to reach an efficient dissociation by molecular collisions. While in principle any highly excited mode can lead to dissociation, the excited vibrational levels corresponding to the asymmetric stretch has a relatively long lifetime and its relaxation is thus much slower than that of the symmetric stretch and bending modes [84].

To establish a reaction mechanism for vibrationally excited $CO_2$-$N_2$ plasmas a DC glow discharge is used here as it generates a stable (axially) homogeneous plasma (in the positive column) and is accessible to different diagnostics [21]. Despite its evident limitations, namely, fairly low energy efficiency and $CO_2$ dissociation fractions, these plasma sources are optimal for validation of volume averaged 0D self-consistent kinetic models with detailed and complex kinetics and therefore for the fundamental study of $CO_2$-containing plasmas. The $CO_2$, CO and NO densities and the vibrational kinetics are diagnosed by FTIR absoprtion spectroscopy, and actinometry is used to determine the O atom density and loss frequency.

The paper is structured as follows. After the introduction, Section 2 briefly gives information on the experimental set-up and the diagnostics used. This is followed in Section 3 by a description of the model and the kinetic scheme used to study the $CO_2$-$N_2$ discharge. Moreover, in this section we report rate coefficients for electron-impact reactions and vibration-translation and vibration-vibration exchanges involving $CO_2$, $N_2$ and the dissociation products. The comparison between the experiments and the simulations is presented and discussed in section 4 to gain further insight into the underlying kinetics. Finally, perspectives and conclusions are reported in section 5.

## II. Experiments

Experiments are performed in non-thermal plasmas sustained by a continuous DC glow discharge, operating at pressures in the range P = 0.6 - 4 Torr and discharge current I = 50 mA, in a cylindrical Pyrex tube of radius R = 1 cm. Two tube lengths are used throughout this study, 23 cm for the Fourier transform infrared (FTIR) measurements and 67 cm for the actinometry and reduced field measurements. The experimental setup and diagnostics used are very similar to those described in [36,88,89]. In this section we provide a less detailed but essential description of the experiments.

For all measurements, the pressure, flow and current are imposed. The experiments are conducted both in pure $CO_2$ and in a $CO_2$-$N_2$ mixture (Air Liquide Alphagaz 1 standard for both $CO_2$ and $N_2$). The gas flows are controlled using mass flow controllers (Bronkhorst F-210CV) and a total gas flow of 7.4 sccm has been used in the present experiments as done previously in [31,34,36,38,88,89]. The electric field in the positive column of the discharge is estimated by measuring the voltage drop between two tungsten pins placed 20 cm apart and pointing radially inside the reactor. Since the positive column can be considered homogeneous, the electric field measurement is representative of the average field in the plasma bulk. The error on the reduced electric field, E/N, was calculated taking into account the uncertainty on the $T_{rot}$ (see below) used for the calculation of the plasma density N.

In situ FTIR absorption spectroscopy was used to determine the different vibrationally excited states densities of $CO_2$ and CO from which are deduced the characteristic vibrational temperatures of CO and of the different modes of $CO_2$, namely, bending, symmetric stretching and asymmetric stretching modes, corresponding to $T_1$, $T_2$ and $T_3$ respectively [20]. The detected IR spectra contain several lines of CO and $CO_2$ vibrational transitions which are fitted using a MATLAB® algorithm, according to the procedure described in [36,88]. The measurements further give information on the rotational temperature of both $CO_2$ and CO which is assumed to be representative of the gas temperature [89]. By default, the fit assumes a Treanor vibrational distribution for the three vibrational modes of $CO_2$ [36,88,90] and bending and symmetric modes are assumed to be in equilibrium i.e. $T_1 = T_2 = T_{1,2}$ and this assumption was confirmed in [33,37,89]. During the experiments, an FTIR resolution of 0.2 cm$^{-1}$ is used and the error on the different temperatures was estimated to be 30 K and 27 K for $T_{rot}$ and $T_{1,2}$ respectively, 67 K for $T_3$ and 357 K for $T_{CO}$ at 5 Torr, 50 mA and in pure $CO_2$ [36]. As an outcome of the fitting procedure, the dissociation fraction

$$\alpha = \frac{[CO]}{[CO] + [CO_2]}, \qquad (1)$$

where [CO] and [$CO_2$] represent the gas phase concentrations of CO and $CO_2$ molecules, respectively, is also obtained. Due to the small dimensions of the sample chamber of the FTIR spectrometer (Bruker V70??), these measurements are done in the 23 cm long reactor. The temperatures for the same pressure and current are assumed to be the same in both reactors due to the fast timescales of temperature evolution in comparison with the residence times in our experimental conditions [89]. Finally, note that the measured vibrational temperatures are only representative of the populations of the lower vibrational levels and do not bring any information regarding the populations of the higher ones, which may deviate from equilibrium Boltzmann and/or Treanor distributions and are modified by the presence of $N_2$.

Downstream measurements, in which the same plasma reactor is placed upstream a measurement cell installed in the FTIR sample compartment, were also performed, similarly to what has been done by Morillo-Candas *et al.* [91]. Indeed, due to the small fraction of $N_xO_y$ species in the discharge, we had to use a multipass cell placed after the plasma reactor, to increase the optical path to 5 m. Thanks to this configuration, we could reach sufficient sensitivity to observe NO and nitrogen dioxide ($NO_2$) absorption lines and capturing a wavenumber range of 950–3350 cm$^{-1}$ ensured the simultaneous measurement of NO, $NO_2$ as well as $CO_2$ and CO absorption bands. The $NO_2$ signal was at the noise level, so that the uncertainty obtained on the $NO_2$ density was so large that the data obtained are not reliable and therefore not provided here.

O atom densities and wall loss frequencies are measured in pure $CO_2$ by actinometry in a 67 cm length tube [89]. The absolute O atom densities obtained rely on several parameters for which large discrepancies exist in the literature which leads to an error estimation for this specie density above 30% [89,91]. The loss probability of O atoms at the wall, $\gamma_O$, is given in [89] as a function of pressure, for a current of 50 mA, and is deduced from the experimental determination of O atom loss frequencies, in pure $CO_2$, and was extrapolated at lower pressure. The values are summarized in Table 1.

*Table 1: O loss probabilities from [89] obtained experimentally, in pure $CO_2$, for a discharge current of 50 mA and used in our calculations. *Extrapolated value.*

| Pressure (Torr) | O loss probability $\gamma_O$ |
|---|---|
| 0.6 | $8.87 \cdot 10^{-4}$* |
| 0.8 | $6.50 \cdot 10^{-4}$ |
| 1 | $5.33 \cdot 10^{-4}$ |
| 2 | $4.50 \cdot 10^{-4}$ |
| 3 | $4.91 \cdot 10^{-4}$ |
| 4 | $5.33 \cdot 10^{-4}$ |

Overall, the set of measurements provides the gas temperature, vibrational temperatures of CO and the various modes of $CO_2$, the reduced field E/N, the fraction of atomic oxygen, [O]/N, and the densities of $O(^3P)$, $CO(X^1\Sigma^+)$, $CO_2(X^1\Sigma^+_g)$ and $NO(X^2\Pi_r)$ which are compared with the simulation results for the validation of the model described in section IV. The density of the gas, N, is calculated from the ideal gas law with the pressure and gas temperature obtained with the FTIR measurements.

## III. Model

The development and validation of self-consistent models constitute a powerful tool to access quantities that are not available experimentally, to obtain insight into the processes occurring in the plasma, to interpret the experimental results and to predict the behaviour of important physical and chemical quantities in different plasma systems. In particular, 0D kinetic models can describe a detailed plasma chemistry with a significant number of processes included and therefore can provide very useful information on the main source and loss mechanisms of the species of interest like CO and NO. However, the reliability of these models depends strongly on the accuracy of the cross sections and/or rate coefficients used to describe the energy transfers, electron-impact mechanisms and heavy-particle chemical reactions.

The calculations are done with the LisbOn KInetics (LoKI) simulation tool developed in-house under MATLAB® and solving a Boltzmann-chemistry global model [92,93]. Since electrons play an important role in low-temperature plasmas, an accurate description of the electron energy distribution function (EEDF) and macroscopic electron parameters, including rate coefficients of electron-impact processes, is required. This is achieved here using the open-source Boltzmann solver LoKI-B, which numerically solves a space-independent form of the two-term electron Boltzmann equation (EBE) for non-magnetised non-equilibrium low-temperature plasmas, excited by DC or HF electric fields [92] or time-dependent (non-oscillatory) electric fields [94], in different gases or gas mixtures. Moreover, to get a self-consistent description of both electron and heavy species, the homogeneous EBE is coupled with the Chemical solver LoKI-C corresponding to a system of zero-dimensional state-to-state rate-balance equations describing the creation and loss of the most important neutral and charged heavy-particles. The relevant information regarding the model is summarized in this section and a figure of the workflow can be found in [42].

The electron, chemical and vibrational kinetics are coupled into a self-consistent scheme for which the reduced electric field, E/N, corresponds to steady-state conditions where the total rate of production of electrons in ionization events compensates exactly their total loss rate due to ambipolar diffusion to the wall and electron-ion recombination or, in other words, when the densities of negative and positive charges are the same, satisfying the quasi-neutrality condition. Moreover, it is guaranteed, thanks to the coupling between LoKI-B and LoKI-C, that the EEDF and electron parameters are obtained taking into account the self-consistently calculated concentrations of the parent gas and for a value of reduced electric field also self-consistently obtained. Vice-versa, all the rate coefficients of the electron-impact reactions depend on this self-consistent EEDF.

The electron density, $n_e$, is such that the current (I) given as input calculated with:

$$I = \pi R^2 |q_e| n_e v_e, \qquad (2)$$

must give the experimental current intensity of 50 mA; where $q_e$ is the elementary charge of electrons and $v_e$ is the electron drift velocity obtained from the EBE solution. The diffusion of neutral species, including the vibrationally excited species, is taken as in [34,95], and is based on the formula of Chantry [96] to obtain the loss rate of species interacting with the wall due to the combined effect of transport (due to diffusion) and reaction at the wall (with a certain wall recombination/deactivation probability γ). The diffusion of charged-particles is described by classical ambipolar diffusion [95] and the effect of the negative ions on the electron density radial profiles is taken into account [97,98]. The gas flow is used as input parameter to the model, where we assume a simple mechanism of renewal of gas: new $CO_2$/$N_2$ particles enter the reactor while the species produced in the plasma exit at the outlet assuming conservation of atoms in the plasma mixture [42].

Together with the discharge current (I) given in (2), the input parameters of the model are the gas pressure (P), the flow, the initial gas mixture and reactor dimensions (tube length and radius). The loss probability of O atoms at the wall, $\gamma_O$, is also an input parameter and is deduced from the experimental determination of O atom loss frequencies (cf. Table 1). Additionally, in the present simulations the gas temperature obtained experimentally is also used as input parameter. The self-consistent calculation of $T_g$ is possible, using a thermal balance equation [35,95,99] but the study of heat transfer mechanisms is not the focus of this work.

### III.1 Chemistry

Most of the required electron scattering cross-section data can be found at the open-access website LXCat [100,101]. For this work, we use the IST-Lisbon database [100] including complete sets of electron scattering cross sections (CS) with neutral species in the ground-state (GS) based on [95,102,103] for $N_2$, [104] for N, [38,40] for $CO_2$, [41] for CO, and [105,106] for $O_2$ and O. They are defined as sets of cross sections giving a good description of the main electron energy and momentum losses, yielding electron swarm parameters in agreement with available measured data, when used in a two-term Boltzmann solver [100]. In $CO_2$-$N_2$ plasmas, CO, $O_2$ and O may be present with significant concentrations, therefore, besides $CO_2$ and $N_2$, cross sections for electron impact on these three plasma species are of importance. The CS describe the elastic momentum-transfer due to electron collision with the GS, the excitation of electronic and vibrational states from GS and ionisation. The sets of electron scattering cross sections from ground-states $CO_2$, $N_2$, CO and $O_2$ are complemented by additional collisional data, essentially for collisions from their vibrationally excited states, as well as from some of their electronically excited states. They can be found in the IST-Lisbon database [100] but are not part of the complete sets. The list of all the e-V processes and corresponding cross sections for $CO_2$ are reported in the Supplementary Information of [38]. Notice that the superelastic processes are considered and the corresponding cross sections are deduced from the direct processes using the Klein-Rosseland relation [107]. Due to the lack of data, the dissociation cross sections via electron impact from vibrationally excited states are considered with a threshold shift, while keeping the same amplitude as for dissociation from the ground-state [31]. The same procedure is used for ionization from vibrationally excited CO and $CO_2$ molecules as well as attachment for $CO_2$.

The vibrational excitation of oxygen is included in the calculation of the EEDF, assuming a Boltzmann distribution at $T_g$ for vibrational levels up to 4, but not in the Chemistry module as a strong depopulation of the Vibrational Distribution Function (VDF) in the first vibrational levels was observed experimentally [108] and in the calculations of [109,110]. Vibrationally excited $O_2$ molecules do not seem to have a direct influence neither in the EEDF nor in the chemistry [111] but may have a relevant contribution to gas heating, due to V–T deactivation by atomic oxygen, affecting indirectly the discharge [98].

For the description of the rotational excitation and deexcitation of the ground-state $N_2$ and $O_2$ by electron impact, we adopt the continuous (CAR) approach, using the GCAR rotational operator [95,112], whereas a discrete approach that considers a set of cross sections for rotational transitions involving the states CO ($X^1\Sigma^+$, v = 0, J = 0-17) published also in the IST-Lisbon database is used for CO.

The species considered in the plasma chemistry of the model include: ground-state, vibrationally and/or electronically excited CO, $CO_2$, $O_2$, $N_2$, NO, $NO_2$ and CN molecules $CO(X^1\Sigma^+,v=0{:}10)$, $CO(a^3\Pi_r)$, $CO_2(X^1\Sigma^+_g, v_1^{max}=2; v_2^{max}=v_3^{max}=5)$, $O_2(X^3\Sigma_g^-)$, $O_2(a^1\Delta_g)$, $O_2(b^1\Sigma_g^+)$, $O_2(A'^3\Delta_u, A^3\Sigma_u^+, c^1\Sigma_u^-)$, $N_2(X\ ^1\Sigma_g^+,v=0{:}59)$, $N_2(A\ ^3\Sigma_u^+)$, $N_2(B^3\Pi_g)$, $N_2(C\ ^3\Pi_u)$, $N_2(w\ ^1\Delta_u)$, $N_2(a\ ^1\Pi_g)$, $N_2(a'\ ^1\Sigma_u^-)$, $NO(X\ ^2\Pi_r)$, $NO(A\ ^2\Sigma^+)$, $NO(B^2\Pi_r)$, $NO_2(X^2A_1)$, $NO_2(a^4A_2)$, $CN(X^2\Sigma^+)$ and $CN(B^2\Sigma^+)$; ground-state and electronically excited oxygen and nitrogen atoms, $O(^3P)$, $O(^1D)$, $N(^4S)$, $N(^2D)$, $N(^2P)$, ground-state carbon $C(^3P)$, ground-state ozone and vibrationally excited ozone, $O_3$, $O_3^*$; and positive and negative ions, $O^+$, $O_2^+$, $O^-$, $CO_2^+$, $CO^+$, $N_2^+$, $N_2^+(B\ ^2\Sigma_u^+)$, $N^+$, $N_4^+$, $N_3^+$ and $NO^+$. For $O_3^*$ we consider a single effective vibrationally excited state [113]. $O_2(A'^3\Delta_u, A^3\Sigma_u^+, c^1\Sigma_u^-)$ is an effective sum of the Herzberg states. Our model does not take into account directly the kinetics of $N_2(W^3\Delta_u)$. However, it is considered indirectly in the kinetics of the $N_2(B^3\Pi_g)$ state since the collisional data measured in flow systems actually provide the relaxation coefficients for the coupled states as a whole [114]. Besides, the $N_2(B^3\Pi_g)$ state is populated by the radiative cascade $N_2(B'\ ^3\Sigma_u^-) \rightarrow N_2(B^3\Pi_g)$, on the assumption that the $N_2(B'\ ^3\Sigma_u^-)$ is only populated by electron impact and radiatively depopulated (i.e., the populating rate of the $N_2(B^3\Pi_g)$ state due to this mechanism is that for creating the $N_2(B'\ ^3\Sigma_u^-)$ by electron impact) [114,115]. The rate coefficients of heavy-particle reactions are adopted from literature and based on the kinetic schemes of $O_2$ [98] based on [111,116,117], $CO_2$ [42], $N_2$ [114], $N_xO_y$ [95] and CN [118] while the rate coefficients of the electron impact reactions are calculated from the energy-dependent cross sections, in combination with the EEDF. For the kinetics of oxygen, the set proposed in [98] is adopted without modifications except for the exclusion of vibrational states in the heavy-species chemistry. We also added the three-body reaction:

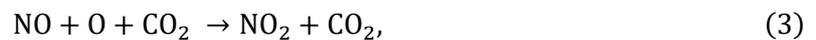

$$\text{NO} + \text{O} + \text{CO}_2 \rightarrow \text{NO}_2 + \text{CO}_2, \tag{3}$$

with the rate coefficients taken from [68].

### III.1.1 Quenching of $N_2(A)$ and $N_2(B)$

Reactions involving $N_2(A\ ^3\Sigma_u^+)$, referred to as $N_2(A)$, and the triplet levels $N_2(B'\ ^3\Sigma_u^-)$, $N_2(B^3\ \Pi_g)$, and $N_2(W\ ^3\Delta_u)$, considered here as a single effective state $N_2(B)$ [119], are relevant for the study of active nitrogen, shock waves and flames [120,121]. Several authors claim that the presence of these species improves the $CO_2$ dissociation [66–68,71,121] and therefore these studies regained interest more recently for the study of $N_2$-containing $CO_2$ discharges. According to [66,121], the most important electronically excited nitrogen state for $CO_2$ conversion is $N_2(B)$ with the rate coefficients being roughly gas kinetic [122] for $N_2(B)$ quenched by $CO_2$ and CO, as can be seen in Table 2, where a brief survey of reported rate coefficients for reactions involving $N_2(B)$ and $CO_2$ or CO is given.

The rate coefficient for $CO_2$ dissociation due to quenching with $N_2(A)$ is much smaller than with $N_2(B)$ and is two orders lower than those by products of $CO_2$ dissociation like CO [123]. Indeed, CO is a more efficient acceptor of the electronic energy from $N_2(A)$ [124–127]. Several rate coefficients found in the literature for the quenching of $N_2(A)$ with CO and $CO_2$ are compiled in Table 3.

The relative importance of $N_2(A)$ and $N_2(B)$ metastable states for $CO_2$ dissociation seem to depend on the operating conditions. For discharges with high electron energy like DBD $N_2(A)$ was claimed to be important [67,71] for $CO_2$ dissociation while for a microwave discharge, with lower electron average energy, it was shown that $N_2(A)$ was of minor importance [68]. However, the study done in the DBD [67] used a value for the rate coefficient for the $N_2(A)$ quenching by $CO_2$ taken from [71], which is two orders of magnitude higher than what is commonly found in literature (see Table 3).

It is worth noting that in the collisional quenching of electronically excited $N_2$ molecules the branching ratio of the quenching products remains uncertain and can affect the modeling predictions in the most significant way [80]. Finally, it is necessary to note that an important feature of many reaction involving $N_2(A,B)$ is a dependence of the rate coefficients and products ratios on the vibrational levels [120,122,123,128] and that the actual measurements of rate coefficients can be a mix of vibrational levels [123].

| process | [120,129] | [71,121,129] | [66,122] | [71] | [120] |
|---|---|---|---|---|---|
| $N_2(B) + CO_2 \rightarrow N_2+CO+O$ | $1.5 \cdot 10^{-10}$ (a) | $8.5 \cdot 10^{-11}$ (b) | | | |
| $N_2(B) + CO_2 \rightarrow$ products | | | **$2 \cdot 10^{-10}$** (a) | | |
| $N_2(B) + CO_2 \rightarrow N_2+CO_2$ | | | | $1 \cdot 10^{-11}$ | |
| $N_2(B) + CO_2 \rightarrow N_2+CO_2$ | | | | $1.9 \cdot 10^{-10}$ | |
| $N_2(B)+CO \rightarrow$ products | | | **$2 \cdot 10^{-10}$** (a) | | $8.5 \cdot 10^{-11}$ |

*Table 2: Rate coefficients in $cm^3 s^{-1}$ for processes involving the quenching of $N_2(B)$, measured at (a) 300 K and (b) 196 K. In bold, values used in the model for section IV.2.*

| process | [71] | [123] | [126] | [120] |
|---|---|---|---|---|
| $N_2(A) + CO_2 \rightarrow N_2+CO+O$ | $1.54 \cdot 10^{-12}$ | | | |
| $N_2(A) + CO_2 \rightarrow$ products | | **$1.98 \cdot 10^{-14}$** (c) | | |
| $N_2(A) + CO_2 \rightarrow N_2+CO_2$ | $9.9 \cdot 10^{-15}$ | | | |
| $N_2(A) + CO \rightarrow$ products | | **$1.7 \cdot 10^{-12}$** (c) | $2.5 \cdot 10^{-11}$ | $2.5 \cdot 10^{-12}$ |

*Table 3: Rate coefficients in $cm^3 s^{-1}$ for processes involving the quenching of $N_2(A)$, (c) measured at 298 K. In bold, values used in the model for section IV.2.*

The quenching processes of $N_2(A)$ and $N_2(B)$ are investigated in section IV.2. The values of the rate coefficients used for this study corresponding to the quenching of $N_2(A)$ with $CO_2$ and CO are, respectively, $1.98 \cdot 10^{-14}$ cm$^3$s$^{-1}$ and $1.7 \cdot 10^{-12}$ cm$^3$s$^{-1}$ [123] and are assumed to occur via the following processes:

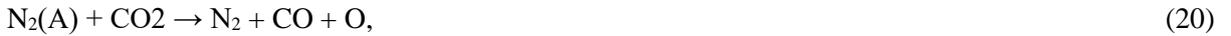
$$N_2(A) + CO2 \rightarrow N_2 + CO + O, \quad (20)$$

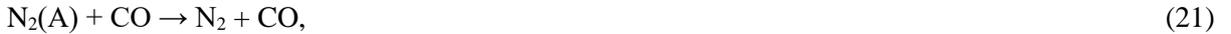
$$N_2(A) + CO \rightarrow N_2 + CO, \quad (21)$$

For the quenching of $N_2(B)$ with $CO_2$ and CO, considered as follows:

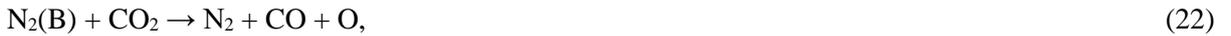
$$N_2(B) + CO_2 \rightarrow N_2 + CO + O, \quad (22)$$

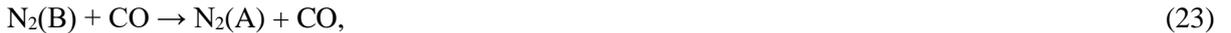
$$N_2(B) + CO \rightarrow N_2(A) + CO, \quad (23)$$

we use the values reported in Piper [122] and used as well in Naidis & Babaeva [66], namely $2 \cdot 10^{-10}$ cm$^3$s$^{-1}$ for both processes.

### III.2 Vibrational kinetics

The vibrational kinetics include excitation/de-excitation mechanisms due to electron-impact collisions (e-V), energy transfer via vibrational–translational (V-T), vibrational–vibrational (V-V) interactions and wall loss of vibrational quanta (W). The VDFs of $CO_2$, $N_2$ and CO are obtained from the solution of the rate-balance equations describing the creation and loss of each individual vibrational level, accounting for these e-V, V-V, V-T, W processes, as well as chemical reactions like electron-impact dissociation from vibrationally-excited molecules and dissociation in pure vibrational mechanisms. The pure $CO_2$ subsystem was recently addressed in [38], where expressions for the V-T and V-V involving $CO_2$, CO, $O_2$ and O, rate coefficients are suggested. The vibrational energy transfers in the nitrogen system have been extensively studied in the past, due to the key role of vibrations in nitrogen plasmas [95,97,109,110,114,115,130]. The rate coefficients used for the pure $N_2$ subsystem are given in [34,95] and the calculation of the V-T and V-V rate coefficients were benchmarked from the calculations of Billing *et al.* [131,132]. The reader is referred to these papers for further details and figures representing these rate coefficients. The rate coefficients were scaled using either the Schwartz, Slawsky, and Herzfeld (SSH) theory [133] (SSH) or the theory from Sharma and Brau (SB) [134,135] (SB), which are summarized in [32] for the case of $CO_2$. A comprehensive description of the SSH scaling law can be found in [16]. In regards of rate coefficients used for inverse reactions, we have used the principle of detailed balance [136] for the various V-T and V-V transitions. We only consider dissociation by electron impact from vibrationally excited $CO_2$ molecules as the higher levels are not significantly populated in our conditions. In other words, we do not consider dissociation through the pure vibrational path ($CO_2(v^*) + CO_2 \rightarrow CO + O + CO_2$), but only stepwise dissociation by electron impact on vibrationally excited $CO_2$. The validity of this assumption is shown and discussed in [38,42].

In the following subsections, we start by discussing the description of the vibrational levels considered and the e-V processes. We then review our current choice of rate coefficients for the relevant mechanisms in $CO_2$-$N_2$ plasmas, namely $CO_2$-$N_2$, $CO_2$-CO and $N_2$-CO V-V exchanges and different V-T processes. Finally, we report the probabilities of deactivation of vibrationally excited species and atomic oxygen recombination at the walls.

### III.2.1 Energy levels and e-V processes

The model considers 59 vibrational levels of $N_2$, up to the energy of ~10 eV. They are included in the model following the formula given in [137]:

$$\frac{E_{N_2}}{hc} = \sum_{i=1}^{5} a_i \left(v + \frac{1}{2}\right)^i, \tag{4}$$

with the coefficients $a_i$ in cm$^{-1}$ equal to 2378.1, -18.516, 0.26662, $-6.2127 \cdot 10^{-3}$ and $3.4215 \cdot 10^{-5}$, respectively, for i=1-5. The energies of the individual $CO_2$ vibrational levels are calculated according to the anharmonic oscillator approximation [138] and more details can be found in [38]. The CO vibrational energy levels are calculated by the formula of the anharmonic Morse oscillator [114] as done in [38]. By convention, the ground-state energy is set to 0 eV for $CO_2$, CO and $N_2$.

As carefully described in [33,40] and briefly reviewed in [38,99], the cross sections for electron-impact excitation (e-V) of $CO_2$ vibrations are obtained from a direct deconvolution of the available lumped cross sections, according to the statistical weights of the various levels, and the missing cross sections are generated following the approach from [38], based on a modified form of the semi-empirical Fridman's approximation [21]. For the electron-impact excitation of the CO vibrations, we have adopted the cross sections from [139] for the vibrational excitation and de-excitation which are largely based on resonant excitation data from Laporta and co-workers [140], and where contributions from non-resonant collisions for the transition e + CO(*v*=0) ↔ e + CO(*v*=1), taken from [141], are also included. Finally, the e–V mechanisms for $N_2$ are described as follows: (i) in the electron kinetics calculations, e.g. when solving the electron Boltzmann equation, by adopting electron-scattering cross sections for the reactions e + $N_2$(X,v) → e + $N_2$(X,v′) (1 ≤ v ≤ v′ ≤10), obtained by applying a threshold shift to the excitation cross sections for the corresponding vibrational transitions from the ground-state $N_2$(X,v = 0); (ii) in the chemistry model calculations, by using electron rate coefficients satisfying the scaling law:

$$k_{v,v+v'} = \frac{k_{0,v'}}{1 + 0.15v}, \qquad (5)$$

For $1 \leq v' \leq 10, 1 \leq v \leq 59$ and where $k_{0,v'}$ are the rate coefficients for vibrational transition from $N_2$ in the ground state to the first 10 vibrational states. This scaling law given by Colonna in [142] was validated in [103] and used in [34,95,143].

### III.2.2 $CO_2$-$N_2$ and $CO_2$-CO V-V exchanges

The transfer between vibrationally excited nitrogen and the asymmetric stretching mode of $CO_2$, is very efficient and may potentially have a positive effect on $CO_2$ dissociation via the ladder climbing mechanism [21]. $CO_2$ is known to couple very strongly to $N_2$ via

$$N_2(0) + CO_2(00^011) \leftrightarrow N_2(1) + CO_2(00^001), \qquad (6)$$

since the energy difference between the first vibrationally excited levels is only 18 cm$^{-1}$ [135]. For details on the notation used for the $CO_2$ vibrations, the reader is referred to [32,37,38]. The hypothesis regarding the selective excitation of the upper laser level of $CO_2$ lasers by resonant transfer of vibrational energy from $N_2$ molecules, necessary to obtain the population inversion in the $CO_2$-$N_2$ laser system, is widely accepted by the scientific community [82,83,85]. Moreover, even though the equivalent reaction between $CO_2$ and CO is not as close to energy resonance as for $N_2$, CO molecules can transfer a considerable amount of energy to the $v_3$ vibration via (7) because the difference between the energies of the vibrational level of CO and the ($00^011$) level of $CO_2$ is 170 cm$^{-1}$ which is typically smaller than the average kinetic energy kT [81],

$$CO(0) + CO_2(00^011) \leftrightarrow CO(1) + CO_2(00^001). \qquad (7)$$

The importance of these two processes comes from the large populations of vibrationally excited CO and $N_2$ molecules present in the discharge, most likely due to direct electronic excitation, followed by the transfer of energy to the vibrational levels of $CO_2$. This is supported by experimental data on the cross sections of collisions between electrons and heavy species (CO and $N_2$), reported in [144]. Indeed, according to Schulz, the electron-impact excitation cross sections of the vibrational levels of CO and $N_2$ are unusually large because of the resonance effect of the short-lived negative CO$^-$ and $N_2^-$ ions [145]. The excitation cross section of the first levels of CO and $N_2$ ($\sim 5 \cdot 10^{-16}$ cm$^{-2}$) is one order of magnitude higher compared to the excitation of $CO_2$ in the first asymmetric stretch mode level by electron impact [33,40], for electrons of 2 eV energy, typical of our discharge.

Since the energy defect is about 10 times larger for CO than for $N_2$ it might be anticipated that CO is not as effective as $N_2$ in exciting the vibration of $CO_2$. Furthermore since the CO molecule has a dipole moment, it has also a spontaneous decay whereas $N_2$ can only decay via collisions with the wall or with other molecules [85]. It appears, however, that the additional dipole-dipole attractive forces between $CO_2$ and CO more than compensate for this disparity, as the $CO_2$-CO transfer can even be more efficient than the $CO_2$-$N_2$ transfer. This could come from the observation that near resonant V-V transfer processes are more likely to occur when the vibrational modes involved are infrared active [146].

The $CO_2$-CO V-V rate coefficient comes from our previous work [38] and is based on [147], while the rate coefficient for $CO_2$-$N_2$ V-V used in the present work is taken from [148]. We compare in Figure 1 the rate coefficients used in this work for the processes (6) and (7) with experimental data available in literature obtained with laser-excited vibrational fluorescence and shock tube techniques by Taylor *et al.* [149], Rosser *et al.* [150], Taylor & Bitterman [151], Stephenson *et al.* [146], Moore *et al.* [152], Starr & Hancock [153], Rosser *et al.* [154], Blauer and Nickelson [148]. Although there are considerable data for $CO_2$-$N_2$, a large scatter can be observed. Above 1000 K there are similarities between the $CO_2$-CO and $CO_2$-$N_2$ V-V rate coefficients as might be expected based on the similarities between $N_2$ and CO molecules. At lower temperatures, however, the situation may seem puzzling.

The positive temperature dependence for the $CO_2$-CO V-V rate coefficient is the reverse of that found for $CO_2$-$N_2$. It is clear that the Schwartz, Slawsky, and Herzfeld (SSH) theory [133] cannot explain the low-temperature behaviour for $CO_2$-$N_2$, as this theory predicts an increase of the reaction rates with temperature. Sharma and Brau (SB) [134,135] have explained the negative temperature dependence of the $CO_2$-$N_2$ V-V energy exchange, at low temperatures, on the basis of the long range force between the dipole moment of $CO_2$ and the quadrupole moment of $N_2$, being more important than the short range repulsive forces normally used in vibrational relaxation theories. Such an interaction is more important at low temperatures and for processes of near energy resonance.

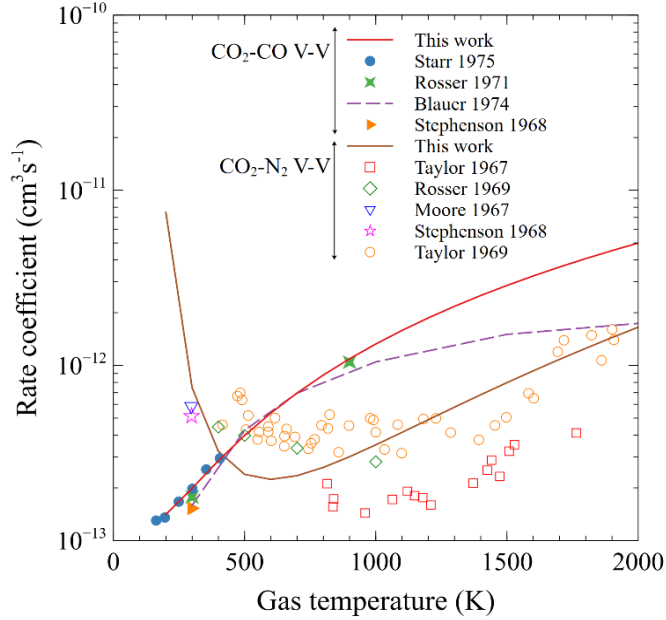

*Figure 1: Comparison of the rate coefficients used in this work with experimental data for the near-resonant V-V exchange between the mode $v_3$ of $CO_2$ and $N_2$ and CO, $CO_2(00^011) + N_2/CO(v=0) \rightarrow CO_2(00^001) + N_2/CO(v=1)$.*

Only a few vibrational levels of the $CO_2$ asymmetric stretching mode are considered in the present model ($v_3 \leq 5$) and for simplicity the rate coefficients for the process

$$N_2(w_i) + CO_2(00^0v_i1) \rightarrow N_2(w_f = w_i + 1) + CO_2(00^0(v_f = v_i - 1)1), \quad (8)$$

can be considered with the same value as for the low lying levels ($i.e.\, w_i = 0, v_i = 1$) as done in [34]. However, in this work, we scale the rate coefficients for the upper levels since they have a noticeable influence on the results, especially the vibrational temperatures, as verified in section IV (cf. Fig. 12). The rate coefficient for the process (6) with $w_i = 0, v_i = 1$, is described by:

$$k(cm^3s^{-1}) = 1{,}66 \cdot 10^{-24} \cdot \exp(a + b \cdot T^{-1/3} + c \cdot T^{-2/3}), \quad (9)$$

where a = 43.8, b = −306 and c = 1288 [148] and is represented in Figure 1. The rate coefficient of the reverse process is calculated from the principle of the detailed balance [136], by multiplying the coefficients of the direct process with the Boltzmann factor, $\exp\left(\frac{-\Delta E}{T_g}\right)$, where $T_g$ is the gas temperature and ΔE the energy difference in K.

To be able to use the SB and SSH theories separately for the scaling of the rate coefficient from [148] (plotted in Fig. 1), we separate it into two parts, one for the long range corresponding to the 'SB part', $L_{SB}$, and one for the short range corresponding to the 'SSH part', $S_{SSH}$. To facilitate the scaling of rate coefficients at low and high temperatures with both theories, the function shown in (9) is then split into SSH and SB parts. The sum of these contributions ($L_{SB}+S_{SSH}$) for the mechanism $N_2(0) + CO_2(00^011) \rightarrow N_2(1) + CO_2(00^001)$ is given in figure 2 to show that we can recover the original rate coefficient for the process (6) after the fitting procedure. Note that in this case we have $a_{SB}$ = -6.28496, $b_{SB}$ = 379.838, $c_{SB}$ = -1059.2, for $L_{SB}$, and $a_{SSH}$ = 27.221, $b_{SSH}$ = 10.8178, $c_{SSH}$ = -224.158 for $S_{SSH}$. We then scale separately the rate coefficients with SB and SSH and add the two contributions to recover the characteristic 'U shaped' curve. All the rate coefficients are given in the Supplementary Information and a comprehensive description of the SSH scaling law can be found in [16].

According to the SB theory and retaining only the terms involving the vibrational levels of $CO_2$ and $N_2$ (and not the rotational distributions) for the only non-vanishing contribution, namely the interaction of the $N_2$ quadrupole moment, Q, with the $CO_2$ dipole moment, μ, an analytical expression for the probability of the transition (8) is [155]:

$$\frac{k_{v_i,w_i}^{v_f,w_f}}{k_{1,0}^{0,1}} = \frac{|<w_f|Q|w_i>|^2 * |<v_f|\mu|v_i>|^2 * I(z(\Delta E))}{|<1|Q|0>|^2 * |<0|\mu|1>|^2 * I(z(\Delta E_{10}))}, \quad (10)$$

with $k_{1,0}^{0,1}$ the rate coefficient for the forward direction of the transition (1) and $|<w_f|\mu/Q|w_i>|^2$ the matrix elements of the dipole moment of $CO_2$ (μ) and quadrupole moments of $N_2$ (Q). They can be approximated by the harmonic oscillator relation [155]:

$$|<w_f = n|\mu/Q|w_i = n-1>|^2_{CO_2/N_2} = n * |<w_f = 1|\mu/Q|w_i = 0>|^2_{CO_2/N_2}, \quad (11)$$

$I(z)$ is the resonance function defined as follows [155]:

$$I(z) = \exp(-z)(0.1339 + 01223z + 0.1477z^2 - 0.0283z^3 + 0.0078z^4 - 0.0007z^5), \quad (12)$$

with $z = \Delta E d \left(\frac{M}{2k_B T_g}\right)^{1/2}$ and $\Delta E = |E_f - E_i|/\hbar$ is the energy mismatch, d is the average of the hard sphere diameters of the two molecules, $E_i$ and $E_f$ the total energies of the initial states and final states, respectively, of $CO_2$ and $N_2$, M the reduced mass of the collision pair and $\Delta E_{10}$ is the energy mismatch in process (6). By substitution of (11) into (12), we obtain:

$$\frac{k_{v_i,w_i}^{v_f,w_f}}{k_{1,0}^{0,1}} = \frac{w_f * |<1|Q|0>|^2 * v_i * |<0|\mu|1>|^2 * I(w(\Delta E))}{|<1|Q|0>|^2 * |<0|\mu|1>|^2 * I(w(\Delta E_{10}))} = k_{1,0}^{0,1} * \frac{w_f * v_i * I(w(\Delta E))}{I(w(\Delta E_{10}))}. \quad (13)$$

The SB theory is only used for near resonant processes as the resonance function, I(w), has a fast decay with increasing $\Delta E$. Considering that the resonance function for values above 50 cm$^{-1}$ provides ~$8 \cdot 10^{-3}$, which leads to rate coefficients below $6 \cdot 10^{-14}$cm$^3$s$^{-1}$, we used the SB theory only for the processes which fulfil $\Delta E < 50\ cm^{-1}$. A few rate coefficients are given in Figure 3. When $\Delta E > 50\ cm^{-1}$ we only consider the $S_{SSH}$ contribution to the rate coefficient and we scale using the harmonic oscillator.

Our scaling approach is only valid for 300 < $T_g$ < 1000 K as the fitting of $L_{SB}$ and $S_{SSH}$ with expression (9) fails outside of this range. Besides, the V-V transfer involving vibrationally excited $N_2$ molecules and the bending and symmetric modes of $CO_2$ is not included in the model as the corresponding rate coefficients are three orders of magnitude lower than the transfer with the asymmetric stretching mode in the gas temperature range of interest [300 K; 1000 K] [156,157].

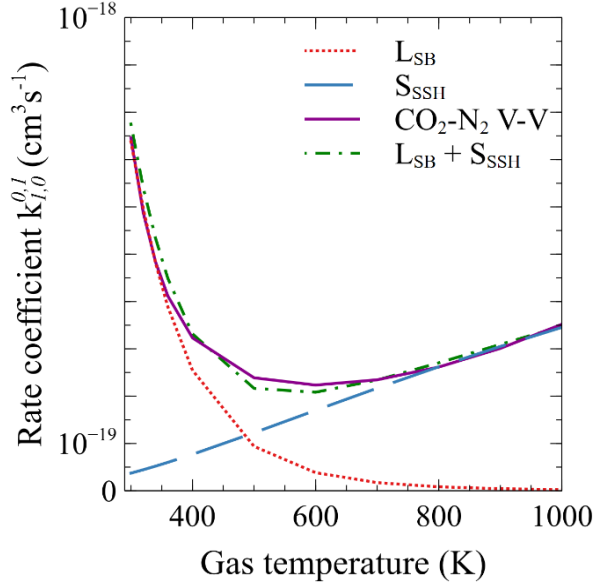

*Figure 2: Fitting of the low temperature (⋯) and high temperature ranges (--) of the rate coefficient corresponding to the $CO_2$-$N_2$ V-V transfer, $N_2(0) + CO_2(00^011) \rightarrow N_2(1) + CO_2(00^001)$ (—) from [148] and the sum of these two contributions ($L_{SB}+S_{SSH}$) (-·-).*

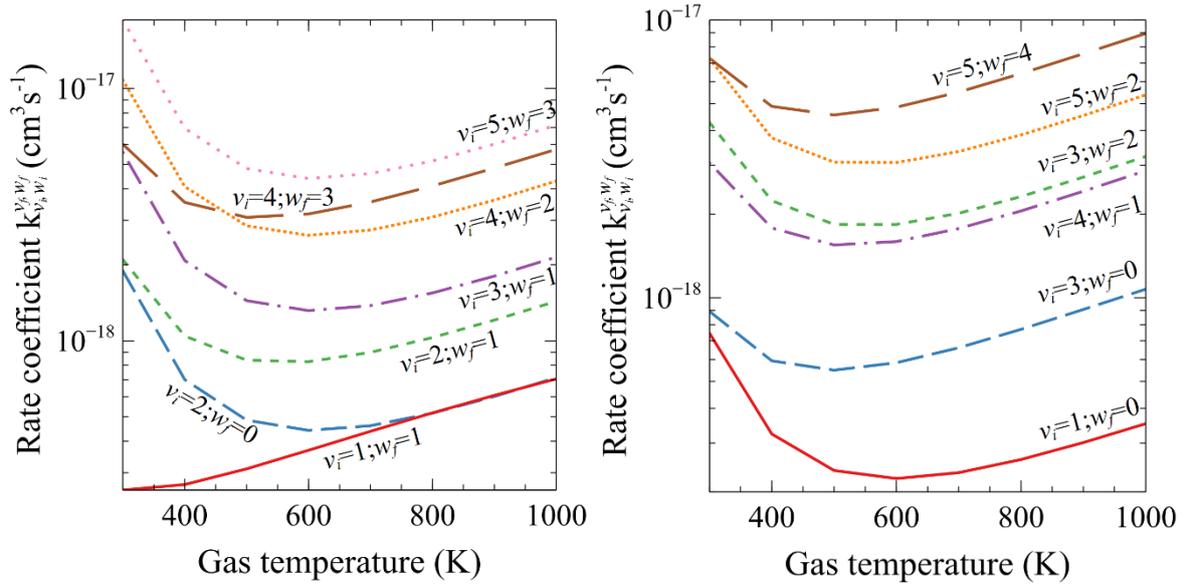

*Figure 3: Rate coefficients ($k_{v_i,w_i}^{v_f,w_f}$) of the process $N_2(w_i) + CO_2(00^0v_i1) \rightarrow N_2(w_f = w_i + 1) + CO_2(00^0(v_f = v_i - 1)1)$ scaled with the procedure described in this section as a function of $T_g$.*

### III.2.3 $N_2$-CO V-V exchanges

The V–V exchange $N_2(1) + CO(0) \rightarrow N_2(0) + CO(1)$ has an equivalent importance to the V–V exchange $N_2(1) + CO_2(00^001) \rightarrow N_2(0) + CO_2(00^011)$ due to the high density of vibrationally excited CO and $N_2$ molecules in the plasma, despite a lower rate coefficient. To obtain accurate rate coefficients we use the analytical expression from Kurnosov *et al.* [158] which seems to be a good approximation to experimental and semiclassical rate coefficients for a large range of near resonant vibrational transitions in the $N_2$–CO system. This expression takes into account short range repulsive and van der Waals forces (first term) as well as quadrupole-quadrupole interaction (second term):

$$k(v, u | v - 1, u + 1; T_g) = a \cdot Z \cdot T_g \cdot z(v) \cdot z(u + 1) \cdot \exp\left(\frac{\Delta E}{2T_g}\right) \cdot F(y) \cdot F_S +$$

$$\frac{b}{T_g} \cdot Z \cdot z(v) \cdot |\langle u + 1|q|u\rangle|^2 \cdot \exp\left(\frac{\Delta E}{2T_g}\right) \cdot \exp\left(-\frac{\Delta E^2}{CT_g}\right), \qquad (20)$$

with $Z = (\pi\sigma^2)V_M$ the gas kinetic collision rate coefficient, with $V_M$ the average relative translational velocity, $\pi\sigma^2$ the gas kinetic collision cross section. $\Delta E$ is defined as $[E_{CO}(u+1) + E_{N_2}(v-1)] - [E_{CO}(u) + E_{N_2}(v)]$ and should be in Kelvin. $z(v) = \frac{v}{1-v\cdot\chi_e}$, with the anharmonicity factor $\chi_{e_{CO}} = 0.00612$ and $\chi_{e_{N_2}} = 0.00607$ and $F(y)$ and the argument of the adiabatic function $y$ are defined as (18) and (19). $F_S$ is Shin's factor

$$F_S = \exp\left[\frac{4}{\pi \cdot \sqrt{T^*}} \cdot y^{\frac{1}{3}} + \frac{16}{3 \cdot \pi^2 \cdot T^*}\right], \qquad (21)$$

with $T^* = \frac{T}{\varepsilon_{ST}}$, $|\langle u + 1|q|u\rangle|^2$ is the matrix element of the quadrupole moment of N$_2$ and $\varepsilon_{ST}$ is the well depth taken to be equal to the Lennard-Jones potential well depth. The semi-empirical parameters in equation (20) were obtained by fitting the rate coefficients from [159,160] in the temperature range 80 K < $T_g$ < 700 K and assuming that $|\langle u + 1|q|u\rangle|^2 \approx (u+1)$, $a = 6.6 \times 10^{-8} K^{-1}, b = 0.04\, K, C = 145\, K$. According to [158], the best-fitted interaction length L used in $y$ is equal to 0.185 Å. With this value, equation (20) adequately reproduces all the semiclassical data for resonant and non-resonant transitions [158].

The various coefficients are fitted through the exponential expression (9) and the fitting parameters can be found in the Supplementary Information. In Figure 4, we represent the rate coefficients obtained in this work following the procedure from Kurnosov *et al.* [158] for the process N$_2$(1) + CO(0) → N$_2$(0) + CO(1) as a function of the gas temperature. Figure 4 shows as well the rate coefficients calculated using a mixed quantum-classical method by Hong *et al.* [161], a fitting of experimental data from Sato *et al.* [162] found in a survey of vibrational relaxation rate coefficients from Blauer & Nickerson [148], calculated by Shin [163] and other rate coefficients determined experimentally by Zittel & Moore [164], Stephenson & Mosburg [165], Starr *et al.* [166], Allen & Simpson [159] Green & Hancock [167] and Mastrocinque *et al.* [160] and found a good agreement with the experiments. Our rate coefficients are valid for $T_g$ < 1000 K. For higher gas temperatures the Quantum-classical rate coefficients calculated by Hong *et al.* [161] should be used instead, as they are in good agreement with the experimental data from Sato *et al.* [162] (not shown here) measured for $T_g$ > 1600 K.

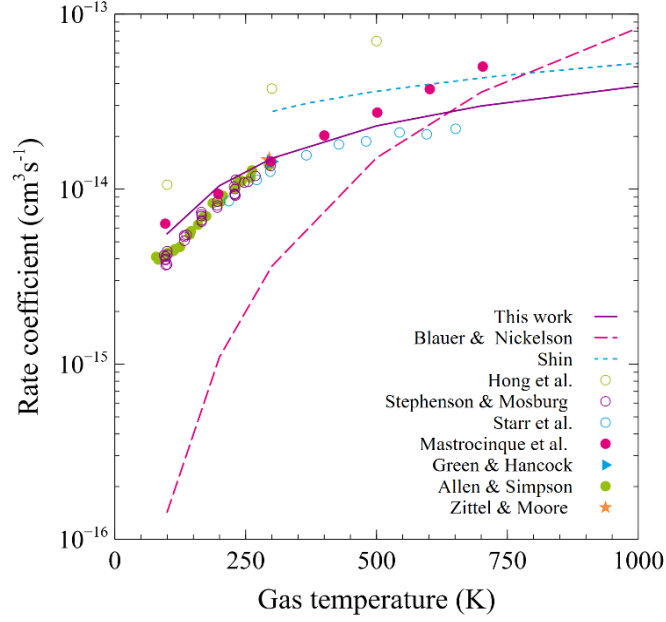

*Figure 4: Comparison of the rate coefficients used in this work (—) with experimental data (symbols), fitting at high $T_g$ (— —) and calculation (--) for the near resonant VV exchange between $N_2$ and CO, $N_2(1) + CO(0) \rightarrow N_2(0) + CO(1)$.*

### III.2.4 M(v)-M' V-T

The rate coefficients for the V-T relaxation from level $v$ to $v$-1 of $N_2$ by collision with CO are obtained following the work of Plonjes *et al.* [168]. For the $N_2(v)$-$N_2$ V-T, the expressions recommended in [95,115,130] were favoured in the present paper as they lead to rate coefficients in a good agreement with the semiclassical calculations from [131,132] in the range 300 K < $T_g$ < 1000 K. Vibrational dissociation by V-T processes is included as a transition from the last bound vibrational level of $N_2$ (v = 59 here) to a pseudo-level in the continuum as done in [95,102,130] for $N_2$ and originally for $H_2$ in [169]. Note that, Terraz *et al.* [34] extended the applicability to temperatures in the range [200 K; 300 K] compared with the rate coefficients previously used in [95,115,130]. To do so the results from the semi-classical model of [131,132] available only for a few temperatures in the range [200 K ; 8000 K] were fitted as a function of the gas temperature and over two different temperature regimes, according to the exponential form (9). All the details for the determination of the $N_2(v)$-$N_2/CO_2$ rate coefficients using this procedure can be found in [34] as well as all the necessary coefficients to calculate the rate coefficients until v = 20 for $N_2$ (supplementary appendix).

For the other M(v)-M' V-T rate coefficients we scale them according to the theoretical dependences from the SSH theory on the vibrational levels and gas temperature, as done in [95] for $N_2(v)$-$O_2$ and in [34] for $N_2(v)$-$CO_2$ energy transfers, calculated from the V-T $N_2(v)$-$N_2$. To illustrate this method, the rate coefficient $k_{v,v-1}$ ($N_2$) for deactivation of a vibrational level of CO by $N_2$, corresponding to the process $CO(v) + N_2 \rightarrow CO(v-1) + N_2$, is estimated from the rate coefficient for the relaxation of CO due to CO(v)-CO collisions, $k_{v,v-1}$ (CO) [38,170].

$$k_{v,v-1}(N_2)(T_g) = C_{radius} \times C_{mass} \times k_{v,v-1}(CO)(T_g) \times \frac{F(Y_{v,v-1}^{CO-N_2})}{F(Y_{v,v-1}^{CO-CO})}, \qquad (14)$$

$$C_{radius} = \left(\frac{r_{CO-N_2}}{r_{CO-CO}}\right)^2 \text{ where } r_{i-j} = \frac{1}{2}(r_i + r_j), \qquad (15)$$

with $r_i$ the diameter of the molecule i in the hard sphere collision model

$$C_{mass} = \left(\frac{\mu_{CO-N_2}}{\mu_{CO-CO}}\right)^{\frac{1}{2}}, \tag{16}$$

where $\mu_{i-j} = \frac{m_i \times m_j}{m_i + m_j}$ is the reduced mass of the collision and $m_i$ the mass of molecule i.

*F(y)* is the adiabacity function given by

$$F(y) \begin{cases} \frac{1}{2}\left[3 - e^{\frac{2y}{3}}\right]e^{-\frac{2y}{3}}, y < 21.622 \\ 8\left(\frac{\pi}{3}\right)^{\frac{1}{2}} y^{\frac{7}{3}} e^{-\frac{2y}{3}}, y \geq 21.622 \end{cases}, \tag{17}$$

with *y* the adiabacity factor

$$y_{v,v-1} = \Delta E_{v,v-1} \times \left(\frac{\pi L}{\hbar}\right) \times \sqrt{\frac{\mu}{2k_B T_g}}, \tag{18}$$

with $\Delta E_{v,v-1}$ the energy difference between the vibrational levels *v* and *v*-1 of $N_2$, $k_B$ the Boltzmann constant, $T_g$ the gas temperature and L is the 'interaction length' calculated from the expression [16]:

$$L = \frac{1}{\frac{17.5}{r_0}}, \tag{19}$$

with $r_0$ the Lennard Jones potential distance from [171]. The same approach was adopted in our previous work [38] for CO(*v*)-M (M=$CO_2$, $O_2$) using the rate coefficient from CO(*v*)-CO V-T and $CO_2$(*v*)-M' (M'=CO, $O_2$) using the rate coefficient from $CO_2$(v)-$CO_2$ V-T.

III.3 Surfaces processes

An important process in the vibrational kinetics, in our experimental conditions, is the deactivation of vibrationally excited $CO_2$, CO and $N_2$ molecules through collisions at the reactor wall. This deactivation of vibrationally excited states is shown to have a significant influence on the vibrational characteristic temperatures, especially for pressures below 1 Torr [38,172]. Due to the lack of experimental values, we set the same value of deactivation probability, $\gamma_v$, for any mode of $CO_2$ i.e. $\gamma_v(CO_2(v>0)) = 0.2$ for a Pyrex surface (average value from Table 1 of [173]) where the $CO_2$ molecule is assumed to deactivate to the vibrational ground state. We use a constant value of $4 \cdot 10^{-2}$ for the deactivation probability for all levels of CO and $4.9 \cdot 10^{-4}$ for the deactivation probability for all levels of $N_2$ both taken from [173]. As opposed to $CO_2$, we consider single-quantum transitions for CO and $N_2$ as done for $N_2$ and $O_2$ in [95], where only one vibrational quantum is lost upon collision with the wall. We also investigate the impact of changing the deactivation probability for the bending, symmetric stretch and mixed modes of $CO_2$, from 0.2 to 0.05, on the vibrational temperatures (cf. Section (IV.1, Fig. 9 and 10). Note that the gas mixture might influence the wall de-excitation probability of vibrationally excited molecules. For instance, a linear enhancement of the $N_2(v)$ wall deactivation probability from $1.3 \cdot 10^{-3}$ to $2.6 \cdot 10^{-3}$ with an increase in the admixture of $CO_2$ (from 0.066 % to 0.5 %) in a low-pressure $N_2$ plasma was observed for a Pyrex surface [174].

The importance of the $CO_2$-O V-T exchanges calls for a verification of the accuracy of the calculation of the atomic oxygen concentration. In turn, the loss of O atoms is mainly controlled by recombination at the walls to form $O_2$ molecules characterized by the wall O loss/recombination probability, $\gamma_O$. In this study, $\gamma_O$ is assumed to remain constant for the different mixture compositions, equal to the probability measured in pure $CO_2$ [89] as a function of pressure and at 50 mA (cf. Table 1). Changes in $\gamma_O$ are not expected to significantly modify the $CO_2$ dissociation fraction, as shown in [38,175], but could lead to an increase/decrease in O atom fractions affecting the vibrational temperatures because of the strong $CO_2$-O V-T deactivation. It is known that the recombination probability depends on the nature of the gas ignited in the reactor [89,176]. In $CO_2$-$N_2$ plasmas, CO molecules, C and N atoms or even $CO_2$ molecules are likely to be chemisorbed or physisorbed on the reactor walls, changing the surface configuration and competing for the adsorption sites with O atoms limiting the recombination into $O_2$. The possibility of CO adsorbed at the walls was already suggested by Cenian *et al.* [177] where the rate for the O recombination at the wall is assumed to depend on the CO concentration in the gas phase (see reaction 11 in Table 3.1 of [177]). We thus expect $\gamma_O$ to be lower for a $CO_2$-$N_2$ plasma than for a pure $O_2$ plasma. However, at present it is difficult to draw any conclusions on the values of $\gamma_O$ in $CO_2$-$N_2$ with regards to the pure $CO_2$ case.

## IV. Results

### IV.1 Model validation

In this section, the results of our model are compared with experimental data measured in a $CO_2$-$N_2$ DC discharge for pressures between 0.6 and 4 Torr and a current of 50 mA. The calculated reduced electric field, vibrational temperatures, atomic oxygen and NO fractions and $CO_2$ dissociation fraction are compared with the ones obtained experimentally. For completeness, tables with the calculated concentrations of all the species considered in the model are given in the Supplementary Information for a few selected conditions.

The model shows a satisfactory quantitative agreement and reproduces very well the dependencies of the measured quantities for different pressures and initial $CO_2$ fractions in the different $CO_2$-$N_2$ mixtures. Once validated against the benchmark experiments, the model is used for the interpretation of the measured quantities and the identification of the main processes ruling the discharge.

In Figures 5 and 6 we represent E/N and [O]/N, respectively, as a function of pressure for a pure $CO_2$ plasma and for a $CO_2$-$N_2$ plasma with a ratio of 1:1. A decrease of E/N with increasing pressure can be observed in experiments and simulations for pure $CO_2$ and for the mixture. There is, in general, a quite good agreement between the model predictions and the experimental values, although the absolute value of E/N is somewhat overestimated in the calculations at the low pressures, especially for the pure $CO_2$ discharge. A maximum of [O]/N around 1 Torr can be observed in experiments and simulations for pure $CO_2$ which is mostly due to the O loss frequency showing a minimum at 1 Torr [89]. However, the maximum of experimental [O]/N seems to be shifted for $CO_2$-$N_2$ which is not reproduced in the simulations. Since the simulations consider the same recombination probability for atomic oxygen at the wall via O + wall $\xrightarrow{\gamma_O}$ $O_2$, $\gamma_O$, for pure $CO_2$ and for the $CO_2$-$N_2$ mixture (cf. section III.2.5), the results suggest that the dependence on pressure of $\gamma_O$ might be different in the $CO_2$-$N_2$ mixtures than in pure $CO_2$. To investigate the importance of this quantity on the [O]/N simulation results, we multiply by two the $\gamma_O$ and the resulting [O]/N are given in Figures 6 and 7.

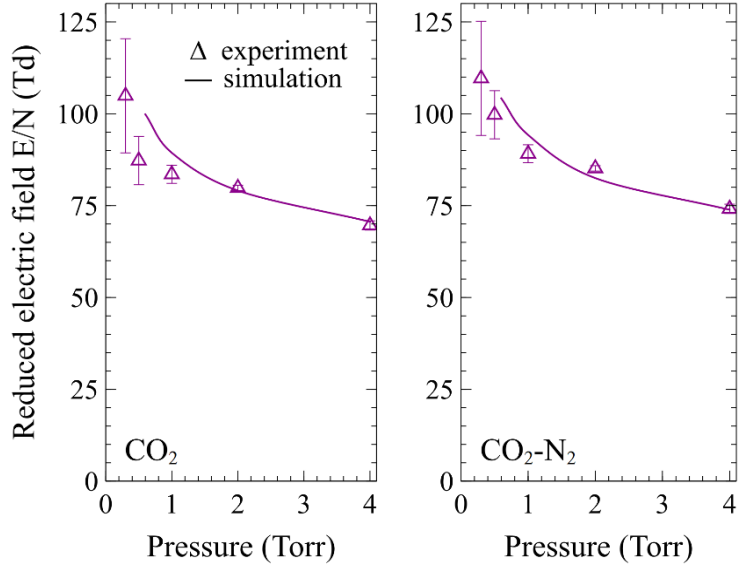

*Figure 5: Reduced electric field E/N as a function of pressure for pure $CO_2$ (left) and a 50/50 $CO_2$-$N_2$ mixture, at a current of 50 mA: experiment (∆) and model calculations (—).*

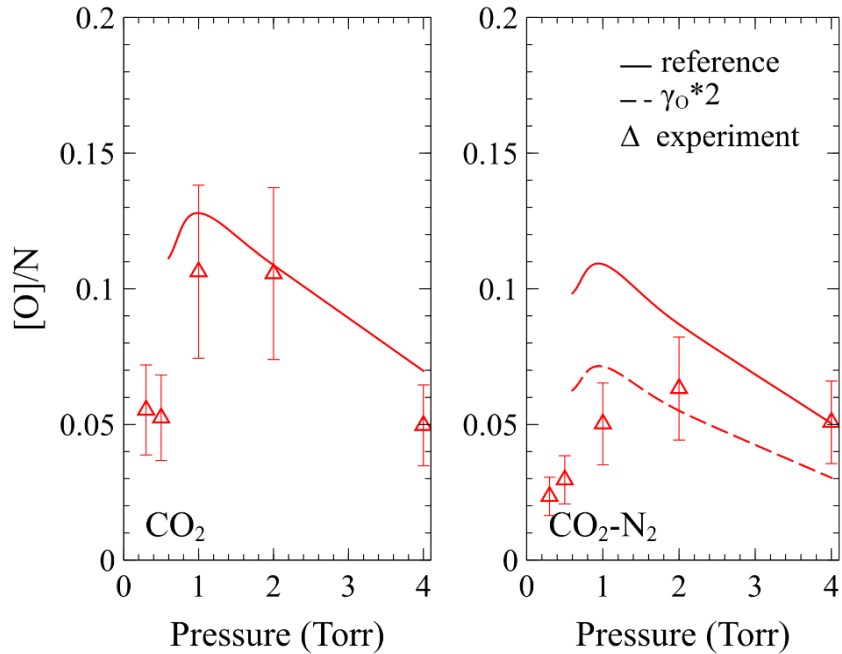

*Figure 6: Atomic oxygen fraction, [O]/N, as a function of pressure for pure $CO_2$ (left) and a 50/50 $CO_2$-$N_2$ mixture, at a current of 50 mA: experiment (∆), model calculations with the default $\gamma_O$ values from [89] (—) and $\gamma_O$ values multiplied by two (– –).*

Figure 7 shows the measured and calculated values of the reduced electric field, E/N, and the atomic oxygen fraction, [O]/N, as a function of the $CO_2$ initial fraction for a discharge current of 50 mA, at 2 Torr. Overall, the self-consistently calculated reduced electric field variation with the $CO_2$ initial fraction agrees well with the experimentally measured E/N. These results prove that the ionization rates and ion chemistry (transport and charge exchange) are well characterized and that the model can be used as a predictive tool when no experimental data for E/N are available. The atomic oxygen fraction, [O]/N, also shows a proper trend with the $CO_2$ fraction but remains too high for all conditions except for the pure $CO_2$ case. One possibility to justify this discrepancy would be a too strong $O_2$ dissociation by electron impact associated with Phelps cross section, as suggested by several authors [178,179] and further discussed in [38,111]. The agreement could be improved by using Polak's cross section for $O_2$ dissociation by electron impact [180] as discussed in [38] in the context of $CO_2$-$O_2$ plasmas. Another possibility would be a larger recombination probability at the wall for the different mixtures in comparison to the pure $CO_2$ case. A dependence of $\gamma_O$ on the $CO_2$ fraction in the different $CO_2$-$N_2$ mixtures seems plausible, as discussed in section III.2.5. Upon multiplying $\gamma_O$ by two, we can observe that [O]/N decreases by 24 % and 38 % for $CO_2$ initial fractions of 0.1 and 0.9 respectively, leading to a [O]/N too low for $CO_2$ fractions higher than 0.25 and too high for $CO_2$ fractions below. From this analysis it can be inferred that [O]/N is very sensitive to the recombination probability of atomic oxygen $\gamma_O$ at the wall and that to obtain a good agreement between the simulations and experiment the $\gamma_O$ should increase with the $N_2$ fraction.

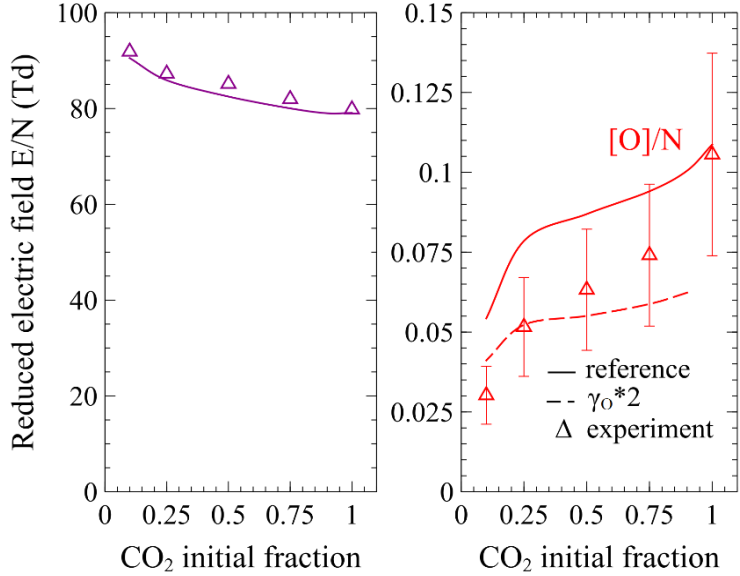

*Figure 7: Measured and calculated reduced electric field, E/N, and atomic oxygen fraction, [O]/N, of a $CO_2$-$N_2$ discharge as a function of the $CO_2$ fraction in the initial mixture, at 2 Torr and 50 mA: from experiment (Δ) and model calculations with the default $\gamma_O$ values from [89] (—) and $\gamma_O$ values multiplied by two (– –).*

Our state-to-state model also provides the populations of each individual vibrational levels of the different modes of $CO_2$ and of CO and $N_2$. The vibrational temperature is then calculated assuming a Treanor distribution [90], as:

$$T_{v,ij} = \left( \frac{E_1}{\ln(p_i/p_j) - \frac{E_j - E_i - E_1}{k_B T_g}} \right) / k_B, \qquad (24)$$

where $E_1$ is the energy of the first level, $p_i$ and $p_j$ and $E_i$ and $E_j$ are the population and energy of levels i and j, respectively, $T_g$ is the gas temperature and $k_B$ the Boltzmann constant. We use an average of the temperatures calculated using the first three vibrational levels. Note that for the specific case of the $CO_2$ bending mode corresponding to $v_2$, we assume a Boltzmann distribution to obtain the vibrational temperature $T_2$ (see equation 5 in [99]). Moreover, for simplicity in the presentation of the results, we define a common temperature of the bending and symmetric stretching modes denoted $T_{1,2}$ [33,89] which is enough for a simple description of the vibrational kinetics even though this is not imposed in the model [38].

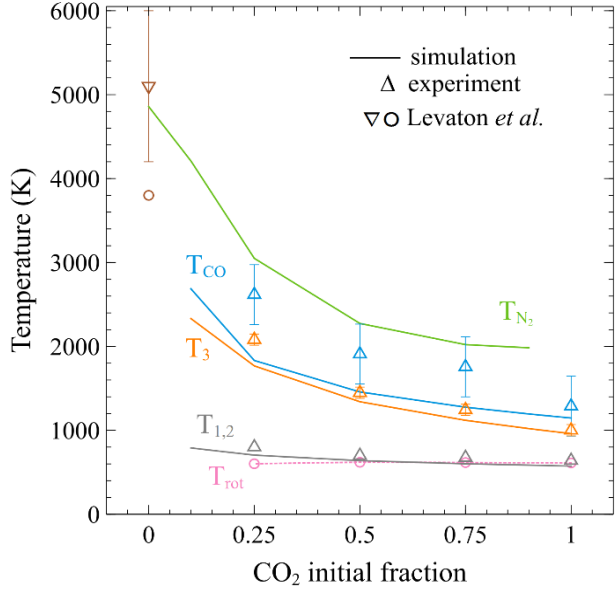

*Figure 8: Experimental values (Δ) and calculated values (line) of the common vibrational temperature of the CO$_2$ bending and symmetric modes T$_{1,2}$, the vibrational temperature of the asymmetric stretching mode T$_3$, the CO and N$_2$ vibrational temperatures T$_{CO}$ and T$_{N2}$ and the rotational temperature T$_{rot}$ (o) (used as input parameter for the model) when a discharge is ignited in different mixtures of CO$_2$-N$_2$, at 50mA, and a pressure of 2 Torr.*

Figure 8 shows the vibrational temperatures of CO and of the different modes of CO$_2$ obtained from the model and experiment as well as the experimental rotational temperature T$_{rot}$ used as input parameter for the model and the calculated vibrational temperature of N$_2$ as a function of the CO$_2$ initial fraction. Note that the vibrational temperatures are calculated with expression (24) using the very first points (v=0, 1 and 2) of the vibrational distributions (see Fig. 11). All the vibrational temperatures increase with the N$_2$ content. This trend can be partly explained by the reduced quenching of vibrations of CO$_2$/CO/N$_2$ by O atoms as [O]/N decreases with decreasing the CO$_2$ initial fraction (cf. Fig. 7). Indeed, O atoms are very efficient quenchers of the CO$_2$ vibrations as already discussed in section III.2.5. Besides, the vibrational deactivation of N$_2$ also takes place essentially by V-T deactivation with O atoms and to a lesser extent through the Zel'dovich reaction forming NO, as also shown in [95]. Note that, the CO$_2$-N$_2$ V-V and CO-N$_2$ V-V transfers play a major role to explain the trends observed and this is discussed in more details at the end of the section (Fig. 12). The rotational temperature T$_{rot}$ remains almost constant with N$_2$ addition as already observed in [70]. The temperature T$_{1,2}$ is very close to T$_{rot}$, as expected from the small energy difference between levels of symmetric and bending modes that favours the V-T deexcitation. The V–V up pumping and CO$_2$-N$_2$ V-V mechanism produce important concentrations of vibrationally excited states of CO$_2$ in the asymmetric stretch mode. The large energy difference between the asymmetric stretch mode levels of CO$_2$ and the anharmonic V-V up pumping and CO$_2$-N$_2$ V-V transfers are beneficial to reach high concentration of vibrationally excited CO$_2$. However, part of the energy stored in ν$_3$ is lost through the intermode V-V transfer (CO$_2$(ν$_3$)+CO$_2$(ν$_{1,2}$)→ CO$_2$(ν$_3$−1)+CO$_2$(ν$_{1,2}$+1)), and through quenching by O atoms. T$_{CO}$ systematically showing larger values than T$_3$ for all the conditions reported can be explained by different rate coefficients for V-T relaxation, the lack of inter-mode V-V relaxation processes for CO, and more efficient vibrational excitation through electron-to-vibrational energy transfers for CO [38,89]. Finally, T$_{N2}$ is higher than T$_{CO}$ and two main reasons can be evoked. First, the quenching of vibrationally excited molecules by O atoms is more efficient for CO than N$_2$. Indeed, the rate coefficient included in this work for the N$_2$-O V-T, as described in [22, 53], is at least 10 times lower than for CO [38,136] in the temperature range [300 K;1000 K]. Second, the deactivation of vibrationally excited states at the wall is more important in the case of CO than N$_2$ as the deactivation probability is two orders of magnitude lower for N$_2$ (cf. section III.2.5).

The calculated and measured vibrational temperatures of N$_2$ obtained for a flowing DC glow discharge in pure N$_2$ by Levaton *et al.* [181] are given on Fig. 8 as an indicator since we do not have access to experimental values for $T_{N_2}$ with FTIR spectrometry. The experimental conditions are very similar to the ones used in the present work in terms of current, radius of the glass tube, pressure, and type of discharge. The main difference comes from the flow with a rate around 10 times higher in [181]. Since the $T_{rot}$ obtained by FTIR does not change for the different mixtures studied in this work (cf. Fig. 10), we use the same for the pure $N_2$ simulation. The $T_{N_2}$ calculated with our model is in very good agreement with the data point obtained in [181] by Optical Emission Spectroscopy, but higher than the value calculated using the state-to-state kinetic model from the same author [181]. To further validate the vibrational kinetics in $CO_2$-$N_2$ plasmas, future measurements (similar to what was done in [75]) require the detection of the vibrational temperature of $N_2$.

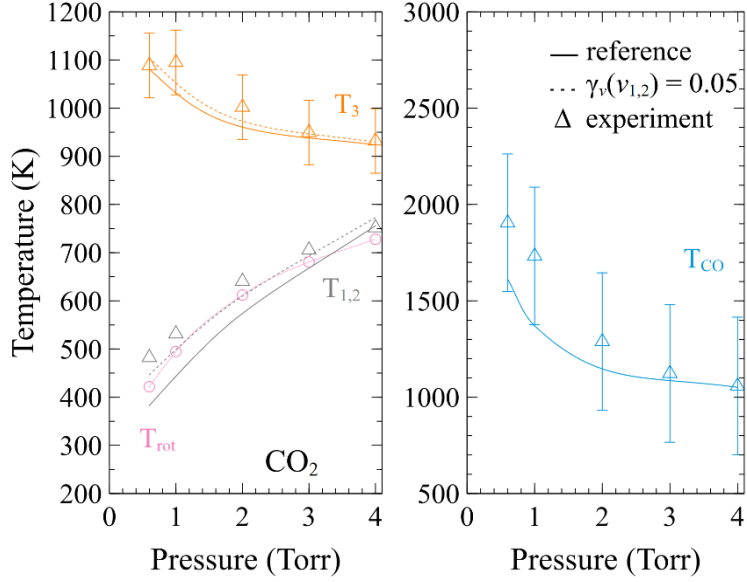

*Figure 9: Experimental values (Δ) and calculated values (line) of the common vibrational temperature of the $CO_2$ bending and symmetric modes $T_{1,2}$, the vibrational temperature of the asymmetric stretching mode $T_3$, the CO and $N_2$ vibrational temperatures $T_{CO}$ and $T_{N_2}$ and the rotational temperature $T_{rot}$ (o) (used as input parameter for the model) when a discharge is ignited in pure $CO_2$, at 50mA, as a function of pressure. The model calculations were done including with the default probabilities of deactivation of the vibrationally excited states of CO and $N_2$ and the different modes of $CO_2$, at the wall from section IV.6. (—) and $\gamma_v(v_{1,2}) = 0,05$ (···)*

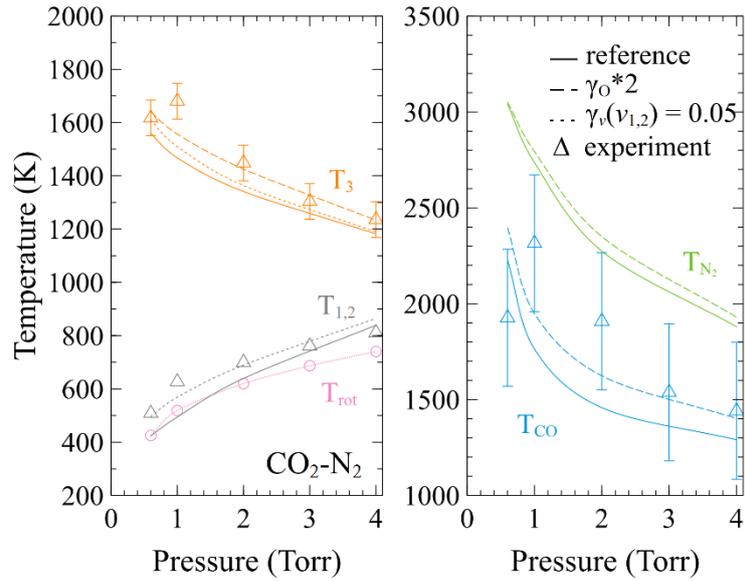

*Figure 10: Experimental values (Δ) and calculated values (line) of the common vibrational temperature of the $CO_2$ bending and symmetric modes $T_{1,2}$, the vibrational temperature of the asymmetric stretching mode $T_3$, the CO and $N_2$ vibrational*

*temperatures $T_{CO}$ and $T_{N2}$ and the rotational temperature $T_{rot}$ (○) (used as input parameter for the model) when a discharge is ignited in $CO_2$-$N_2$ (50 % of each) at 50mA, as a function of pressure. The model calculations were done including with the default probabilities of deactivation of the vibrationally excited states of CO and $N_2$ and the different modes of $CO_2$, at the wall from section IV.6. (—), $\gamma_v(v_{1,2})$ = 0,05 (⋯) and multiplying the default $\gamma_O$ values from [89] by two (– –).*

Figures 9 and 10 show the vibrational temperatures of CO and of the different modes of $CO_2$ obtained from the model and experiment as well as the experimental rotational temperature $T_{rot}$ used as input parameter for the model and the calculated vibrational temperature of $N_2$ as a function of pressure for pure $CO_2$ and a $CO_2$-$N_2$ mixture with 50 % $CO_2$, respectively. The increase of $T_{rot}$ with the pressure can be explained by the higher power dissipated in the plasma because of the higher voltage (and therefore power) required to maintain the current as the pressure is increased. The exact mechanisms responsible for the conversion of electrical energy into gas heating are complex, but include enhanced V-T transfers at higher pressure and possible exothermic reactions [35,70,99]. The influence of deactivation of vibrationally excited $CO_2$ and CO at the walls, described in section III.2.5., is illustrated in Figure 9 for the case of pure $CO_2$ and Figure 10 for the $CO_2$-$N_2$ mixture. Including the wall deactivation for vibrationally excited $CO_2$, CO and $N_2$ molecules mostly affects the vibrational temperatures of these molecules at lower pressures as observed in [38,172]. The default deactivation probabilities of the different modes of $CO_2$ are 0.2 (Section III.2.5.) and lead to simulated $T_{1,2}$ values lower than in the experiment. The results labelled $\gamma_v(v_{1,2})$ = 0.05 correspond to simulations with a deactivation probability for the bending, symmetric stretch and mixed modes of $CO_2$ of 0.05, instead of 0.2, as proposed for pure $CO_2$ in [38] and lead to a better agreement of the calculated $T_{1,2}$ with experiment. Since no experimental values for the wall deactivation probability (on Pyrex) of the bending mode of $CO_2$ was reported in the literature to our knowledge, the present results suggest a value for $\gamma_v(v_{1,2})$ smaller than for $\gamma_v(v_3)$. Finally, we observe that $T_{CO}$ and $T_3$ increase upon multiplying the $\gamma_O$ by two as a consequence of the reduced deactivation through V-T with O. Note that for readability of the figure we do not represent this case for $T_{1,2}$ as it changes by a maximum of 3 % only.

The vibrational distribution functions (VDFs) of the asymmetric stretch mode of $CO_2$, and bending mode of $CO_2$, of $N_2$ and CO, calculated for a DC glow discharge operating at 2 Torr for different $CO_2$-$N_2$ gas mixtures are represented in Figure 11. The simulations evince a remarkable deviation from equilibrium for $N_2$ and CO, and to a lesser extent for $CO_2$, emphasizing the importance of detailed state-to-state models to understand the vibrational energy transfers taking place in the system, as already pointed out in [54]. To maintain this non-equilibrium, the gas temperature must be kept low to suppress the quenching of vibrations. Indeed, V-T relaxation poses an inherent loss mechanism of vibrational energy and increases the gas temperature, which in turn enhances the V-T relaxation creating a positive feedback loop between gas heating and quenching [8]. The VDFs for 90 % and 75% of $CO_2$ are only represented for the asymmetric stretch of $CO_2$ for readability of the figure as they almost overlap with each other in the other cases.

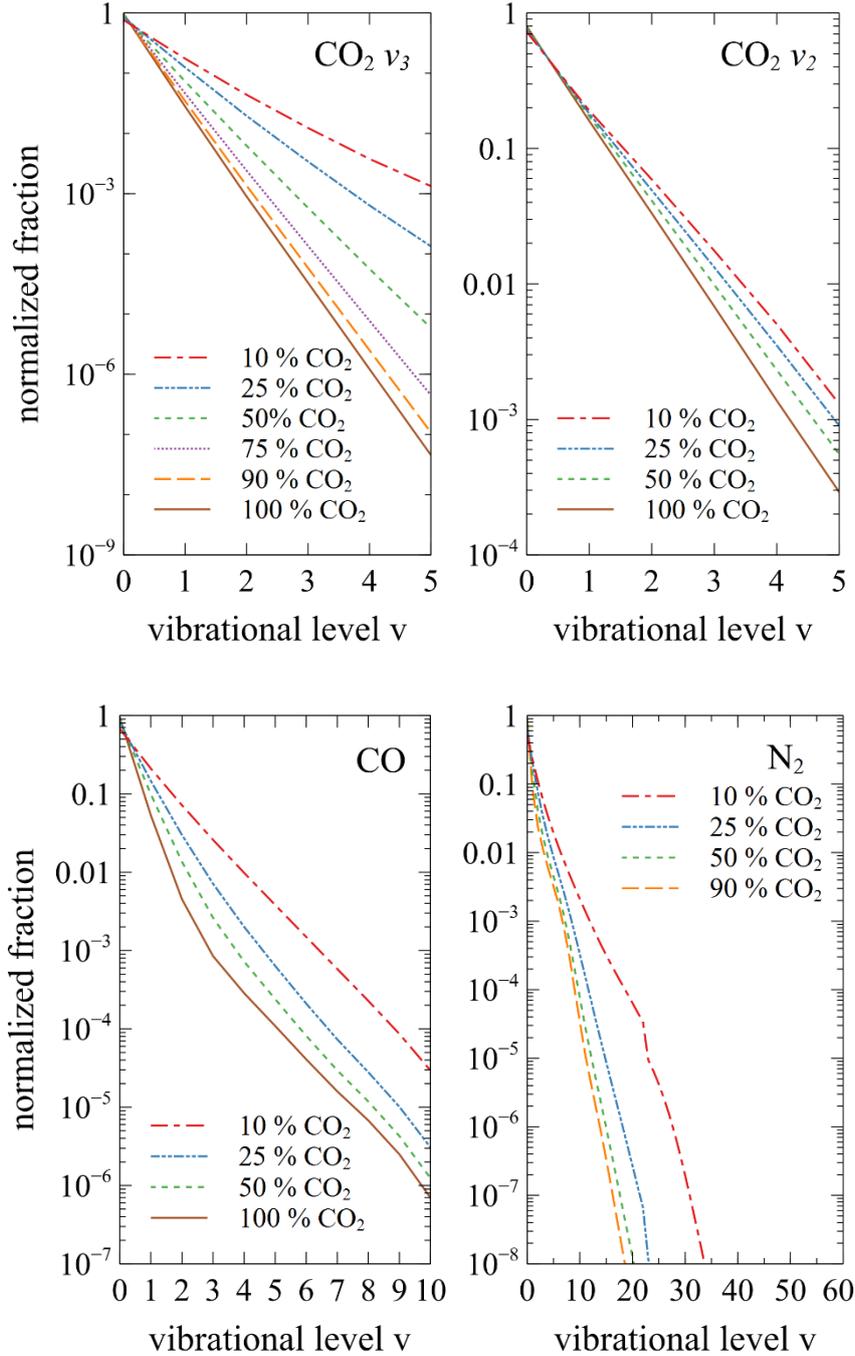

*Figure 11: Vibrational distribution functions of the asymmetric stretch mode of $CO_2$ ($v_3$), bending mode of $CO_2$ ($v_2$), of $N_2$ and CO, calculated for a DC glow discharge operating at 2 Torr and I = 50 mA for different $CO_2$-$N_2$ gas mixtures.*

It is well-known that an important part of electron energy is transferred into the vibrational excitation of $CO_2$, CO and $N_2$, in the range of E/N between 1 and 100 Td [172,182]. This emphasizes the necessity of having a correct description of the vibrational energy transfers between these species. In particular, it is expected that near resonant $CO_2$-$N_2$ and CO-$N_2$ transfers may have an impact shaping the VDFs. To quantify the importance of these transfers, we removed them from the simulations and checked the changes in the vibrational temperatures. The vibrational temperatures $T_3$, $T_{CO}$ and $T_{N_2}$ are represented as a function of the $CO_2$ initial fraction in Figure 12. We can notice that without the $CO_2$-$N_2$ V-V transfer $T_3$ is almost constant, and far from experimental values, for all the mixtures considered and the same can be said for $T_{CO}$ and the CO-$N_2$ V-V transfer. In the case of $T_{N_2}$ both transfers are also playing a significant role but the increase of $T_{N_2}$ with decreasing the $CO_2$ initial fraction also comes from the enhancement of the electron-impact excitation of vibrations of $N_2$ as the fraction of $N_2$ increases in the mixture.

The scaling of $CO_2$-$N_2$ V-V, according to the Sharma Brau theory, explained in detail in section III.2.2., also influences $T_3$ and $T_{N2}$. Therefore, we represent these vibrational temperatures as a function of the $CO_2$ initial fraction keeping the rate coefficient for the $CO_2$-$N_2$ V-V as represented in Figure 1 regardless of the vibrational levels involved (no SB scaling) and scaling this rate coefficient (reference). From this examination it can be concluded that the increase of $T_3$ with $N_2$ is mostly due to the $CO_2$-$N_2$ V-V and similarly for $T_{CO}$ with CO-$N_2$ V-V. Both processes (and the SB scaling) significantly improve the agreement between the simulations and experiment for $T_3$ and $T_{CO}$ further conveying the importance of including these processes in the model.

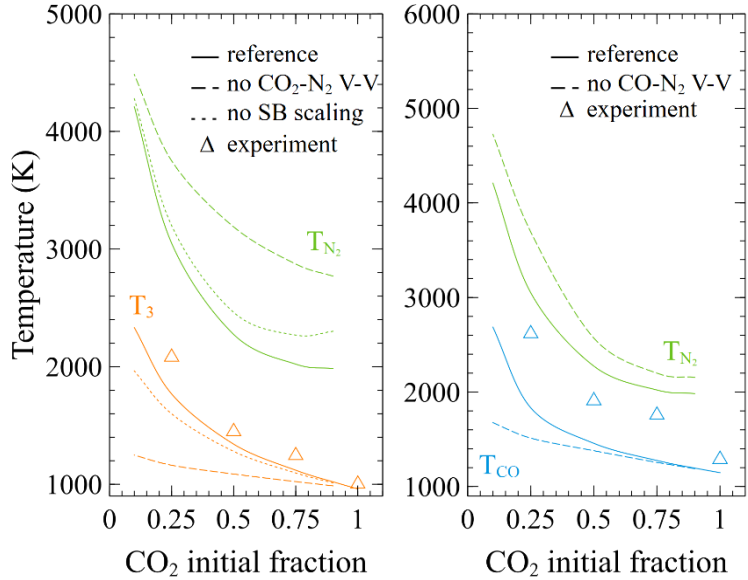

*Figure 12: Experimental values (Δ) and calculated values (line) of the common vibrational temperature of the $CO_2$ of the asymmetric stretching mode $T_3$, the CO and $N_2$ vibrational temperatures $T_{CO}$ and $T_{N2}$ as a function of the $CO_2$ initial fraction, at 50mA. The model calculations were done excluding (– –), including the $CO_2$-$N_2$ V-V transfer with (—) and without scaling (⋯) (left) and including (—) and excluding (– –) the CO-$N_2$ V-V transfer (right). The error bars are not represented here but are the same as in Fig. 13.*

### IV.2 Effect of $N_2$ on $CO_2$ dissociation

In this section the beneficial effect of admixture of $N_2$ on the $CO_2$ dissociation, already observed experimentally and by modeling in [34,68,70,71,183], is investigated. Several reasons can be assigned to this effect. For instance, the enhancement of the reduced electric field, the effect of electronically excited states on the $CO_2$ dissociation directly or via changes of the EEDF, the vibrational excitation of nitrogen and its transfer to the $CO_2$ asymmetric stretching mode and the dilution with $N_2$ limiting the influence of back reaction mechanisms are possibilities often advanced [8,66,172].

Figure 13 shows the absolute (α) and effective ($α_{eff}$) $CO_2$ dissociation fraction as a function of the $CO_2$ initial fraction at 2 Torr. The latter takes into account the initial fraction of $CO_2$ in the gas mixture (for instance, when only 10 % $CO_2$ is present in the gas mixture, the absolute conversion needs to be multiplied by a factor 0.1). The increase of the absolute $CO_2$ dissociation fraction, α, (cf. equation 1) with $N_2$, both in the experimental data and the calculations, indicates that $N_2$ has a beneficial effect on $CO_2$ splitting. Dissociation fractions up to 70 % are observed, a very encouraging result considering that the present setup is designed for fundamental studies only. However, it might be more relevant to use the effective dissociation instead as it takes into account the initial fraction of $CO_2$ in the gas mixture. The effective $CO_2$ conversion decreasing linearly when adding $N_2$ can be explained by the lower $CO_2$ fraction in the mixture, which compensates the enhancement of the absolute conversion of $CO_2$ in the presence of $N_2$. the decrease of $α_{eff}$ was already observed in the simulations of an unpacked

DBD reactor [183] and of a MW discharge [68]. Note that no significant dissociation of $N_2$ was observed for all the simulations conducted (below 0.05 %).

Overall, a good agreement is obtained between the experimental and calculated trend for the $CO_2$ dissociation even when reactions (20)-(23) (cf. section III.1.1) are not included in the simulations (reference case).

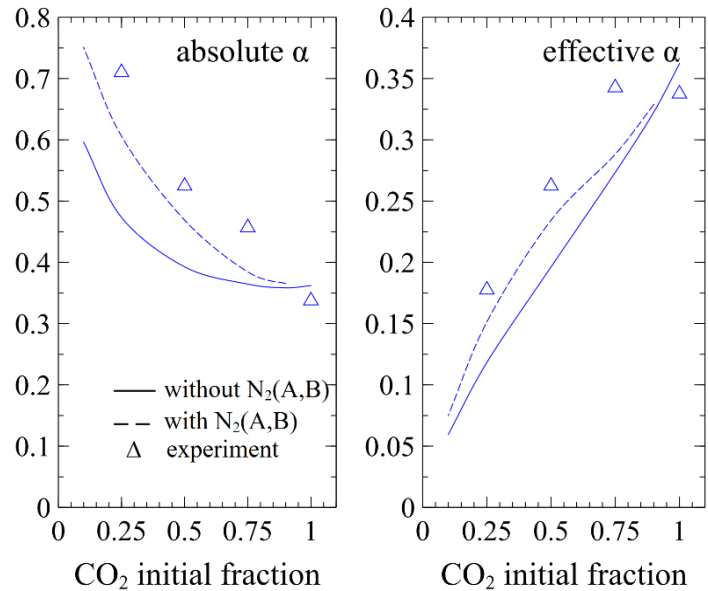

*Figure 13: Absolute and effective dissociation fraction for different $CO_2/N_2$ mixtures for a discharge current of 50 mA and a pressure of 2 Torr. The curve labelled 'with $N_2(A,B)$' corresponds to the reference case to which we added the processes from Tables 2 and 3.*

The agreement is further improved when the dissociative quenching processes with the two metastable states, $N_2(A)$ and $N_2(B)$, are included in the model, as described in section III.1.1. Indeed, these states open a new $CO_2$ dissociation channel and are claimed to be responsible for the increase of α in the presence of $N_2$ by several authors [63,66,71,120,121,129,183]. As can be seen in Figure 13, adding these processes noticeably improves the absolute value and trend of the $CO_2$ dissociation fraction. This shows that the dissociation of $CO_2$ molecules in collisions with electronically excited states of $N_2$ gives a substantial contribution to the conversion rate. We have verified that considering the $CO_2$ dissociation by $N_2(A)$ does not change appreciably the results, which shows that $N_2(B)$ is the main contributor. It is worth pointing out that the dissociation fraction is also increasing when these processes are not accounted for, suggesting that other phenomena are also responsible for the beneficial effect of $N_2$ on the dissociation of $CO_2$.

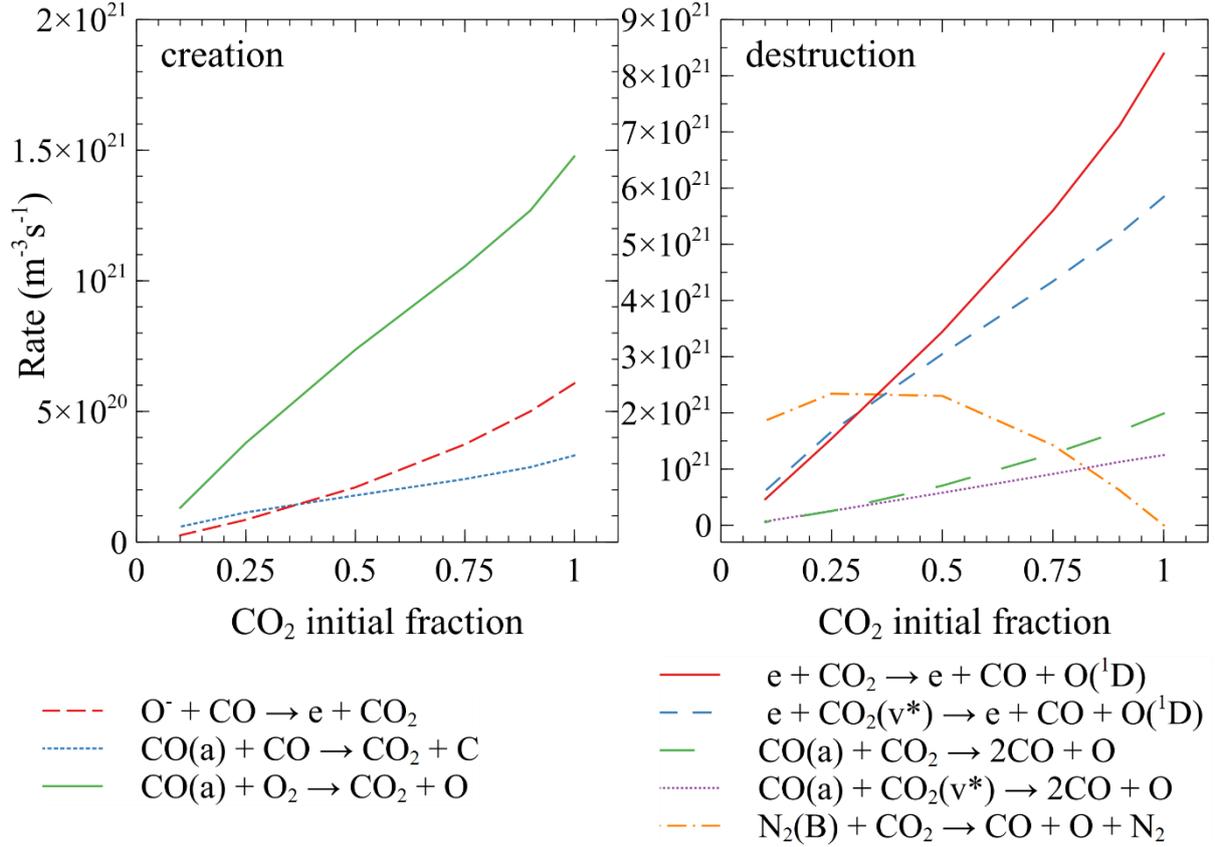

*Figure 14: Rate of the most important creation (left) and destruction (right) processes of $CO_2$, as a function of $CO_2$ fraction in the $CO_2$-$N_2$ gas mixture, at 2 Torr.*

To explain the trend of the $CO_2$ dissociation fraction with the addition of $N_2$, we plot in Figure 14 the rates of the most important creation and destruction processes of $CO_2$, as a function of the $CO_2$ initial fraction for P = 2 Torr and I = 50 mA. In the system under study, $CO_2$ can be decomposed by direct electron impact, both on molecules in the vibrational ground-state (GS) and in vibrational excited states (VES), noted $CO_2(v^*)$, through $e + CO_2 \rightarrow CO + O(^1D)$. The contribution of the vibrational states comes mainly from the lower-laying levels $(01^101)$, $(02^201)$, $(10^002)$ and $(00^011)$. For high $N_2$ content, corresponding to a higher vibrational temperature of the asymmetric mode of $CO_2$ (cf. Fig. 10), the electron-impact dissociation on the VES surpasses the one from the GS. However, this does not have any impact on the $CO_2$ dissociation fraction because overall the rates of electron impact from GS and VES decrease with the $N_2$ content due to the decrease of $CO_2$ density and rate coefficient (cf. discussion of figure 15 concerning the EEDF).

At high $N_2$ fractions, the contribution of $CO_2$ dissociation in quenching collisions with $N_2(B)$ via:

$$N_2(B) + CO_2 \rightarrow CO + O + N_2, \qquad (25)$$

even exceeds that of dissociation of $CO_2$ molecules by electron impact. However, this is accompanied by a lower $CO_2$ content in the mixture, leading to a drop in effective $CO_2$ conversion (Fig. 13). Above 75% $N_2$, both the electron-impact dissociation and the dissociation by $N_2$ metastable molecules rates decrease due to the lower $CO_2$ concentration, which is not compensated by the higher $N_2$ concentration (and thus higher dissociation by $N_2$ metastable molecules). In addition, for small $O_2$ and CO concentrations, $CO_2$ dissociation can be stimulated through:

$$CO(a^3\Pi_r) + CO_2 \rightarrow 2\,CO + O. \qquad (26)$$

The importance of these two mechanisms shows that despite their low density, electronically excited species can have a large influence on the overall plasma chemistry. The calculated densities of these electronically excited states are given in the Supplementary Information for several conditions. We observe that both the calculated $N_2(A)$ and $N_2(B)$ densities increase with the $N_2$ content, going from $1.45 \cdot 10^{16}$ m$^{-3}$ and $1.55 \cdot 10^{15}$ m$^{-3}$ at 90 % $CO_2$ to $1.6 \cdot 10^{18}$ m$^{-3}$ and $6.9 \cdot 10^{16}$ m$^{-3}$ at 10 % $CO_2$, respectively for $N_2(A)$ and $N_2(B)$ while the $CO(a)$ density remains relatively constant ($3.85 \pm 0.3 \cdot 10^{16}$ m$^{-3}$) for the different mixtures, at 2 Torr and 50 mA. Even though the density of $N_2(A)$ is always, at least, one order of magnitude higher than $N_2(B)$, the same process as (25) but involving $N_2(A)$ contributes to less than 0.2 % of the $CO_2$ dissociation for all conditions represented in Fig. 14.

Concerning the creation of $CO_2$, the electronically excited state $CO(a^3\Pi_r)$, noted $CO(a)$ thereafter, promotes the CO recombination into $CO_2$ due to bimolecular reactions with $O_2$ and CO as follows:

$$CO(a) + O_2 \rightarrow CO_2 + O, \tag{27}$$

$$CO(a) + CO \rightarrow CO_2 + C, \tag{28}$$

attesting the complex role that this electronically excited state has on the overall kinetics, as already pointed out in [38,182,184,185]. Finally, another main creation mechanism involving O$^-$ ions is :

$$O^- + CO \rightarrow e + CO_2, \tag{29}$$

playing a significant role for high $CO_2$ fractions as already mentioned in [38].

As can be observed in Figure 14, the rates of the back reaction processes (27), (28) and (29) forming $CO_2$ decrease with increasing $N_2$ content. Indeed, adding $N_2$ dilutes the dissociation products which limits the back reaction mechanisms and leads to an increase of the $CO_2$ dissociation fraction. Some studies of $CO_2$ dissociation by non-thermal plasma have been performed using not only $N_2$ but also rare gases like Argon or Helium as a diluent [69,74,186–188]. However, investigating how the dilution affects the $CO_2$ dissociation fraction is not straightforward as adding a diluent changes the plasma electrical properties.

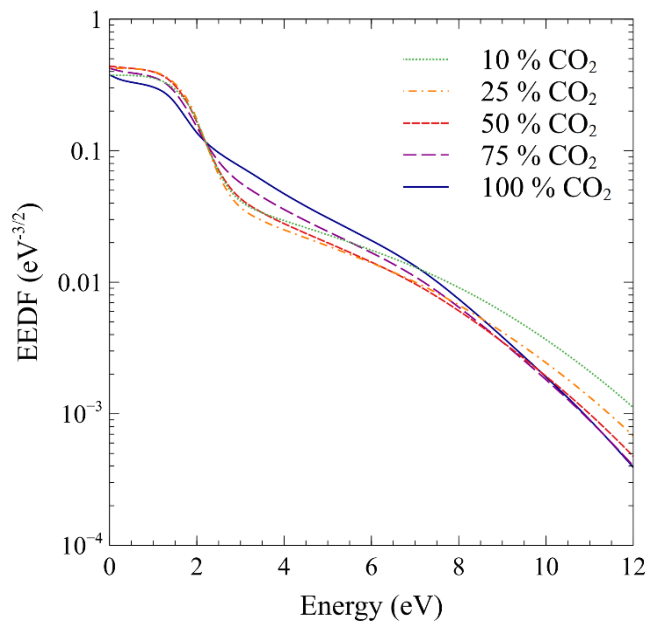

*Figure 15: Electron Energy Distribution Function (EEDF) calculated for different $CO_2/N_2$ mixtures for a discharge current of 50 mA and a pressure of 2 Torr (self-consistent simulations).*

Finally, we analyse the changes in the Electron Energy Distribution Function (EEDF) upon addition of $N_2$. The self-consistently calculated EEDFs for different $CO_2/N_2$ mixtures for a discharge current of 50 mA and a pressure of 2 Torr are shown in Figure 15. Note that changes in the EEDF are associated with several factors, including the modifications in E/N, mixture composition ($N_2$, CO, $O_2$, O) and/or changes in $T_v$ and superelastic collisions with vibrationally and electronically excited states.

Two main effects explain the shape and dependence (with the $N_2$ content) of the EEDFs we observe here. On the one hand, the self-consistently calculated reduced electric field increases (from 79 to 90.5 Td) with the nitrogen fraction (cf. Fig. 7) and on the other hand, as the mixture is changing, the effect of $N_2$ and of the dissociation products, namely CO, $O_2$ and O is more pronounced in mixtures with higher $N_2$ content than in pure $CO_2$. The EEDFs derived for the $CO_2$-$N_2$ (above 25% of $CO_2$) plasmas have a lower population of electrons around the $CO_2$ dissociation energies (7.5 and 11.9 eV [180]) than in a pure $CO_2$ plasma, as already observed in [34] for a $CO_2$-$N_2$ mixture with 50 % $CO_2$. Accordingly, the direct electron dissociation rate coefficient for the GS is lower for 50 % $CO_2$ ($9.4 \cdot 10^{-17}$ $m^3s^{-1}$) and 75 % $CO_2$ ($8.9 \cdot 10^{-17}$ $m^3s^{-1}$), than for pure $CO_2$ ($9.5 \cdot 10^{-17}$ $m^3s^{-1}$). Therefore, the enhancement of the reduced electric field is not sufficient to increase the contribution of direct electron-impact dissociation for low $N_2$ content. We can infer that the increase in the $CO_2$ dissociation when $N_2$ is added into the mixture is not due to an enhancement of the direct electron-impact dissociation, but rather to the enhanced contribution of reaction (25) and to the reduction of back reactions as a consequence of dilution with $N_2$.

## IV.3 $NO_x$ formation

Earlier studies in DBD reactors showed that the presence of $N_2$ during $CO_2$ splitting, along with the main $CO_2$ decomposition products (CO and oxygen species), leads to the formation of $NO_x$ compounds, with concentrations in the range of several hundreds of ppm [71,183]. Furthermore, if $NO_x$ compounds are produced, it is important to know whether high enough concentrations might be obtained for two reasons. On the one hand, $NO_x$ can be considered relevant for nitrogen fixation on earth and in the context of ISRU on Mars; on the other hand, these species have a severe negative impact on air quality giving rise to several environmental and health issues [20,21] leading to restriction of their emissions. The mechanisms of $NO_x$ formation in cold plasmas at reduced pressure are relatively well known and have been described in various modeling works [35,191–194].

In this work, downstream measurements were performed to investigate the composition and concentration of the produced $NO_x$ species in the effluent of our $CO_2$-$N_2$ DC glow discharge, using FTIR spectroscopy. We could detect NO and nitrogen dioxide ($NO_2$) but the amount of $NO_2$ was too low to be quantified.

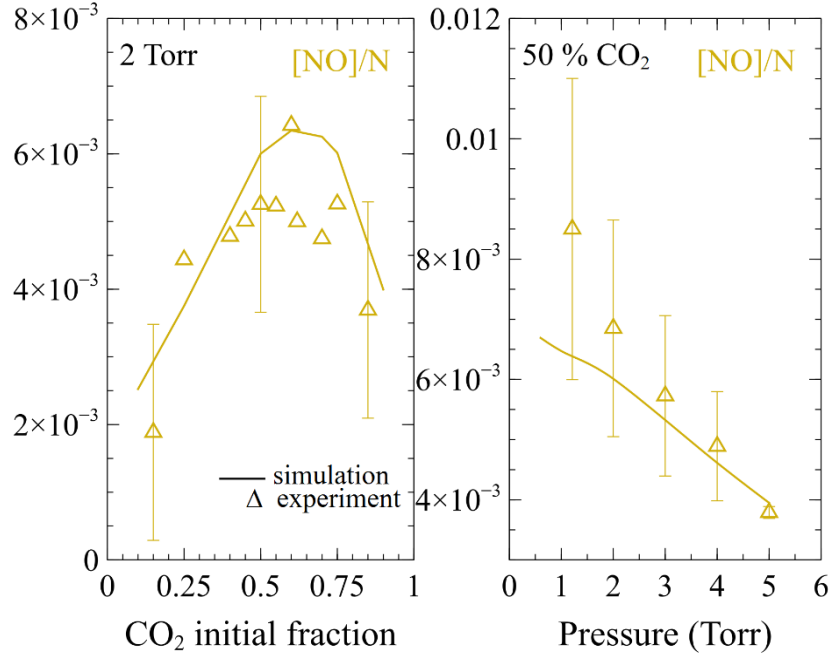

*Figure 16: Measured (Δ) and calculated (—) NO fraction, [NO]/N, of a $CO_2$-$N_2$ discharge as a function of the $CO_2$ initial fraction (left) and as a function of pressure for 50% $CO_2$ (right).*

The measured and calculated NO fractions are plotted as a function of the $CO_2$ initial fraction in the gas mixture and as a function of pressure at a fixed $N_2$ content of 50% in Figure 16. Experimentally and in the simulations the NO fraction as a function of the $CO_2$ initial fraction follows a parabolic trend with a maximum around 50% $N_2$. Concerning the pressure dependence, we can observe that both the measured and calculated [NO]/N decreases with rising pressure. Note that the FTIR measurements were realized in the afterglow while the model results are obtained for the active part of the discharge. However, it was verified that this does not have a significant effect on the NO density measured. Since we find a good agreement between the experiment and simulations (Fig. 16), within the experimental reproducibility error, we believe that our model can be used to explain the observed trends. Therefore, we investigate the contributions of main creation and destruction mechanisms of NO for the different gas mixtures studied at 2 Torr and this is shown in Figure 17. Note that in the simulations the obtained NO fraction is about four orders of magnitude higher than the $NO_2$ fraction and that the NO density was affected by up to 30% and the maximum was slightly shifted towards 50% when the O recombination probability is multiplied by two in the model (see Section IV.1.).

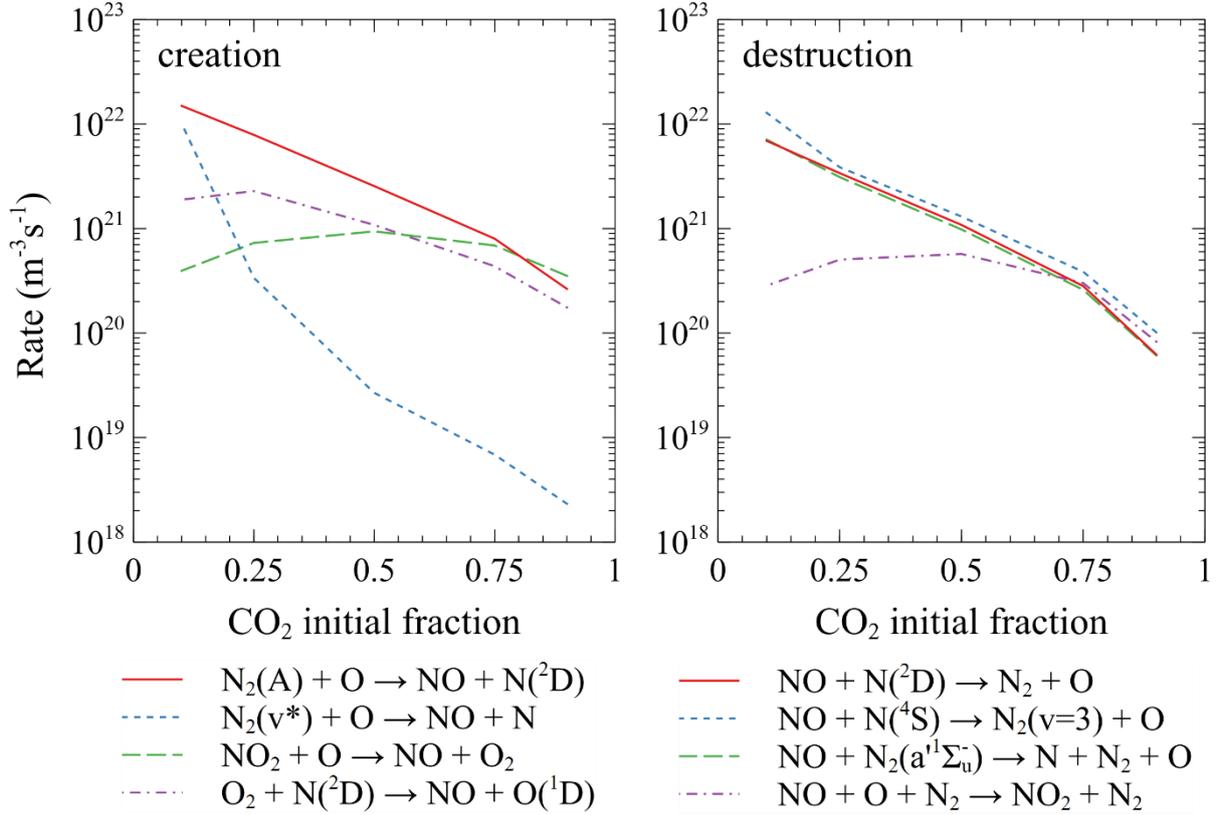

*Figure 17: Rate of the most important creation (left) and destruction (right) processes of NO, as a function of $CO_2$ initial fraction in the $CO_2$-$N_2$ gas mixture, at 2 Torr.*

In our $CO_2$-$N_2$ plasma the NO species are formed out of N reacting with $O_2$ and $N_2(A^3\Sigma_u^+)$ and $N_2(v\geq13)$ reacting with O atoms, originating from $N_2$ and $CO_2$. Thus, it is not unexpected that the maximum of the NO fraction is achieved when both reactants are present in approximately equal concentrations. The dominant formation mechanism of NO at low $N_2$ content is the reaction between O radicals and $NO_2$ molecules, forming NO and $O_2$ molecules. The NO formation via collisions between $N(^2D)$ atoms and $O_2$ is important even if the density of N atoms is much lower than O (typically, by several orders of magnitude in our conditions). On the other hand, nitrogen atoms also play an important role in the destruction of NO. An important NO loss mechanism is the recombination with O atoms into $NO_2$ through a three-body reaction (see reaction given in (3)). This third body can be either $CO_2$ (mainly important above 75 % $CO_2$), $N_2$ (especially for $N_2$ content above 25 %) or $O_2$. At higher $N_2$ fractions, the concentrations of N and $N_2(A)$ will rise giving increasing importance to their reactions whilst the amount of atomic and molecular oxygen will decrease. This will encourage the loss of NO through processes involving species coming from $N_2$. The three-body reaction $NO + O + CO_2 \rightarrow NO_2 + CO_2$ is not represented in Fig. 17 as it only contributes significantly to the NO destruction above 90% of $CO_2$ (12 % maximum contribution). Regarding charge transfers, it is worth noticing that the process $O_2^+ + NO \rightarrow O_2 + NO^+$ does not constitute a real creation process for NO as more than 99% of $NO^+$ is transformed back to NO at the wall and is thus not represented in Figure 17.

$NO_2$ is a significant source of NO production and vice versa. The only important process for $NO_2$ production is the three-body recombination between NO and O, mostly with $N_2$ (for $CO_2$ fraction below 75 %) and with $CO_2$ and $O_2$ as a third body. The main destruction mechanism accounting for more than 99% of the $NO_2$ loss is the reaction of $NO_2$ with O atoms giving NO and $O_2$ molecules.

The mechanism of the plasma chemistry of $CO_2/N_2$ mixtures has been discussed in detail by Snoeckx *et al.* for a DBD reactor [183] and by Heijkers *et al.* for a MW discharge [68]. In both cases the formation of NO was also observed, albeit through a different mechanism. In a microwave plasma, the lower energy of the electrons causes vibrational excitation to become more important than electronic excitation and the vibrationally excited $N_2$ molecules (with v>13) react with O atoms to form N and NO through the Zel'dovich reaction [95,109], instead of the electronically excited $N_2$ in the DBD. In our conditions, this mechanism is becoming important only for high $N_2$ content. In Figure 17 we represent the sum of the rates for all the vibrational states of $N_2$ above v=13 represented as $N_2(v^*)$.

# V. Conclusions

This work reports the validation of a 0D self-consistent kinetic model of $CO_2$-$N_2$ plasmas, from the comparison of the model predictions with recent experimental data obtained in a continuous DC glow discharge ignited in different $CO_2$-$N_2$ gas mixtures. A good agreement is obtained between the calculated vibrational temperatures of $CO_2$ and CO, [O]/N, [NO]/N, E/N and dissociation fractions, and the corresponding experimental data measured by in situ FTIR spectroscopy and actinometry. The reaction mechanism (validated set of reactions and corresponding rate coefficients) we propose predicts the quantities mentioned above for pressures between 0.6 and 4 Torr, discharge current of 50 mA and for different compositions ranging from 100 % to only 25 % of $CO_2$ in a $CO_2$-$N_2$ mixture.

Our model includes a self-consistent description of the $CO_2$, CO and $N_2$ vibrational kinetics and allows us to evince a remarkable deviation from equilibrium for $N_2$ and CO vibrational distribution functions, and to a lesser extent for $CO_2$. This result emphasizes the importance of detailed state-to-state kinetic models and of an understanding of the vibrational energy transfers taking place. In particular, the transfer of vibrational energy from nitrogen to the asymmetric stretching mode of $CO_2$, described by the $CO_2$-$N_2$ V-V transfer in the simulations, greatly influences the vibrational temperature associated to $v_3$. However, the dissociation fraction does not depend on it and therefore, we can conclude that the degree of vibrational excitation reached remains insufficient for an effective dissociation via the pure vibrational mechanism in our conditions.

The experimental data show an enhanced dissociation of $CO_2$ when $N_2$ is added to the plasma. The effect can be attributed to the presence of the metastable $N_2(B^3\Pi_g)$ (creating a new channel for $CO_2$ dissociation) and to the dilution of dissociation products (limiting the back reactions). The electronically excited state $CO(a^3\Pi_r)$ also influences the plasma chemistry even with the addition of a significant amount of $N_2$, showing the importance of electronically excited states in the dissociation kinetics. The studies of the self-consistently calculated EEDFs showed that the enhancement of the reduced electric field is not sufficient to increase the contribution of direct electron-impact dissociation for $N_2$ content below 75 % of the initial mixture. Overall, the admixture of $N_2$ has a beneficial impact on $CO_2$ decomposition but it does not compensate for the decrease of the $CO_2$ initial fraction.

Two main surface processes play a significant role in the plasma kinetics, namely, the recombination of atomic oxygen to form $O_2$ and the deactivation of the vibrations of CO, $N_2$ and the different modes of $CO_2$ at the reactor walls. The O atom densities were shown to strongly depend on the former surface loss process [89]. In turn the vibrational temperatures are greatly affected by the O atoms as they are strong quenchers of vibrations [38,195]. In light of the comparison between the simulations and experimental results in section IV.1., we can infer that the recombination probability of atomic oxygen, $\gamma_O$, should increase with the $N_2$ content. The present results corroborate previous studies that have shown the important role of wall deactivation of vibrationally excited $CO_2$ and $N_2$ at pressures around 1 Torr and below, where it changes the vibrational kinetics [34,38] and has a major contribution to gas heating [35]. Similarly, O recombination at the wall and V-T transfers with O atoms greatly contribute to the thermal balance in oxygen discharges [98]. Therefore, an accurate determination of the experimental $\gamma_O$ and $\gamma_v$ in realistic gas mixtures, containing $CO_2$ and $N_2$ but also the dissociation products like CO, is therefore needed for further studies of plasmas in $CO_2$-$N_2$ mixtures.

It is important to analyse the formation of $NO_x$ compounds in $CO_2$-$N_2$ plasmas as $NO_x$ may have a severe negative impact on air quality giving rise to several environmental and health issues, but high enough concentrations can be considered relevant for nitrogen fixation on earth [49,50] and in the context of In Situ Resource and Utilization on Mars [54]. If necessary, one way to prevent the formation of $NO_x$ compounds is to use the separation membrane technology where the oxygen species can be transported away from the reaction mixture and therefore prevent oxygen species to react with $N_2$. Moreover, an oxygen-permeable membrane placed in the plasma will shift the equilibrium of the $CO_2$ dissociation reaction towards increased fractions of CO and O [52,196]. However, membranes have only been applied, up to now, to separate $O_2$ from the gas mixture, and not the O atoms which participate as well in the $NO_x$ production. One solution would be to modify the surface of the plasma reactor to increase the recombination probability of O atoms at the walls as already done in [195] and modeled in [38]. In contrast, to obtain high concentrations of $NO_x$ one solution would be exposing the $O_2$ coming from the membrane to a $N_2$ plasma leading to the formation of $NO_x$, essential for the synthesis of fertilizers and nitrogen fixation [49,50,54].

Overall, the present work adds one more layer to the systematic set-by-step validation strategy of $CO_2$ non-equilibrium kinetics initiated in [37]. By developing joint experimental, modelling and simulation activities, it has been possible to gather a significant amount of experimental data and simulation results that allow a thorough model validation and can be used to validate or benchmark other models. A similar endeavour is carried out in [98] for pure $O_2$ plasmas, where the benefits of adopting such paradigm in the development of reaction mechanisms for plasma chemistry is highlighted. Future developments may include a detailed analysis of the surface kinetics, the validation of the kinetics of other species with new measurements performed in the same conditions, such as vibrationally temperatures (or distributions) of $N_2$ molecules, nitrogen atoms, or molecular nitrogen electronically excited states, or the extension of the domain of validity of the present kinetic scheme to other pressure or excitation conditions.

## VI. Acknowledgments

This work was partially supported by the European Union's Horizon 2020 research and innovation programme under grant agreement MSCA ITN 813393, and by Portuguese FCT-Fundação para a Ciência e a Tecnologia, under projects UIDB/50010/2020, UIDP/50010/2020, PTDC/FIS-PLA/1616/2021, EXPL/FIS-PLA/0076/2021 and PD/BD/150414/2019 (PD-F APPLAuSE).

# Supporting Information

# Validation of non-equilibrium kinetics in $CO_2$-$N_2$ plasmas

C. Fromentin[1], T. Silva[1], T. C. Dias[1], E. Baratte[2], O. Guaitella[2] and V. Guerra[1]

[1] Instituto de Plasmas e Fusão Nuclear, Instituto Superior Técnico, Universidade de Lisboa, Portugal

[2] Laboratoire de Physique des Plasmas (UMR 7648), CNRS, Univ. Paris Saclay, Sorbonne Université, École Polytechnique, France

## A. Rate coefficients of some V-V processes included in the model

### i. $N_2$-CO V-V:

The process:

$N_2$(v) + CO(w) ↔ $N_2$(v') + CO(w')    | nitrogenCO2VV| a, b, c |

has a rate coefficient of $k = 1{,}66 \cdot 10^{-30} \cdot \exp(a + b \cdot T^{-1/3} + c \cdot T^{-2/3})$ in m³s⁻¹, with T the gas temperature.

| | | |
|---|---|---|
| N2(X,v=1) + CO(X,v=0) ↔ N2(X,v=0) + CO(X,v=1) | \| nitrogenCO2VV\| | 26.3332, -28.231, 36.2268\| |
| N2(X,v=2) + CO(X,v=0) ↔ N2(X,v=1) + CO(X,v=1) | \| nitrogenCO2VV\| | 20.3092, 68.1801, -287.656\| |
| N2(X,v=3) + CO(X,v=0) ↔ N2(X,v=2) + CO(X,v=1) | \| nitrogenCO2VV\| | 22.0824, 44.2821, -175.498\| |
| N2(X,v=4) + CO(X,v=0) ↔ N2(X,v=3) + CO(X,v=1) | \| nitrogenCO2VV\| | 24.0239, 18.1109, -65.108\| |
| N2(X,v=5) + CO(X,v=0) ↔ N2(X,v=4) + CO(X,v=1) | \| nitrogenCO2VV\| | 25.8676, -6.42702, 32.3927\| |
| N2(X,v=6) + CO(X,v=0) ↔ N2(X,v=5) + CO(X,v=1) | \| nitrogenCO2VV\| | 27.7473, -31.3048, 125.122\| |
| N2(X,v=7) + CO(X,v=0) ↔ N2(X,v=6) + CO(X,v=1) | \| nitrogenCO2VV\| | 29.5121, -54.0773, 203.928\| |
| N2(X,v=8) + CO(X,v=0) ↔ N2(X,v=7) + CO(X,v=1) | \| nitrogenCO2VV\| | 30.0891, -58.7989, 212.465\| |
| N2(X,v=9) + CO(X,v=0) ↔ N2(X,v=8) + CO(X,v=1) | \| nitrogenCO2VV\| | 30.881, -69.8849, 256.471\| |
| N2(X,v=10) + CO(X,v=0) ↔ N2(X,v=9) + CO(X,v=1) | \| nitrogenCO2VV\| | 29.5312, -45.2338, 149.915\| |
| N2(X,v=11) + CO(X,v=0) ↔ N2(X,v=10) + CO(X,v=1) | \| nitrogenCO2VV\| | 29.5352, -42.901, 129.101\| |
| N2(X,v=12) + CO(X,v=0) ↔ N2(X,v=11) + CO(X,v=1) | \| nitrogenCO2VV\| | 29.8874, -45.2534, 120.008\| |
| N2(X,v=13) + CO(X,v=0) ↔ N2(X,v=12) + CO(X,v=1) | \| nitrogenCO2VV\| | 29.9643, -42.9403, 89.0297\| |
| N2(X,v=14) + CO(X,v=0) ↔ N2(X,v=13) + CO(X,v=1) | \| nitrogenCO2VV\| | 29.3396, -29.103, 9.7148\| |
| N2(X,v=15) + CO(X,v=0) ↔ N2(X,v=14) + CO(X,v=1) | \| nitrogenCO2VV\| | 27.8155, -0.646867, -129.407\| |
| N2(X,v=16) + CO(X,v=0) ↔ N2(X,v=15) + CO(X,v=1) | \| nitrogenCO2VV\| | 27.1068, 12.9426, -203.402\| |
| N2(X,v=17) + CO(X,v=0) ↔ N2(X,v=16) + CO(X,v=1) | \| nitrogenCO2VV\| | 28.6016, -12.1581, -111.344\| |
| N2(X,v=18) + CO(X,v=0) ↔ N2(X,v=17) + CO(X,v=1) | \| nitrogenCO2VV\| | 29.7772, -30.2926, -58.3128\| |
| N2(X,v=19) + CO(X,v=0) ↔ N2(X,v=18) + CO(X,v=1) | \| nitrogenCO2VV\| | 29.17, -17.065, -143.787\| |
| N2(X,v=20) + CO(X,v=0) ↔ N2(X,v=19) + CO(X,v=1) | \| nitrogenCO2VV\| | 34.6908, -104.556, 179.69\| |
| N2(X,v=21) + CO(X,v=0) ↔ N2(X,v=20) + CO(X,v=1) | \| nitrogenCO2VV\| | 32.109, -54.5671, -74.0396\| |
| N2(X,v=22) + CO(X,v=0) ↔ N2(X,v=21) + CO(X,v=1) | nitrogenCO2VV | 24.7211, 79.8345, -695.697 |
| N2(X,v=23) + CO(X,v=0) ↔ N2(X,v=22) + CO(X,v=1) | nitrogenCO2VV | 24.9209, 73.9707, -674.065 |
| N2(X,v=24) + CO(X,v=0) ↔ N2(X,v=23) + CO(X,v=1) | nitrogenCO2VV | 28.5622, 5.81752, -373.63 |
| N2(X,v=25) + CO(X,v=0) ↔ N2(X,v=24) + CO(X,v=1) | nitrogenCO2VV | 31.154, -41.5012, -176.977 |
| N2(X,v=26) + CO(X,v=0) ↔ N2(X,v=25) + CO(X,v=1) | nitrogenCO2VV | 32.2306, -60.632, -111.204 |
| N2(X,v=27) + CO(X,v=0) ↔ N2(X,v=26) + CO(X,v=1) | nitrogenCO2VV | 33.0182, -74.8529, -65.8755 |
| N2(X,v=28) + CO(X,v=0) ↔ N2(X,v=27) + CO(X,v=1) | nitrogenCO2VV | 33.9564, -91.9747, -6.32267 |
| N2(X,v=29) + CO(X,v=0) ↔ N2(X,v=28) + CO(X,v=1) | nitrogenCO2VV | 43.4796, -254.273, 663.788 |
| N2(X,v=30) + CO(X,v=0) ↔ N2(X,v=29) + CO(X,v=1) | nitrogenCO2VV | 30.564, -17.9899, -424.744 |
| N2(X,v=31) + CO(X,v=0) ↔ N2(X,v=30) + CO(X,v=1) | nitrogenCO2VV | 27.7379, 28.5226, -625.886 |
| N2(X,v=32) + CO(X,v=0) ↔ N2(X,v=31) + CO(X,v=1) | nitrogenCO2VV | 30.829, -33.3464, -334.196 |
| N2(X,v=33) + CO(X,v=0) ↔ N2(X,v=32) + CO(X,v=1) | nitrogenCO2VV | 33.6843, -88.9491, -81.1479 |
| N2(X,v=34) + CO(X,v=0) ↔ N2(X,v=33) + CO(X,v=1) | nitrogenCO2VV | 34.7668, -110.18, 5.64985 |
| N2(X,v=35) + CO(X,v=0) ↔ N2(X,v=34) + CO(X,v=1) | nitrogenCO2VV | 34.8102, -111.565, -1.97593 |
| N2(X,v=36) + CO(X,v=0) ↔ N2(X,v=35) + CO(X,v=1) | nitrogenCO2VV | 34.7416, -110.751, -20.2669 |
| N2(X,v=37) + CO(X,v=0) ↔ N2(X,v=36) + CO(X,v=1) | nitrogenCO2VV | 34.4341, -105.579, -58.3648 |
| N2(X,v=38) + CO(X,v=0) ↔ N2(X,v=37) + CO(X,v=1) | nitrogenCO2VV | 45.5932, -299.167, 761.766 |
| N2(X,v=39) + CO(X,v=0) ↔ N2(X,v=38) + CO(X,v=1) | nitrogenCO2VV | 26.2462, 59.0595, -902.365 |
| N2(X,v=40) + CO(X,v=0) ↔ N2(X,v=39) + CO(X,v=1) | nitrogenCO2VV | 27.0993, 36.8597, -782.868 |
| N2(X,v=41) + CO(X,v=0) ↔ N2(X,v=40) + CO(X,v=1) | nitrogenCO2VV | 31.73, -54.0204, -354.017 |
| N2(X,v=42) + CO(X,v=0) ↔ N2(X,v=41) + CO(X,v=1) | nitrogenCO2VV | 34.2438, -102.633, -136.098 |
| N2(X,v=43) + CO(X,v=0) ↔ N2(X,v=42) + CO(X,v=1) | nitrogenCO2VV | 34.5945, -109.5, -119.431 |
| N2(X,v=44) + CO(X,v=0) ↔ N2(X,v=43) + CO(X,v=1) | nitrogenCO2VV | 34.6051, -109.844, -134.055 |
| N2(X,v=45) + CO(X,v=0) ↔ N2(X,v=44) + CO(X,v=1) | nitrogenCO2VV | 34.6302, -110.461, -147.463 |
| N2(X,v=46) + CO(X,v=0) ↔ N2(X,v=45) + CO(X,v=1) | nitrogenCO2VV | 34.4717, -107.741, -176.064 |
| N2(X,v=47) + CO(X,v=0) ↔ N2(X,v=46) + CO(X,v=1) | nitrogenCO2VV | 48.1873, -349.272, 867.79 |
| N2(X,v=48) + CO(X,v=0) ↔ N2(X,v=47) + CO(X,v=1) | nitrogenCO2VV | 22.0091, 138.858, -1412.78 |
| N2(X,v=49) + CO(X,v=0) ↔ N2(X,v=48) + CO(X,v=1) | nitrogenCO2VV | 27.3199, 31.3902, -886.6 |
| N2(X,v=50) + CO(X,v=0) ↔ N2(X,v=49) + CO(X,v=1) | nitrogenCO2VV | 33.2078, -83.5415, -343.361 |
| N2(X,v=51) + CO(X,v=0) ↔ N2(X,v=50) + CO(X,v=1) | nitrogenCO2VV | 34.7509, -113.508, -215.362 |
| N2(X,v=52) + CO(X,v=0) ↔ N2(X,v=51) + CO(X,v=1) | nitrogenCO2VV | 34.8601, -115.775, -221.044 |
| N2(X,v=53) + CO(X,v=0) ↔ N2(X,v=52) + CO(X,v=1) | nitrogenCO2VV | 34.8929, -116.58, -233.768 |
| N2(X,v=54) + CO(X,v=0) ↔ N2(X,v=53) + CO(X,v=1) | nitrogenCO2VV | 34.9307, -117.489, -245.996 |
| N2(X,v=55) + CO(X,v=0) ↔ N2(X,v=54) + CO(X,v=1) | nitrogenCO2VV | 34.7726, -114.8, -274.767 |
| N2(X,v=56) + CO(X,v=0) ↔ N2(X,v=55) + CO(X,v=1) | nitrogenCO2VV | 50.7052, -398.104, 965.111 |
| N2(X,v=57) + CO(X,v=0) ↔ N2(X,v=56) + CO(X,v=1) | nitrogenCO2VV | 19.6094, 183.904, -1763. |

| N2(X,v=58) + CO(X,v=0) ↔ N2(X,v=57) + CO(X,v=1) | nitrogenCO2VV| 28.7204, 2.91625, -881.976|
| N2(X,v=59) + CO(X,v=0) ↔ N2(X,v=58) + CO(X,v=1) | nitrogenCO2VV| 34.3987, -107.96, -358.729|

| N2(X,v=1) + CO(X,v=1) ↔ N2(X,v=0) + CO(X,v=2) | nitrogenCO2VV| 26.981, -26.0999, 7.89671|
| N2(X,v=2) + CO(X,v=1) ↔ N2(X,v=1) + CO(X,v=2) | nitrogenCO2VV| 19.4259, 95.0473, -412.61|
| N2(X,v=3) + CO(X,v=1) ↔ N2(X,v=2) + CO(X,v=2) | nitrogenCO2VV| 21.9281, 58.5236, -244.966|
| N2(X,v=4) + CO(X,v=1) ↔ N2(X,v=3) + CO(X,v=2) | nitrogenCO2VV| 24.1953, 26.5833, -107.814|
| N2(X,v=5) + CO(X,v=1) ↔ N2(X,v=4) + CO(X,v=2) | nitrogenCO2VV| 26.2081, -1.14478, 6.12999|
| N2(X,v=6) + CO(X,v=1) ↔ N2(X,v=5) + CO(X,v=2) | nitrogenCO2VV| 28.2131, -28.4666, 112.38|
| N2(X,v=7) + CO(X,v=1) ↔ N2(X,v=6) + CO(X,v=2) | nitrogenCO2VV| 30.1963, -55.0413, 209.764|
| N2(X,v=8) + CO(X,v=1) ↔ N2(X,v=7) + CO(X,v=2) | nitrogenCO2VV| 31.1801, -65.8525, 243.291|
| N2(X,v=9) + CO(X,v=1) ↔ N2(X,v=8) + CO(X,v=2) | nitrogenCO2VV| 31.6112, -70.6277, 262.664|
| N2(X,v=10) + CO(X,v=1) ↔ N2(X,v=9) + CO(X,v=2) | nitrogenCO2VV| 30.3535, -47.7734, 166.84|
| N2(X,v=11) + CO(X,v=1) ↔ N2(X,v=10) + CO(X,v=2) | nitrogenCO2VV| 30.0098, -40.4913, 132.572|
| N2(X,v=12) + CO(X,v=1) ↔ N2(X,v=11) + CO(X,v=2) | nitrogenCO2VV| 30.4767, -44.8218, 134.47|
| N2(X,v=13) + CO(X,v=1) ↔ N2(X,v=12) + CO(X,v=2) | nitrogenCO2VV| 30.6932, -45.0098, 116.586|
| N2(X,v=14) + CO(X,v=1) ↔ N2(X,v=13) + CO(X,v=2) | nitrogenCO2VV| 30.3889, -36.6968, 62.4105|
| N2(X,v=15) + CO(X,v=1) ↔ N2(X,v=14) + CO(X,v=2) | nitrogenCO2VV| 29.2469, -14.809, -47.3793|
| N2(X,v=16) + CO(X,v=1) ↔ N2(X,v=15) + CO(X,v=2) | nitrogenCO2VV| 28.1446, 5.66111, -149.719|
| N2(X,v=17) + CO(X,v=1) ↔ N2(X,v=16) + CO(X,v=2) | nitrogenCO2VV| 28.9285, -6.99532, -110.216|
| N2(X,v=18) + CO(X,v=1) ↔ N2(X,v=17) + CO(X,v=2) | nitrogenCO2VV| 30.6683, -35.2702, -10.7146|
| N2(X,v=19) + CO(X,v=1) ↔ N2(X,v=18) + CO(X,v=2) | nitrogenCO2VV| 29.8027, -17.6694, -113.42|
| N2(X,v=20) + CO(X,v=1) ↔ N2(X,v=19) + CO(X,v=2) | nitrogenCO2VV| 34.4967, -90.5394, 146.727|
| N2(X,v=21) + CO(X,v=1) ↔ N2(X,v=20) + CO(X,v=2) | nitrogenCO2VV| 32.998, -59.8838, -20.9005|
| N2(X,v=22) + CO(X,v=1) ↔ N2(X,v=21) + CO(X,v=2) | nitrogenCO2VV| 26.3412, 61.4596, -584.133|
| N2(X,v=23) + CO(X,v=1) ↔ N2(X,v=22) + CO(X,v=2) | nitrogenCO2VV| 25.2668, 79.482, -673.535|
| N2(X,v=24) + CO(X,v=1) ↔ N2(X,v=23) + CO(X,v=2) | nitrogenCO2VV| 28.4635, 19.8308, -413.254|
| N2(X,v=25) + CO(X,v=1) ↔ N2(X,v=24) + CO(X,v=2) | nitrogenCO2VV| 31.2705, -31.1343, -201.283|
| N2(X,v=26) + CO(X,v=1) ↔ N2(X,v=25) + CO(X,v=2) | nitrogenCO2VV| 32.3347, -49.5656, -141.075|
| N2(X,v=27) + CO(X,v=1) ↔ N2(X,v=26) + CO(X,v=2) | nitrogenCO2VV| 33.133, -63.5476, -99.1671|
| N2(X,v=28) + CO(X,v=1) ↔ N2(X,v=27) + CO(X,v=2) | nitrogenCO2VV| 34.1563, -81.8531, -36.3461|
| N2(X,v=29) + CO(X,v=1) ↔ N2(X,v=28) + CO(X,v=2) | nitrogenCO2VV| 43.6842, -243.848, 630.203|
| N2(X,v=30) + CO(X,v=1) ↔ N2(X,v=29) + CO(X,v=2) | nitrogenCO2VV| 31.3319, -17.9028, -411.373|
| N2(X,v=31) + CO(X,v=1) ↔ N2(X,v=30) + CO(X,v=2) | nitrogenCO2VV| 28.0961, 36.2131, -647.92|
| N2(X,v=32) + CO(X,v=1) ↔ N2(X,v=31) + CO(X,v=2) | nitrogenCO2VV| 31.2603, -26.9969, -350.44|

| Reaction | Type | Parameters |
|---|---|---|
| N2(X,v=33) + CO(X,v=1) ↔ N2(X,v=32) + CO(X,v=2) | nitrogenCO2VV | 34.1982, -84.1272, -90.6923 |
| N2(X,v=34) + CO(X,v=1) ↔ N2(X,v=33) + CO(X,v=2) | nitrogenCO2VV | 35.4413, -108.427, 10.453 |
| N2(X,v=35) + CO(X,v=1) ↔ N2(X,v=34) + CO(X,v=2) | nitrogenCO2VV | 35.5695, -111.486, 10.8357 |
| N2(X,v=36) + CO(X,v=1) ↔ N2(X,v=35) + CO(X,v=2) | nitrogenCO2VV | 35.5216, -111.153, -4.9239 |
| N2(X,v=37) + CO(X,v=1) ↔ N2(X,v=36) + CO(X,v=2) | nitrogenCO2VV | 35.231, -106.389, -40.79 |
| N2(X,v=38) + CO(X,v=1) ↔ N2(X,v=37) + CO(X,v=2) | nitrogenCO2VV | 46.1449, -295.323, 757.31 |
| N2(X,v=39) + CO(X,v=1) ↔ N2(X,v=38) + CO(X,v=2) | nitrogenCO2VV | 27.5905, 47.765, -834.771 |
| N2(X,v=40) + CO(X,v=1) ↔ N2(X,v=39) + CO(X,v=2) | nitrogenCO2VV | 27.916, 35.5083, -762.061 |
| N2(X,v=41) + CO(X,v=1) ↔ N2(X,v=40) + CO(X,v=2) | nitrogenCO2VV | 32.3243, -51.1791, -352.959 |
| N2(X,v=42) + CO(X,v=1) ↔ N2(X,v=41) + CO(X,v=2) | nitrogenCO2VV | 34.9206, -101.417, -127.061 |
| N2(X,v=43) + CO(X,v=1) ↔ N2(X,v=42) + CO(X,v=2) | nitrogenCO2VV | 35.3166, -109.18, -105.987 |
| N2(X,v=44) + CO(X,v=1) ↔ N2(X,v=43) + CO(X,v=2) | nitrogenCO2VV | 35.3177, -109.364, -121.263 |
| N2(X,v=45) + CO(X,v=1) ↔ N2(X,v=44) + CO(X,v=2) | nitrogenCO2VV | 35.3346, -109.84, -135.264 |
| N2(X,v=46) + CO(X,v=1) ↔ N2(X,v=45) + CO(X,v=2) | nitrogenCO2VV | 35.1675, -106.973, -164.474 |
| N2(X,v=47) + CO(X,v=1) ↔ N2(X,v=46) + CO(X,v=2) | nitrogenCO2VV | 48.6326, -343.759, 856.996 |
| N2(X,v=48) + CO(X,v=1) ↔ N2(X,v=47) + CO(X,v=2) | nitrogenCO2VV | 23.0774, 132.474, -1366.88 |
| N2(X,v=49) + CO(X,v=1) ↔ N2(X,v=48) + CO(X,v=2) | nitrogenCO2VV | 27.9304, 33.803, -882.896 |
| N2(X,v=50) + CO(X,v=1) ↔ N2(X,v=49) + CO(X,v=2) | nitrogenCO2VV | 33.7625, -80.0616, -344.732 |
| N2(X,v=51) + CO(X,v=1) ↔ N2(X,v=50) + CO(X,v=2) | nitrogenCO2VV | 35.4266, -112.355, -205.541 |
| N2(X,v=52) + CO(X,v=1) ↔ N2(X,v=51) + CO(X,v=2) | nitrogenCO2VV | 35.5522, -114.938, -209.702 |
| N2(X,v=53) + CO(X,v=1) ↔ N2(X,v=52) + CO(X,v=2) | nitrogenCO2VV | 35.5844, -115.731, -222.468 |
| N2(X,v=54) + CO(X,v=1) ↔ N2(X,v=53) + CO(X,v=2) | nitrogenCO2VV | 35.6222, -116.641, -234.682 |
| N2(X,v=55) + CO(X,v=1) ↔ N2(X,v=54) + CO(X,v=2) | nitrogenCO2VV | 35.4659, -113.988, -263.261 |
| N2(X,v=56) + CO(X,v=1) ↔ N2(X,v=55) + CO(X,v=2) | nitrogenCO2VV | 51.1906, -393.34, 957.882 |
| N2(X,v=57) + CO(X,v=1) ↔ N2(X,v=56) + CO(X,v=2) | nitrogenCO2VV | 20.5333, 180.254, -1729.91 |
| N2(X,v=58) + CO(X,v=1) ↔ N2(X,v=57) + CO(X,v=2) | nitrogenCO2VV | 29.2418, 7.06244, -886.597 |
| N2(X,v=59) + CO(X,v=1) ↔ N2(X,v=58) + CO(X,v=2) | nitrogenCO2VV | 35.0137, -105.633, -354.509 |
| N2(X,v=1) + CO(X,v=2) ↔ N2(X,v=0) + CO(X,v=3) | nitrogenCO2VV | 27.4833, -26.3038, -12.1825 |
| N2(X,v=2) + CO(X,v=2) ↔ N2(X,v=1) + CO(X,v=3) | nitrogenCO2VV | 17.3403, 137.944, -608.467 |
| N2(X,v=3) + CO(X,v=2) ↔ N2(X,v=2) + CO(X,v=3) | nitrogenCO2VV | 21.2886, 76.4103, -332.498 |
| N2(X,v=4) + CO(X,v=2) ↔ N2(X,v=3) + CO(X,v=3) | nitrogenCO2VV | 24.0462, 35.9698, -157.594 |
| N2(X,v=5) + CO(X,v=2) ↔ N2(X,v=4) + CO(X,v=3) | nitrogenCO2VV | 26.2951, 4.01149, -23.4525 |
| N2(X,v=6) + CO(X,v=2) ↔ N2(X,v=5) + CO(X,v=3) | nitrogenCO2VV | 28.4639, -26.3416, 98.2404 |
| N2(X,v=7) + CO(X,v=2) ↔ N2(X,v=6) + CO(X,v=3) | nitrogenCO2VV | 30.624, -56.0838, 211.466 |
| N2(X,v=8) + CO(X,v=2) ↔ N2(X,v=7) + CO(X,v=3) | nitrogenCO2VV | 31.9653, -72.7365, 270.594 |
| N2(X,v=9) + CO(X,v=2) ↔ N2(X,v=8) + CO(X,v=3) | nitrogenCO2VV | 31.9634, -69.7467, 259.35 |
| N2(X,v=10) + CO(X,v=2) ↔ N2(X,v=9) + CO(X,v=3) | nitrogenCO2VV | 30.8039, -48.7856, 174.562 |
| N2(X,v=11) + CO(X,v=2) ↔ N2(X,v=10) + CO(X,v=3) | nitrogenCO2VV | 30.3042, -39.6059, 137.909 |
| N2(X,v=12) + CO(X,v=2) ↔ N2(X,v=11) + CO(X,v=3) | nitrogenCO2VV | 30.5861, -41.4426, 135.187 |
| N2(X,v=13) + CO(X,v=2) ↔ N2(X,v=12) + CO(X,v=3) | nitrogenCO2VV | 31.0533, -45.6196, 135.52 |
| N2(X,v=14) + CO(X,v=2) ↔ N2(X,v=13) + CO(X,v=3) | nitrogenCO2VV | 31.0201, -41.895, 102.467 |
| N2(X,v=15) + CO(X,v=2) ↔ N2(X,v=14) + CO(X,v=3) | nitrogenCO2VV | 30.2641, -26.5553, 21.6727 |
| N2(X,v=16) + CO(X,v=2) ↔ N2(X,v=15) + CO(X,v=3) | nitrogenCO2VV | 29.0803, -4.68498, -85.0465 |
| N2(X,v=17) + CO(X,v=2) ↔ N2(X,v=16) + CO(X,v=3) | nitrogenCO2VV | 29.1216, -4.43065, -99.578 |
| N2(X,v=18) + CO(X,v=2) ↔ N2(X,v=17) + CO(X,v=3) | nitrogenCO2VV | 31.3302, -41.1274, 38.8394 |
| N2(X,v=19) + CO(X,v=2) ↔ N2(X,v=18) + CO(X,v=3) | nitrogenCO2VV | 30.217, -19.4877, -78.8056 |
| N2(X,v=20) + CO(X,v=2) ↔ N2(X,v=19) + CO(X,v=3) | nitrogenCO2VV | 34.1178, -78.5852, 123.126 |
| N2(X,v=21) + CO(X,v=2) ↔ N2(X,v=20) + CO(X,v=3) | nitrogenCO2VV | 33.4137, -62.0127, 18.0351 |
| N2(X,v=22) + CO(X,v=2) ↔ N2(X,v=21) + CO(X,v=3) | nitrogenCO2VV | 27.9888, 37.15, -445.385 |
| N2(X,v=23) + CO(X,v=2) ↔ N2(X,v=22) + CO(X,v=3) | nitrogenCO2VV | 25.6785, 78.1875, -640.839 |
| N2(X,v=24) + CO(X,v=2) ↔ N2(X,v=23) + CO(X,v=3) | nitrogenCO2VV | 28.1667, 31.8651, -442.514 |
| N2(X,v=25) + CO(X,v=2) ↔ N2(X,v=24) + CO(X,v=3) | nitrogenCO2VV | 31.136, -21.8395, -218.864 |
| N2(X,v=26) + CO(X,v=2) ↔ N2(X,v=25) + CO(X,v=3) | nitrogenCO2VV | 32.2214, -40.2223, -160.912 |
| N2(X,v=27) + CO(X,v=2) ↔ N2(X,v=26) + CO(X,v=3) | nitrogenCO2VV | 32.9528, -52.5048, -129.149 |
| N2(X,v=28) + CO(X,v=2) ↔ N2(X,v=27) + CO(X,v=3) | nitrogenCO2VV | 33.9497, -69.923, -72.576 |
| N2(X,v=29) + CO(X,v=2) ↔ N2(X,v=28) + CO(X,v=3) | nitrogenCO2VV | 43.4483, -230.927, 587.018 |
| N2(X,v=30) + CO(X,v=2) ↔ N2(X,v=29) + CO(X,v=3) | nitrogenCO2VV | 31.7653, -17.1373, -399.931 |
| N2(X,v=31) + CO(X,v=2) ↔ N2(X,v=30) + CO(X,v=3) | nitrogenCO2VV | 28.0068, 46.7731, -682.424 |
| N2(X,v=32) + CO(X,v=2) ↔ N2(X,v=31) + CO(X,v=3) | nitrogenCO2VV | 31.2746, -18.3129, -376.844 |
| N2(X,v=33) + CO(X,v=2) ↔ N2(X,v=32) + CO(X,v=3) | nitrogenCO2VV | 34.335, -77.6478, -107.59 |
| N2(X,v=34) + CO(X,v=2) ↔ N2(X,v=33) + CO(X,v=3) | nitrogenCO2VV | 35.7497, -105.163, 8.2543 |
| N2(X,v=35) + CO(X,v=2) ↔ N2(X,v=34) + CO(X,v=3) | nitrogenCO2VV | 35.9973, -110.513, 19.3345 |
| N2(X,v=36) + CO(X,v=2) ↔ N2(X,v=35) + CO(X,v=3) | nitrogenCO2VV | 35.994, -111.097, 8.02783 |
| N2(X,v=37) + CO(X,v=2) ↔ N2(X,v=36) + CO(X,v=3) | nitrogenCO2VV | 35.7519, -107.318, -23.0184 |
| N2(X,v=38) + CO(X,v=2) ↔ N2(X,v=37) + CO(X,v=3) | nitrogenCO2VV | 46.4312, -291.792, 753.901 |
| N2(X,v=39) + CO(X,v=2) ↔ N2(X,v=38) + CO(X,v=3) | nitrogenCO2VV | 28.6614, 36.2975, -766.758 |
| N2(X,v=40) + CO(X,v=2) ↔ N2(X,v=39) + CO(X,v=3) | nitrogenCO2VV | 28.4726, 33.6972, -739.334 |
| N2(X,v=41) + CO(X,v=2) ↔ N2(X,v=40) + CO(X,v=3) | nitrogenCO2VV | 32.6487, -48.6264, -350.742 |
| N2(X,v=42) + CO(X,v=2) ↔ N2(X,v=41) + CO(X,v=3) | nitrogenCO2VV | 35.3147, -100.26, -117.921 |
| N2(X,v=43) + CO(X,v=2) ↔ N2(X,v=42) + CO(X,v=3) | nitrogenCO2VV | 35.762, -109.041, -91.8179 |
| N2(X,v=44) + CO(X,v=2) ↔ N2(X,v=43) + CO(X,v=3) | nitrogenCO2VV | 35.7518, -109.04, -107.828 |
| N2(X,v=45) + CO(X,v=2) ↔ N2(X,v=44) + CO(X,v=3) | nitrogenCO2VV | 35.7581, -109.334, -122.586 |
| N2(X,v=46) + CO(X,v=2) ↔ N2(X,v=45) + CO(X,v=3) | nitrogenCO2VV | 35.5787, -106.254, -152.683 |
| N2(X,v=47) + CO(X,v=2) ↔ N2(X,v=46) + CO(X,v=3) | nitrogenCO2VV | 48.7883, -338.208, 845.993 |
| N2(X,v=48) + CO(X,v=2) ↔ N2(X,v=47) + CO(X,v=3) | nitrogenCO2VV | 23.884, 125.604, -1318.74 |
| N2(X,v=49) + CO(X,v=2) ↔ N2(X,v=48) + CO(X,v=3) | nitrogenCO2VV | 28.2683, 35.9281, -877.83 |
| N2(X,v=50) + CO(X,v=2) ↔ N2(X,v=49) + CO(X,v=3) | nitrogenCO2VV | 34.0246, -76.4881, -346.563 |
| N2(X,v=51) + CO(X,v=2) ↔ N2(X,v=50) + CO(X,v=3) | nitrogenCO2VV | 35.8121, -111.155, -195.96 |
| N2(X,v=52) + CO(X,v=2) ↔ N2(X,v=51) + CO(X,v=3) | nitrogenCO2VV | 35.9568, -114.104, -198.354 |
| N2(X,v=53) + CO(X,v=2) ↔ N2(X,v=52) + CO(X,v=3) | nitrogenCO2VV | 35.9882, -114.882, -211.178 |
| N2(X,v=54) + CO(X,v=2) ↔ N2(X,v=53) + CO(X,v=3) | nitrogenCO2VV | 36.026, -115.791, -223.388 |
| N2(X,v=55) + CO(X,v=2) ↔ N2(X,v=54) + CO(X,v=3) | nitrogenCO2VV | 35.8714, -113.171, -251.782 |
| N2(X,v=56) + CO(X,v=2) ↔ N2(X,v=55) + CO(X,v=3) | nitrogenCO2VV | 51.3828, -388.47, 950.152 |
| N2(X,v=57) + CO(X,v=2) ↔ N2(X,v=56) + CO(X,v=3) | nitrogenCO2VV | 21.1819, 176.366, -1695.69 |
| N2(X,v=58) + CO(X,v=2) ↔ N2(X,v=57) + CO(X,v=3) | nitrogenCO2VV | 29.4826, 11.0718, -890.557 |
| N2(X,v=59) + CO(X,v=2) ↔ N2(X,v=58) + CO(X,v=3) | nitrogenCO2VV | 35.3341, -103.171, -350.943 |
| N2(X,v=1) + CO(X,v=3) ↔ N2(X,v=0) + CO(X,v=4) | nitrogenCO2VV | 28.0399, -29.5167, -20.2133 |
| N2(X,v=2) + CO(X,v=3) ↔ N2(X,v=1) + CO(X,v=4) | nitrogenCO2VV | 13.8918, 203.073, -904.017 |
| N2(X,v=3) + CO(X,v=3) ↔ N2(X,v=2) + CO(X,v=4) | nitrogenCO2VV | 20.2934, 98.6659, -441.184 |
| N2(X,v=4) + CO(X,v=3) ↔ N2(X,v=3) + CO(X,v=4) | nitrogenCO2VV | 23.7898, 45.528, -211.175 |
| N2(X,v=5) + CO(X,v=3) ↔ N2(X,v=4) + CO(X,v=4) | nitrogenCO2VV | 26.3439, 8.28304, -53.105 |
| N2(X,v=6) + CO(X,v=3) ↔ N2(X,v=5) + CO(X,v=4) | nitrogenCO2VV | 28.7007, -25.4209, 84.7976 |
| N2(X,v=7) + CO(X,v=3) ↔ N2(X,v=6) + CO(X,v=4) | nitrogenCO2VV | 30.9861, -57.514, 210.531 |
| N2(X,v=8) + CO(X,v=3) ↔ N2(X,v=7) + CO(X,v=4) | nitrogenCO2VV | 32.5146, -77.4178, 285.439 |
| N2(X,v=9) + CO(X,v=3) ↔ N2(X,v=8) + CO(X,v=4) | nitrogenCO2VV | 32.1491, -68.2122, 250.454 |
| N2(X,v=10) + CO(X,v=3) ↔ N2(X,v=9) + CO(X,v=4) | nitrogenCO2VV | 31.0571, -48.1596, 172.47 |
| N2(X,v=11) + CO(X,v=3) ↔ N2(X,v=10) + CO(X,v=4) | nitrogenCO2VV | 30.5592, -39.6642, 142.824 |
| N2(X,v=12) + CO(X,v=3) ↔ N2(X,v=11) + CO(X,v=4) | nitrogenCO2VV | 30.5894, -38.1716, 132.761 |
| N2(X,v=13) + CO(X,v=3) ↔ N2(X,v=12) + CO(X,v=4) | nitrogenCO2VV | 31.0996, -43.231, 140.519 |
| N2(X,v=14) + CO(X,v=3) ↔ N2(X,v=13) + CO(X,v=4) | nitrogenCO2VV | 31.4071, -44.8769, 130.811 |
| N2(X,v=15) + CO(X,v=3) ↔ N2(X,v=14) + CO(X,v=4) | nitrogenCO2VV | 31.019, -35.6442, 76.8818 |
| N2(X,v=16) + CO(X,v=3) ↔ N2(X,v=15) + CO(X,v=4) | nitrogenCO2VV | 29.9767, -16.1616, -18.1923 |
| N2(X,v=17) + CO(X,v=3) ↔ N2(X,v=16) + CO(X,v=4) | nitrogenCO2VV | 29.4407, -5.96878, -73.3649 |
| N2(X,v=18) + CO(X,v=3) ↔ N2(X,v=17) + CO(X,v=4) | nitrogenCO2VV | 31.9056, -47.3071, 87.4179 |

| N2(X,v=19) + CO(X,v=3) ↔ N2(X,v=18) + CO(X,v=4) | nitrogenCO2VV | 30.5841, -22.437, -40.94 |
| N2(X,v=20) + CO(X,v=3) ↔ N2(X,v=19) + CO(X,v=4) | nitrogenCO2VV | 33.7437, -68.9434, 109.336 |
| N2(X,v=21) + CO(X,v=3) ↔ N2(X,v=20) + CO(X,v=4) | nitrogenCO2VV | 33.5605, -61.5366, 45.0187 |
| N2(X,v=22) + CO(X,v=3) ↔ N2(X,v=21) + CO(X,v=4) | nitrogenCO2VV | 29.6205, 10.961, -298.823 |
| N2(X,v=23) + CO(X,v=3) ↔ N2(X,v=22) + CO(X,v=4) | nitrogenCO2VV | 26.3567, 69.6395, -574.67 |
| N2(X,v=24) + CO(X,v=3) ↔ N2(X,v=23) + CO(X,v=4) | nitrogenCO2VV | 27.9109, 40.7174, -456.41 |
| N2(X,v=25) + CO(X,v=3) ↔ N2(X,v=24) + CO(X,v=4) | nitrogenCO2VV | 30.9008, -13.1942, -232.125 |
| N2(X,v=26) + CO(X,v=3) ↔ N2(X,v=25) + CO(X,v=4) | nitrogenCO2VV | 32.1043, -33.3844, -167.386 |
| N2(X,v=27) + CO(X,v=3) ↔ N2(X,v=26) + CO(X,v=4) | nitrogenCO2VV | 32.7013, -42.7386, -151.171 |
| N2(X,v=28) + CO(X,v=3) ↔ N2(X,v=27) + CO(X,v=4) | nitrogenCO2VV | 33.5599, -57.2306, -109.976 |
| N2(X,v=29) + CO(X,v=3) ↔ N2(X,v=28) + CO(X,v=4) | nitrogenCO2VV | 42.9428, -215.624, 535.166 |
| N2(X,v=30) + CO(X,v=3) ↔ N2(X,v=29) + CO(X,v=4) | nitrogenCO2VV | 32.1047, -17.1061, -383.413 |
| N2(X,v=31) + CO(X,v=3) ↔ N2(X,v=30) + CO(X,v=4) | nitrogenCO2VV | 27.6726, 59.5068, -725.782 |
| N2(X,v=32) + CO(X,v=3) ↔ N2(X,v=31) + CO(X,v=4) | nitrogenCO2VV | 31.0242, -7.03326, -414.264 |
| N2(X,v=33) + CO(X,v=3) ↔ N2(X,v=32) + CO(X,v=4) | nitrogenCO2VV | 34.2496, -69.2999, -132.463 |
| N2(X,v=34) + CO(X,v=3) ↔ N2(X,v=33) + CO(X,v=4) | nitrogenCO2VV | 35.8407, -100.041, -2.28859 |
| N2(X,v=35) + CO(X,v=3) ↔ N2(X,v=34) + CO(X,v=4) | nitrogenCO2VV | 36.2392, -108.222, 21.7098 |
| N2(X,v=36) + CO(X,v=3) ↔ N2(X,v=35) + CO(X,v=4) | nitrogenCO2VV | 36.3094, -110.233, 17.0567 |
| N2(X,v=37) + CO(X,v=3) ↔ N2(X,v=36) + CO(X,v=4) | nitrogenCO2VV | 36.1555, -108.154, -6.02542 |
| N2(X,v=38) + CO(X,v=3) ↔ N2(X,v=37) + CO(X,v=4) | nitrogenCO2VV | 46.6161, -288.454, 750.973 |
| N2(X,v=39) + CO(X,v=3) ↔ N2(X,v=38) + CO(X,v=4) | nitrogenCO2VV | 29.6106, 25.0124, -700.032 |
| N2(X,v=40) + CO(X,v=3) ↔ N2(X,v=39) + CO(X,v=4) | nitrogenCO2VV | 28.9312, 31.5849, -715.466 |
| N2(X,v=41) + CO(X,v=3) ↔ N2(X,v=40) + CO(X,v=4) | nitrogenCO2VV | 32.8731, -46.3516, -347.456 |
| N2(X,v=42) + CO(X,v=3) ↔ N2(X,v=41) + CO(X,v=4) | nitrogenCO2VV | 35.5957, -99.1396, -108.816 |
| N2(X,v=43) + CO(X,v=3) ↔ N2(X,v=42) + CO(X,v=4) | nitrogenCO2VV | 36.102, -109.098, -76.8904 |
| N2(X,v=44) + CO(X,v=3) ↔ N2(X,v=43) + CO(X,v=4) | nitrogenCO2VV | 36.0795, -108.904, -93.6271 |
| N2(X,v=45) + CO(X,v=3) ↔ N2(X,v=44) + CO(X,v=4) | nitrogenCO2VV | 36.0726, -108.976, -109.297 |
| N2(X,v=46) + CO(X,v=3) ↔ N2(X,v=45) + CO(X,v=4) | nitrogenCO2VV | 35.8766, -105.614, -140.58 |
| N2(X,v=47) + CO(X,v=3) ↔ N2(X,v=46) + CO(X,v=4) | nitrogenCO2VV | 48.8266, -332.654, 834.931 |
| N2(X,v=48) + CO(X,v=3) ↔ N2(X,v=47) + CO(X,v=4) | nitrogenCO2VV | 24.6019, 118.195, -1268.11 |
| N2(X,v=49) + CO(X,v=3) ↔ N2(X,v=48) + CO(X,v=4) | nitrogenCO2VV | 28.5052, 37.735, -871.265 |
| N2(X,v=50) + CO(X,v=3) ↔ N2(X,v=49) + CO(X,v=4) | nitrogenCO2VV | 34.1655, -72.8472, -348.735 |
| N2(X,v=51) + CO(X,v=3) ↔ N2(X,v=50) + CO(X,v=4) | nitrogenCO2VV | 36.0776, -109.909, -186.621 |
| N2(X,v=52) + CO(X,v=3) ↔ N2(X,v=51) + CO(X,v=4) | nitrogenCO2VV | 36.2442, -113.279, -186.977 |
| N2(X,v=53) + CO(X,v=3) ↔ N2(X,v=52) + CO(X,v=4) | nitrogenCO2VV | 36.2745, -114.036, -199.883 |
| N2(X,v=54) + CO(X,v=3) ↔ N2(X,v=53) + CO(X,v=4) | nitrogenCO2VV | 36.312, -114.94, -212.107 |
| N2(X,v=55) + CO(X,v=3) ↔ N2(X,v=54) + CO(X,v=4) | \| nitrogenCO2VV\| | 36.1588, -112.35, -240.332\| |
| N2(X,v=56) + CO(X,v=3) ↔ N2(X,v=55) + CO(X,v=4) | \| nitrogenCO2VV\| | 51.4514, -383.492, 941.914\| |
| N2(X,v=57) + CO(X,v=3) ↔ N2(X,v=56) + CO(X,v=4) | \| nitrogenCO2VV\| | 21.7262, 172.225, -1660.27\| |
| N2(X,v=58) + CO(X,v=3) ↔ N2(X,v=57) + CO(X,v=4) | \| nitrogenCO2VV\| | 29.6133, 14.9291, -893.783\| |
| N2(X,v=59) + CO(X,v=3) ↔ N2(X,v=58) + CO(X,v=4) | \| nitrogenCO2VV\| | 35.5294, -100.57, -348.055\| |
| N2(X,v=1) + CO(X,v=4) ↔ N2(X,v=0) + CO(X,v=5) | \| nitrogenCO2VV\| | 28.7124, -36.0457, -13.9044\| |
| N2(X,v=2) + CO(X,v=4) ↔ N2(X,v=1) + CO(X,v=5) | \| nitrogenCO2VV\| | 9.96555, 276.42, -1240.94\| |
| N2(X,v=3) + CO(X,v=4) ↔ N2(X,v=2) + CO(X,v=5) | \| nitrogenCO2VV\| | 19.0126, 125.06, -570.169\| |
| N2(X,v=4) + CO(X,v=4) ↔ N2(X,v=3) + CO(X,v=5) | \| nitrogenCO2VV\| | 23.5601, 53.8472, -262.334\| |
| N2(X,v=5) + CO(X,v=4) ↔ N2(X,v=4) + CO(X,v=5) | \| nitrogenCO2VV\| | 26.4837, 10.3646, -77.186\| |
| N2(X,v=6) + CO(X,v=4) ↔ N2(X,v=5) + CO(X,v=5) | \| nitrogenCO2VV\| | 29.0255, -26.5243, 75.5598\| |
| N2(X,v=7) + CO(X,v=4) ↔ N2(X,v=6) + CO(X,v=5) | \| nitrogenCO2VV\| | 31.3554, -59.646, 208.363\| |
| N2(X,v=8) + CO(X,v=4) ↔ N2(X,v=7) + CO(X,v=5) | \| nitrogenCO2VV\| | 32.9103, -80.3156, 289.647\| |
| N2(X,v=9) + CO(X,v=4) ↔ N2(X,v=8) + CO(X,v=5) | \| nitrogenCO2VV\| | 32.2215, -65.6863, 234.278\| |
| N2(X,v=10) + CO(X,v=4) ↔ N2(X,v=9) + CO(X,v=5) | \| nitrogenCO2VV\| | 31.1683, -46.4548, 163.069\| |
| N2(X,v=11) + CO(X,v=4) ↔ N2(X,v=10) + CO(X,v=5) | \| nitrogenCO2VV\| | 30.3445, -33.2697, 120.98\| |
| N2(X,v=12) + CO(X,v=4) ↔ N2(X,v=11) + CO(X,v=5) | \| nitrogenCO2VV\| | 30.8673, -39.6111, 142.913\| |
| N2(X,v=13) + CO(X,v=4) ↔ N2(X,v=12) + CO(X,v=5) | \| nitrogenCO2VV\| | 30.9429, -38.807, 134.935\| |
| N2(X,v=14) + CO(X,v=4) ↔ N2(X,v=13) + CO(X,v=5) | \| nitrogenCO2VV\| | 31.536, -44.8697, 145.124\| |
| N2(X,v=15) + CO(X,v=4) ↔ N2(X,v=14) + CO(X,v=5) | \| nitrogenCO2VV\| | 31.5625, -42.1543, 118.813\| |
| N2(X,v=16) + CO(X,v=4) ↔ N2(X,v=15) + CO(X,v=5) | \| nitrogenCO2VV\| | 30.7966, -27.205, 43.9881\| |
| N2(X,v=17) + CO(X,v=4) ↔ N2(X,v=16) + CO(X,v=5) | \| nitrogenCO2VV\| | 29.9453, -11.6386, -31.9725\| |
| N2(X,v=18) + CO(X,v=4) ↔ N2(X,v=17) + CO(X,v=5) | \| nitrogenCO2VV\| | 32.4229, -53.3216, 132.543\| |
| N2(X,v=19) + CO(X,v=4) ↔ N2(X,v=18) + CO(X,v=5) | \| nitrogenCO2VV\| | 30.9588, -26.4456, -0.733776\| |
| N2(X,v=20) + CO(X,v=4) ↔ N2(X,v=19) + CO(X,v=5) | \| nitrogenCO2VV\| | 33.4465, -61.8396, 105.668\| |
| N2(X,v=21) + CO(X,v=4) ↔ N2(X,v=20) + CO(X,v=5) | \| nitrogenCO2VV\| | 33.5614, -59.6581, 65.0434\| |
| N2(X,v=22) + CO(X,v=4) ↔ N2(X,v=21) + CO(X,v=5) | \| nitrogenCO2VV\| | 31.0166, -12.0466, -167.913\| |
| N2(X,v=23) + CO(X,v=4) ↔ N2(X,v=22) + CO(X,v=5) | \| nitrogenCO2VV\| | 27.3369, 54.3095, -478.07\| |
| N2(X,v=24) + CO(X,v=4) ↔ N2(X,v=23) + CO(X,v=5) | \| nitrogenCO2VV\| | 27.8215, 45.1497, -449.872\| |
| N2(X,v=25) + CO(X,v=4) ↔ N2(X,v=24) + CO(X,v=5) | \| nitrogenCO2VV\| | 30.601, -4.80598, -243.414\| |
| N2(X,v=26) + CO(X,v=4) ↔ N2(X,v=25) + CO(X,v=5) | \| nitrogenCO2VV\| | 32.0497, -29.2349, -160.124\| |
| N2(X,v=27) + CO(X,v=4) ↔ N2(X,v=26) + CO(X,v=5) | \| nitrogenCO2VV\| | 32.4764, -35.0556, -161.759\| |
| N2(X,v=28) + CO(X,v=4) ↔ N2(X,v=27) + CO(X,v=5) | \| nitrogenCO2VV\| | 33.1095, -45.0786, -142.571\| |
| N2(X,v=29) + CO(X,v=4) ↔ N2(X,v=28) + CO(X,v=5) | \| nitrogenCO2VV\| | 42.2409, -198.377, 476.918\| |
| N2(X,v=30) + CO(X,v=4) ↔ N2(X,v=29) + CO(X,v=5) | \| nitrogenCO2VV\| | 32.4783, -19.2927, -354.594\| |
| N2(X,v=31) + CO(X,v=4) ↔ N2(X,v=30) + CO(X,v=5) | \| nitrogenCO2VV\| | 27.2177, 73.0085, -771.099\| |
| N2(X,v=32) + CO(X,v=4) ↔ N2(X,v=31) + CO(X,v=5) | \| nitrogenCO2VV\| | 30.5627, 6.75581, -461.952\| |
| N2(X,v=33) + CO(X,v=4) ↔ N2(X,v=32) + CO(X,v=5) | \| nitrogenCO2VV\| | 33.9878, -59.0261, -165.177\| |
| N2(X,v=34) + CO(X,v=4) ↔ N2(X,v=33) + CO(X,v=5) | \| nitrogenCO2VV\| | 35.7541, -92.8773, -21.6812\| |
| N2(X,v=35) + CO(X,v=4) ↔ N2(X,v=34) + CO(X,v=5) | \| nitrogenCO2VV\| | 36.3242, -104.205, 16.3304\| |
| N2(X,v=36) + CO(X,v=4) ↔ N2(X,v=35) + CO(X,v=5) | \| nitrogenCO2VV\| | 36.4985, -108.17, 20.504\| |
| N2(X,v=37) + CO(X,v=4) ↔ N2(X,v=36) + CO(X,v=5) | \| nitrogenCO2VV\| | 36.4777, -108.577, 8.76549\| |
| N2(X,v=38) + CO(X,v=4) ↔ N2(X,v=37) + CO(X,v=5) | \| nitrogenCO2VV\| | 46.7419, -285.105, 747.615\| |
| N2(X,v=39) + CO(X,v=4) ↔ N2(X,v=38) + CO(X,v=5) | \| nitrogenCO2VV\| | 30.4683, 14.3525, -636.677\| |
| N2(X,v=40) + CO(X,v=4) ↔ N2(X,v=39) + CO(X,v=5) | \| nitrogenCO2VV\| | 29.3312, 29.4419, -691.741\| |
| N2(X,v=41) + CO(X,v=4) ↔ N2(X,v=40) + CO(X,v=5) | \| nitrogenCO2VV\| | 33.0479, -44.2884, -343.437\| |
| N2(X,v=42) + CO(X,v=4) ↔ N2(X,v=41) + CO(X,v=5) | \| nitrogenCO2VV\| | 35.8143, -97.9997, -100.055\| |
| N2(X,v=43) + CO(X,v=4) ↔ N2(X,v=42) + CO(X,v=5) | \| nitrogenCO2VV\| | 36.3897, -109.339, -61.2993\| |
| N2(X,v=44) + CO(X,v=4) ↔ N2(X,v=43) + CO(X,v=5) | \| nitrogenCO2VV\| | 36.3557, -108.98, -78.5912\| |
| N2(X,v=45) + CO(X,v=4) ↔ N2(X,v=44) + CO(X,v=5) | \| nitrogenCO2VV\| | 36.3334, -108.799, -95.2798\| |
| N2(X,v=46) + CO(X,v=4) ↔ N2(X,v=45) + CO(X,v=5) | \| nitrogenCO2VV\| | 36.1168, -105.086, -128.024\| |
| N2(X,v=47) + CO(X,v=4) ↔ N2(X,v=46) + CO(X,v=5) | \| nitrogenCO2VV\| | 48.8035, -327.144, 824.006\| |
| N2(X,v=48) + CO(X,v=4) ↔ N2(X,v=47) + CO(X,v=5) | \| nitrogenCO2VV\| | 25.2874, 110.195, -1214.76\| |
| N2(X,v=49) + CO(X,v=4) ↔ N2(X,v=48) + CO(X,v=5) | \| nitrogenCO2VV\| | 28.6964, 39.1873, -863.042\| |
| N2(X,v=50) + CO(X,v=4) ↔ N2(X,v=49) + CO(X,v=5) | \| nitrogenCO2VV\| | 34.2404, -69.1724, -351.095\| |
| N2(X,v=51) + CO(X,v=4) ↔ N2(X,v=50) + CO(X,v=5) | \| nitrogenCO2VV\| | 36.2763, -108.619, -177.511\| |
| N2(X,v=52) + CO(X,v=4) ↔ N2(X,v=51) + CO(X,v=5) | \| nitrogenCO2VV\| | 36.468, -112.471, -175.537\| |
| N2(X,v=53) + CO(X,v=4) ↔ N2(X,v=52) + CO(X,v=5) | \| nitrogenCO2VV\| | 36.4968, -113.2, -188.558\| |
| N2(X,v=54) + CO(X,v=4) ↔ N2(X,v=53) + CO(X,v=5) | \| nitrogenCO2VV\| | 36.5336, -114.091, -200.825\| |
| N2(X,v=55) + CO(X,v=4) ↔ N2(X,v=54) + CO(X,v=5) | \| nitrogenCO2VV\| | 36.3815, -111.526, -228.909\| |
| N2(X,v=56) + CO(X,v=4) ↔ N2(X,v=55) + CO(X,v=5) | \| nitrogenCO2VV\| | 51.4497, -378.406, 933.164\| |
| N2(X,v=57) + CO(X,v=4) ↔ N2(X,v=56) + CO(X,v=5) | \| nitrogenCO2VV\| | 22.2202, 167.809, -1623.57\| |
| N2(X,v=58) + CO(X,v=4) ↔ N2(X,v=57) + CO(X,v=5) | \| nitrogenCO2VV\| | 29.6881, 18.6193, -896.203\| |
| N2(X,v=59) + CO(X,v=4) ↔ N2(X,v=58) + CO(X,v=5) | \| nitrogenCO2VV\| | 35.6527, -97.8252, -345.861\| |
| | | |
| N2(X,v=1) + CO(X,v=5) ↔ N2(X,v=0) + CO(X,v=6) | \| nitrogenCO2VV\| | 29.5142, -45.8604, 7.53336\| |
| N2(X,v=2) + CO(X,v=5) ↔ N2(X,v=1) + CO(X,v=6) | \| nitrogenCO2VV\| | 6.01693, 350.254, -1584.71\| |
| N2(X,v=3) + CO(X,v=5) ↔ N2(X,v=2) + CO(X,v=6) | \| nitrogenCO2VV\| | 17.6738, 152.078, -704.339\| |
| N2(X,v=4) + CO(X,v=5) ↔ N2(X,v=3) + CO(X,v=6) | \| nitrogenCO2VV\| | 23.4978, 58.8402, -301.726\| |
| N2(X,v=5) + CO(X,v=5) ↔ N2(X,v=4) + CO(X,v=6) | nitrogenCO2VV | 26.8328, 8.57754, -88.2173 |
| N2(X,v=6) + CO(X,v=5) ↔ N2(X,v=5) + CO(X,v=6) | nitrogenCO2VV | 29.5105, -30.5336, 74.581 |
| N2(X,v=7) + CO(X,v=5) ↔ N2(X,v=6) + CO(X,v=6) | nitrogenCO2VV | 31.7628, -62.6401, 205.957 |
| N2(X,v=8) + CO(X,v=5) ↔ N2(X,v=7) + CO(X,v=6) | nitrogenCO2VV | 33.1834, -81.5264, 283.71 |
| N2(X,v=9) + CO(X,v=5) ↔ N2(X,v=8) + CO(X,v=6) | nitrogenCO2VV | 32.2106, -62.1688, 210.767 |
| N2(X,v=10) + CO(X,v=5) ↔ N2(X,v=9) + CO(X,v=6) | nitrogenCO2VV | 31.1889, -43.7306, 146.179 |
| N2(X,v=11) + CO(X,v=5) ↔ N2(X,v=10) + CO(X,v=6) | nitrogenCO2VV | 30.1934, -28.4173, 100.894 |
| N2(X,v=12) + CO(X,v=5) ↔ N2(X,v=11) + CO(X,v=6) | nitrogenCO2VV | 30.4178, -30.5468, 112.43 |
| N2(X,v=13) + CO(X,v=5) ↔ N2(X,v=12) + CO(X,v=6) | nitrogenCO2VV | 31.3229, -42.3498, 152.611 |
| N2(X,v=14) + CO(X,v=5) ↔ N2(X,v=13) + CO(X,v=6) | nitrogenCO2VV | 31.349, -40.9694, 142.642 |
| N2(X,v=15) + CO(X,v=5) ↔ N2(X,v=14) + CO(X,v=6) | nitrogenCO2VV | 31.8887, -45.8982, 147.361 |
| N2(X,v=16) + CO(X,v=5) ↔ N2(X,v=15) + CO(X,v=6) | nitrogenCO2VV | 31.508, -36.9472, 97.9109 |
| N2(X,v=17) + CO(X,v=5) ↔ N2(X,v=16) + CO(X,v=6) | nitrogenCO2VV | 30.5881, -20.1511, 18.6202 |
| N2(X,v=18) + CO(X,v=5) ↔ N2(X,v=17) + CO(X,v=6) | nitrogenCO2VV | 32.8848, -58.7855, 172.331 |
| N2(X,v=19) + CO(X,v=5) ↔ N2(X,v=18) + CO(X,v=6) | nitrogenCO2VV | 31.3515, -31.1981, 39.9442 |
| N2(X,v=20) + CO(X,v=5) ↔ N2(X,v=19) + CO(X,v=6) | nitrogenCO2VV | 33.2572, -57.285, 111.454 |
| N2(X,v=21) + CO(X,v=5) ↔ N2(X,v=20) + CO(X,v=6) | nitrogenCO2VV | 33.5127, -57.6153, 83.1573 |
| N2(X,v=22) + CO(X,v=5) ↔ N2(X,v=21) + CO(X,v=6) | nitrogenCO2VV | 32.0087, -28.4057, -68.2942 |
| N2(X,v=23) + CO(X,v=5) ↔ N2(X,v=22) + CO(X,v=6) | nitrogenCO2VV | 28.5656, 33.7739, -359.147 |
| N2(X,v=24) + CO(X,v=5) ↔ N2(X,v=23) + CO(X,v=6) | nitrogenCO2VV | 27.9806, 44.1908, -419.117 |
| N2(X,v=25) + CO(X,v=5) ↔ N2(X,v=24) + CO(X,v=6) | nitrogenCO2VV | 30.2715, 3.2029, -252.75 |
| N2(X,v=26) + CO(X,v=5) ↔ N2(X,v=25) + CO(X,v=6) | nitrogenCO2VV | 32.0498, -27.1048, -142.78 |
| N2(X,v=27) + CO(X,v=5) ↔ N2(X,v=26) + CO(X,v=6) | nitrogenCO2VV | 32.33, -29.924, -159.196 |
| N2(X,v=28) + CO(X,v=5) ↔ N2(X,v=27) + CO(X,v=6) | nitrogenCO2VV | 32.6899, -34.6917, -165.003 |
| N2(X,v=29) + CO(X,v=5) ↔ N2(X,v=28) + CO(X,v=6) | nitrogenCO2VV | 41.4027, -179.9, 415.623 |
| N2(X,v=30) + CO(X,v=5) ↔ N2(X,v=29) + CO(X,v=6) | nitrogenCO2VV | 32.9607, -24.6947, -308.695 |
| N2(X,v=31) + CO(X,v=5) ↔ N2(X,v=30) + CO(X,v=6) | nitrogenCO2VV | 26.7718, 85.2295, -808.576 |
| N2(X,v=32) + CO(X,v=5) ↔ N2(X,v=31) + CO(X,v=6) | nitrogenCO2VV | 29.9399, 22.4977, -517.029 |
| N2(X,v=33) + CO(X,v=5) ↔ N2(X,v=32) + CO(X,v=6) | nitrogenCO2VV | 33.5765, -46.9662, -204.696 |
| N2(X,v=34) + CO(X,v=5) ↔ N2(X,v=33) + CO(X,v=6) | nitrogenCO2VV | 35.5135, -83.7479, -49.1636 |
| N2(X,v=35) + CO(X,v=5) ↔ N2(X,v=34) + CO(X,v=6) | nitrogenCO2VV | 36.2572, -98.173, 2.18915 |
| N2(X,v=36) + CO(X,v=5) ↔ N2(X,v=35) + CO(X,v=6) | nitrogenCO2VV | 36.5624, -104.525, 16.8451 |
| N2(X,v=37) + CO(X,v=5) ↔ N2(X,v=36) + CO(X,v=6) | nitrogenCO2VV | 36.7179, -108.158, 19.524 |
| N2(X,v=38) + CO(X,v=5) ↔ N2(X,v=37) + CO(X,v=6) | nitrogenCO2VV | 46.8157, -281.452, 742.575 |
| N2(X,v=39) + CO(X,v=5) ↔ N2(X,v=38) + CO(X,v=6) | nitrogenCO2VV | 31.2319, 4.81176, -578.966 |
| N2(X,v=40) + CO(X,v=5) ↔ N2(X,v=39) + CO(X,v=6) | nitrogenCO2VV | 29.6753, 27.667, -670.008 |

| N2(X,v=41) + CO(X,v=5) ↔ N2(X,v=40) + CO(X,v=6) | nitrogenCO2VV | 33.1897, -42.2954, -339.355 |
| N2(X,v=42) + CO(X,v=5) ↔ N2(X,v=41) + CO(X,v=6) | nitrogenCO2VV | 35.9893, -96.735, -92.1615 |
| N2(X,v=43) + CO(X,v=5) ↔ N2(X,v=42) + CO(X,v=6) | nitrogenCO2VV | 36.6468, -109.711, -45.3402 |
| N2(X,v=44) + CO(X,v=5) ↔ N2(X,v=43) + CO(X,v=6) | nitrogenCO2VV | 36.6049, -109.269, -62.7512 |
| N2(X,v=45) + CO(X,v=5) ↔ N2(X,v=44) + CO(X,v=6) | nitrogenCO2VV | 36.566, -108.824, -80.4649 |
| N2(X,v=46) + CO(X,v=5) ↔ N2(X,v=45) + CO(X,v=6) | nitrogenCO2VV | 36.3256, -104.713, -114.849 |
| N2(X,v=47) + CO(X,v=5) ↔ N2(X,v=46) + CO(X,v=6) | nitrogenCO2VV | 48.7461, -321.736, 813.453 |
| N2(X,v=48) + CO(X,v=5) ↔ N2(X,v=47) + CO(X,v=6) | nitrogenCO2VV | 25.9671, 101.551, -1158.5 |
| N2(X,v=49) + CO(X,v=5) ↔ N2(X,v=48) + CO(X,v=6) | nitrogenCO2VV | 28.868, 40.2428, -852.982 |
| N2(X,v=50) + CO(X,v=5) ↔ N2(X,v=49) + CO(X,v=6) | nitrogenCO2VV | 34.2751, -65.5053, -353.456 |
| N2(X,v=51) + CO(X,v=5) ↔ N2(X,v=50) + CO(X,v=6) | nitrogenCO2VV | 36.4327, -107.294, -168.597 |
| N2(X,v=52) + CO(X,v=5) ↔ N2(X,v=51) + CO(X,v=6) | nitrogenCO2VV | 36.6527, -111.693, -163.981 |
| N2(X,v=53) + CO(X,v=5) ↔ N2(X,v=52) + CO(X,v=6) | nitrogenCO2VV | 36.6792, -112.382, -177.168 |
| N2(X,v=54) + CO(X,v=5) ↔ N2(X,v=53) + CO(X,v=6) | nitrogenCO2VV | 36.7149, -113.251, -189.519 |
| N2(X,v=55) + CO(X,v=5) ↔ N2(X,v=54) + CO(X,v=6) | nitrogenCO2VV | 36.5633, -110.7, -217.507 |
| N2(X,v=56) + CO(X,v=5) ↔ N2(X,v=55) + CO(X,v=6) | nitrogenCO2VV | 51.4013, -373.211, 923.903 |
| N2(X,v=57) + CO(X,v=5) ↔ N2(X,v=56) + CO(X,v=6) | nitrogenCO2VV | 22.6891, 163.1, -1585.49 |
| N2(X,v=58) + CO(X,v=5) ↔ N2(X,v=57) + CO(X,v=6) | nitrogenCO2VV | 29.7315, 22.1274, -897.746 |
| N2(X,v=59) + CO(X,v=5) ↔ N2(X,v=58) + CO(X,v=6) | nitrogenCO2VV | 35.7277, -94.9358, -344.367 |

| N2(X,v=1) + CO(X,v=6) ↔ N2(X,v=0) + CO(X,v=7) | nitrogenCO2VV | 30.4277, -58.554, 43.123 |
| N2(X,v=2) + CO(X,v=6) ↔ N2(X,v=1) + CO(X,v=7) | nitrogenCO2VV | 2.225, 421.313, -1919.33 |
| N2(X,v=3) + CO(X,v=6) ↔ N2(X,v=2) + CO(X,v=7) | nitrogenCO2VV | 16.604, 174.121, -818.798 |
| N2(X,v=4) + CO(X,v=6) ↔ N2(X,v=3) + CO(X,v=7) | nitrogenCO2VV | 23.7337, 58.3391, -319.318 |
| N2(X,v=5) + CO(X,v=6) ↔ N2(X,v=4) + CO(X,v=7) | nitrogenCO2VV | 27.4781, 1.50919, -79.4283 |
| N2(X,v=6) + CO(X,v=6) ↔ N2(X,v=5) + CO(X,v=7) | nitrogenCO2VV | 30.1914, -37.9909, 84.9749 |
| N2(X,v=7) + CO(X,v=6) ↔ N2(X,v=6) + CO(X,v=7) | nitrogenCO2VV | 32.2119, -66.4584, 203.827 |
| N2(X,v=8) + CO(X,v=6) ↔ N2(X,v=7) + CO(X,v=7) | nitrogenCO2VV | 33.3528, -81.1625, 268.278 |
| N2(X,v=9) + CO(X,v=6) ↔ N2(X,v=8) + CO(X,v=7) | nitrogenCO2VV | 32.1412, -57.8323, 180.698 |
| N2(X,v=10) + CO(X,v=6) ↔ N2(X,v=9) + CO(X,v=7) | nitrogenCO2VV | 31.1498, -40.177, 122.496 |
| N2(X,v=11) + CO(X,v=6) ↔ N2(X,v=10) + CO(X,v=7) | nitrogenCO2VV | 29.9809, -22.7362, 73.7917 |
| N2(X,v=12) + CO(X,v=6) ↔ N2(X,v=11) + CO(X,v=7) | nitrogenCO2VV | 30.1567, -24.6314, 89.1957 |
| N2(X,v=13) + CO(X,v=6) ↔ N2(X,v=12) + CO(X,v=7) | nitrogenCO2VV | 30.7599, -32.2445, 119.402 |
| N2(X,v=14) + CO(X,v=6) ↔ N2(X,v=13) + CO(X,v=7) | nitrogenCO2VV | 31.8423, -46.4704, 166.942 |
| N2(X,v=15) + CO(X,v=6) ↔ N2(X,v=14) + CO(X,v=7) | nitrogenCO2VV | 31.8545, -45.2031, 157.911 |
| N2(X,v=16) + CO(X,v=6) ↔ N2(X,v=15) + CO(X,v=7) | nitrogenCO2VV | 32.0918, -45.0987, 142.771 |
| N2(X,v=17) + CO(X,v=6) ↔ N2(X,v=16) + CO(X,v=7) | nitrogenCO2VV | 31.2948, -30.0207, 71.8323 |
| N2(X,v=18) + CO(X,v=6) ↔ N2(X,v=17) + CO(X,v=7) | nitrogenCO2VV | 33.2876, -63.4307, 205.534 |
| N2(X,v=19) + CO(X,v=6) ↔ N2(X,v=18) + CO(X,v=7) | nitrogenCO2VV | 31.7481, -36.1841, 78.553 |
| N2(X,v=20) + CO(X,v=6) ↔ N2(X,v=19) + CO(X,v=7) | nitrogenCO2VV | 33.1728, -54.891, 124.322 |
| N2(X,v=21) + CO(X,v=6) ↔ N2(X,v=20) + CO(X,v=7) | nitrogenCO2VV | 33.4772, -56.2233, 102.442 |
| N2(X,v=22) + CO(X,v=6) ↔ N2(X,v=21) + CO(X,v=7) | nitrogenCO2VV | 32.6263, -38.4704, 1.79401 |
| N2(X,v=23) + CO(X,v=6) ↔ N2(X,v=22) + CO(X,v=7) | nitrogenCO2VV | 29.8937, 11.108, -232.591 |
| N2(X,v=24) + CO(X,v=6) ↔ N2(X,v=23) + CO(X,v=7) | nitrogenCO2VV | 28.4263, 37.4974, -363.3 |
| N2(X,v=25) + CO(X,v=6) ↔ N2(X,v=24) + CO(X,v=7) | nitrogenCO2VV | 29.9746, 10.0087, -256.959 |
| N2(X,v=26) + CO(X,v=6) ↔ N2(X,v=25) + CO(X,v=7) | nitrogenCO2VV | 32.0364, -25.4241, -123.112 |
| N2(X,v=27) + CO(X,v=6) ↔ N2(X,v=26) + CO(X,v=7) | nitrogenCO2VV | 32.2862, -27.4687, -143.492 |
| N2(X,v=28) + CO(X,v=6) ↔ N2(X,v=27) + CO(X,v=7) | nitrogenCO2VV | 32.3621, -26.8957, -173.96 |
| N2(X,v=29) + CO(X,v=6) ↔ N2(X,v=28) + CO(X,v=7) | nitrogenCO2VV | 40.4871, -161.043, 355.045 |
| N2(X,v=30) + CO(X,v=6) ↔ N2(X,v=29) + CO(X,v=7) | nitrogenCO2VV | 33.5615, -33.2699, -245.985 |
| N2(X,v=31) + CO(X,v=6) ↔ N2(X,v=30) + CO(X,v=7) | nitrogenCO2VV | 26.4747, 93.7493, -826.846 |
| N2(X,v=32) + CO(X,v=6) ↔ N2(X,v=31) + CO(X,v=7) | nitrogenCO2VV | 29.2236, 39.1121, -574.291 |
| N2(X,v=33) + CO(X,v=6) ↔ N2(X,v=32) + CO(X,v=7) | nitrogenCO2VV | 33.0423, -33.4572, -249.135 |
| N2(X,v=34) + CO(X,v=6) ↔ N2(X,v=33) + CO(X,v=7) | nitrogenCO2VV | 35.1482, -73.0539, -82.4635 |
| N2(X,v=35) + CO(X,v=6) ↔ N2(X,v=34) + CO(X,v=7) | nitrogenCO2VV | 36.0438, -90.0706, -20.5866 |
| N2(X,v=36) + CO(X,v=6) ↔ N2(X,v=35) + CO(X,v=7) | nitrogenCO2VV | 36.495, -99.0088, 5.06243 |
| N2(X,v=37) + CO(X,v=6) ↔ N2(X,v=36) + CO(X,v=7) | nitrogenCO2VV | 36.8595, -106.396, 24.1797 |
| N2(X,v=38) + CO(X,v=6) ↔ N2(X,v=37) + CO(X,v=7) | nitrogenCO2VV | 46.8282, -277.12, 734.321 |
| N2(X,v=39) + CO(X,v=6) ↔ N2(X,v=38) + CO(X,v=7) | nitrogenCO2VV | 31.8877, -3.12846, -529.045 |
| N2(X,v=40) + CO(X,v=6) ↔ N2(X,v=39) + CO(X,v=7) | nitrogenCO2VV | 29.9481, 26.7825, -652.65 |
| N2(X,v=41) + CO(X,v=6) ↔ N2(X,v=40) + CO(X,v=7) | nitrogenCO2VV | 33.2985, -40.1377, -336.278 |
| N2(X,v=42) + CO(X,v=6) ↔ N2(X,v=41) + CO(X,v=7) | nitrogenCO2VV | 36.1251, -95.1832, -85.9054 |
| N2(X,v=43) + CO(X,v=6) ↔ N2(X,v=42) + CO(X,v=7) | nitrogenCO2VV | 36.8802, -110.098, -29.5854 |
| N2(X,v=44) + CO(X,v=6) ↔ N2(X,v=43) + CO(X,v=7) | nitrogenCO2VV | 36.8386, -109.739, -46.3032 |
| N2(X,v=45) + CO(X,v=6) ↔ N2(X,v=44) + CO(X,v=7) | nitrogenCO2VV | 36.7837, -109.055, -64.8792 |
| N2(X,v=46) + CO(X,v=6) ↔ N2(X,v=45) + CO(X,v=7) | nitrogenCO2VV | 36.5184, -104.538, -100.884 |
| N2(X,v=47) + CO(X,v=6) ↔ N2(X,v=46) + CO(X,v=7) | nitrogenCO2VV | 48.6711, -316.493, 803.532 |
| N2(X,v=48) + CO(X,v=6) ↔ N2(X,v=47) + CO(X,v=7) | nitrogenCO2VV | 26.6562, 92.2293, -1099.18 |
| N2(X,v=49) + CO(X,v=6) ↔ N2(X,v=48) + CO(X,v=7) | nitrogenCO2VV | 29.0355, 40.8556, -840.891 |
| N2(X,v=50) + CO(X,v=6) ↔ N2(X,v=49) + CO(X,v=7) | nitrogenCO2VV | 34.2851, -61.8945, -355.602 |
| N2(X,v=51) + CO(X,v=6) ↔ N2(X,v=50) + CO(X,v=7) | nitrogenCO2VV | 36.5601, -105.948, -159.832 |
| N2(X,v=52) + CO(X,v=6) ↔ N2(X,v=51) + CO(X,v=7) | nitrogenCO2VV | 36.8119, -110.96, -152.24 |
| N2(X,v=53) + CO(X,v=6) ↔ N2(X,v=52) + CO(X,v=7) | nitrogenCO2VV | 36.8352, -111.593, -165.661 |
| N2(X,v=54) + CO(X,v=6) ↔ N2(X,v=53) + CO(X,v=7) | nitrogenCO2VV | 36.8689, -112.427, -178.155 |
| N2(X,v=55) + CO(X,v=6) ↔ N2(X,v=54) + CO(X,v=7) | nitrogenCO2VV | 36.717, -109.873, -206.119 |
| N2(X,v=56) + CO(X,v=6) ↔ N2(X,v=55) + CO(X,v=7) | nitrogenCO2VV | 51.3191, -367.91, 914.139 |
| N2(X,v=57) + CO(X,v=6) ↔ N2(X,v=56) + CO(X,v=7) | nitrogenCO2VV | 23.1465, 158.071, -1545.9 |
| N2(X,v=58) + CO(X,v=6) ↔ N2(X,v=57) + CO(X,v=7) | nitrogenCO2VV | 29.7568, 25.4383, -898.339 |
| N2(X,v=59) + CO(X,v=6) ↔ N2(X,v=58) + CO(X,v=7) | nitrogenCO2VV | 35.767, -91.902, -343.573 |
| N2(X,v=1) + CO(X,v=7) ↔ N2(X,v=0) + CO(X,v=8) | nitrogenCO2VV | 31.403, -73.1946, 89.3921 |
| N2(X,v=2) + CO(X,v=7) ↔ N2(X,v=1) + CO(X,v=8) | nitrogenCO2VV | -1.09366, 483.763, -2217.19 |
| N2(X,v=3) + CO(X,v=7) ↔ N2(X,v=2) + CO(X,v=8) | nitrogenCO2VV | 15.9983, 187.669, -896.943 |
| N2(X,v=4) + CO(X,v=7) ↔ N2(X,v=3) + CO(X,v=8) | nitrogenCO2VV | 24.3359, 51.1143, -308.805 |
| N2(X,v=5) + CO(X,v=7) ↔ N2(X,v=4) + CO(X,v=8) | nitrogenCO2VV | 28.4318, -11.1314, -48.3867 |
| N2(X,v=6) + CO(X,v=7) ↔ N2(X,v=5) + CO(X,v=8) | nitrogenCO2VV | 31.0518, -48.7321, 107.383 |
| N2(X,v=7) + CO(X,v=7) ↔ N2(X,v=6) + CO(X,v=8) | nitrogenCO2VV | 32.6924, -70.9723, 202.432 |
| N2(X,v=8) + CO(X,v=7) ↔ N2(X,v=7) + CO(X,v=8) | nitrogenCO2VV | 33.4399, -79.4918, 244.777 |
| N2(X,v=9) + CO(X,v=7) ↔ N2(X,v=8) + CO(X,v=8) | nitrogenCO2VV | 32.0391, -52.9934, 145.584 |
| N2(X,v=10) + CO(X,v=7) ↔ N2(X,v=9) + CO(X,v=8) | nitrogenCO2VV | 31.0821, -36.1486, 93.5398 |
| N2(X,v=11) + CO(X,v=7) ↔ N2(X,v=10) + CO(X,v=8) | nitrogenCO2VV | 29.6858, -15.7056, 37.761 |
| N2(X,v=12) + CO(X,v=7) ↔ N2(X,v=11) + CO(X,v=8) | nitrogenCO2VV | 29.8717, -18.3125, 60.406 |
| N2(X,v=13) + CO(X,v=7) ↔ N2(X,v=12) + CO(X,v=8) | nitrogenCO2VV | 30.4619, -26.2484, 96.5757 |
| N2(X,v=14) + CO(X,v=7) ↔ N2(X,v=13) + CO(X,v=8) | nitrogenCO2VV | 31.2853, -37.0287, 137.264 |
| N2(X,v=15) + CO(X,v=7) ↔ N2(X,v=14) + CO(X,v=8) | nitrogenCO2VV | 32.4019, -51.4844, 184.048 |
| N2(X,v=16) + CO(X,v=7) ↔ N2(X,v=15) + CO(X,v=8) | nitrogenCO2VV | 32.4456, -51.1992, 179.52 |
| N2(X,v=17) + CO(X,v=7) ↔ N2(X,v=16) + CO(X,v=8) | nitrogenCO2VV | 32.0192, -40.4821, 124.433 |
| N2(X,v=18) + CO(X,v=7) ↔ N2(X,v=17) + CO(X,v=8) | nitrogenCO2VV | 33.628, -67.1309, 231.619 |
| N2(X,v=19) + CO(X,v=7) ↔ N2(X,v=18) + CO(X,v=8) | nitrogenCO2VV | 32.1268, -40.8926, 112.737 |
| N2(X,v=20) + CO(X,v=7) ↔ N2(X,v=19) + CO(X,v=8) | nitrogenCO2VV | 33.1611, -53.886, 140.427 |
| N2(X,v=21) + CO(X,v=7) ↔ N2(X,v=20) + CO(X,v=8) | nitrogenCO2VV | 33.477, -55.6677, 123.2 |
| N2(X,v=22) + CO(X,v=7) ↔ N2(X,v=21) + CO(X,v=8) | nitrogenCO2VV | 33.0088, -44.6233, 52.9067 |
| N2(X,v=23) + CO(X,v=7) ↔ N2(X,v=22) + CO(X,v=8) | nitrogenCO2VV | 31.091, -9.35236, -118.151 |
| N2(X,v=24) + CO(X,v=7) ↔ N2(X,v=23) + CO(X,v=8) | nitrogenCO2VV | 29.1411, 25.6512, -285.811 |
| N2(X,v=25) + CO(X,v=7) ↔ N2(X,v=24) + CO(X,v=8) | nitrogenCO2VV | 29.7958, 14.281, -250.581 |
| N2(X,v=26) + CO(X,v=7) ↔ N2(X,v=25) + CO(X,v=8) | nitrogenCO2VV | 31.9249, -22.4412, -109.584 |
| N2(X,v=27) + CO(X,v=7) ↔ N2(X,v=26) + CO(X,v=8) | nitrogenCO2VV | 32.3506, -27.5512, -115.949 |
| N2(X,v=28) + CO(X,v=7) ↔ N2(X,v=27) + CO(X,v=8) | nitrogenCO2VV | 32.1526, -21.9488, -168.877 |
| N2(X,v=29) + CO(X,v=7) ↔ N2(X,v=28) + CO(X,v=8) | nitrogenCO2VV | 39.5474, -142.616, 298.567 |
| N2(X,v=30) + CO(X,v=7) ↔ N2(X,v=29) + CO(X,v=8) | nitrogenCO2VV | 34.2218, -43.754, -172.586 |
| N2(X,v=31) + CO(X,v=7) ↔ N2(X,v=30) + CO(X,v=8) | nitrogenCO2VV | 26.4597, 96.2157, -815.085 |
| N2(X,v=32) + CO(X,v=7) ↔ N2(X,v=31) + CO(X,v=8) | nitrogenCO2VV | 28.5025, 55.0507, -626.504 |
| N2(X,v=33) + CO(X,v=7) ↔ N2(X,v=32) + CO(X,v=8) | nitrogenCO2VV | 32.4165, -18.9942, -296.016 |
| N2(X,v=34) + CO(X,v=7) ↔ N2(X,v=33) + CO(X,v=8) | nitrogenCO2VV | 34.6992, -61.4968, -117.97 |
| N2(X,v=35) + CO(X,v=7) ↔ N2(X,v=34) + CO(X,v=8) | nitrogenCO2VV | 35.7018, -80.1762, -50.3015 |
| N2(X,v=36) + CO(X,v=7) ↔ N2(X,v=35) + CO(X,v=8) | nitrogenCO2VV | 36.2946, -91.526, -14.9103 |
| N2(X,v=37) + CO(X,v=7) ↔ N2(X,v=36) + CO(X,v=8) | nitrogenCO2VV | 36.8791, -102.784, 20.7499 |
| N2(X,v=38) + CO(X,v=7) ↔ N2(X,v=37) + CO(X,v=8) | nitrogenCO2VV | 46.7606, -271.676, 721.151 |
| N2(X,v=39) + CO(X,v=7) ↔ N2(X,v=38) + CO(X,v=8) | nitrogenCO2VV | 32.4215, -9.09216, -488.489 |
| N2(X,v=40) + CO(X,v=7) ↔ N2(X,v=39) + CO(X,v=8) | nitrogenCO2VV | 30.1236, 27.3981, -642.395 |
| N2(X,v=41) + CO(X,v=7) ↔ N2(X,v=40) + CO(X,v=8) | nitrogenCO2VV | 33.3636, -37.4769, -335.703 |
| N2(X,v=42) + CO(X,v=7) ↔ N2(X,v=41) + CO(X,v=8) | nitrogenCO2VV | 36.2177, -93.123, -82.2962 |
| N2(X,v=43) + CO(X,v=7) ↔ N2(X,v=42) + CO(X,v=8) | nitrogenCO2VV | 37.0878, -110.31, -14.9449 |
| N2(X,v=44) + CO(X,v=7) ↔ N2(X,v=43) + CO(X,v=8) | nitrogenCO2VV | 37.0604, -110.303, -29.6788 |
| N2(X,v=45) + CO(X,v=7) ↔ N2(X,v=44) + CO(X,v=8) | nitrogenCO2VV | 36.9931, -109.461, -48.7041 |
| N2(X,v=46) + CO(X,v=7) ↔ N2(X,v=45) + CO(X,v=8) | nitrogenCO2VV | 36.7056, -104.603, -85.9879 |
| N2(X,v=47) + CO(X,v=7) ↔ N2(X,v=46) + CO(X,v=8) | nitrogenCO2VV | 48.5902, -311.483, 794.502 |
| N2(X,v=48) + CO(X,v=7) ↔ N2(X,v=47) + CO(X,v=8) | nitrogenCO2VV | 27.3632, 82.2188, -1036.81 |
| N2(X,v=49) + CO(X,v=7) ↔ N2(X,v=48) + CO(X,v=8) | nitrogenCO2VV | 29.2092, 40.9804, -826.585 |
| N2(X,v=50) + CO(X,v=7) ↔ N2(X,v=49) + CO(X,v=8) | nitrogenCO2VV | 34.2811, -58.3949, -357.294 |
| N2(X,v=51) + CO(X,v=7) ↔ N2(X,v=50) + CO(X,v=8) | nitrogenCO2VV | 36.6672, -104.596, -151.149 |
| N2(X,v=52) + CO(X,v=7) ↔ N2(X,v=51) + CO(X,v=8) | nitrogenCO2VV | 36.9543, -110.294, -140.223 |
| N2(X,v=53) + CO(X,v=7) ↔ N2(X,v=52) + CO(X,v=8) | nitrogenCO2VV | 36.9733, -110.85, -153.964 |
| N2(X,v=54) + CO(X,v=7) ↔ N2(X,v=53) + CO(X,v=8) | nitrogenCO2VV | 37.0039, -111.629, -166.685 |
| N2(X,v=55) + CO(X,v=7) ↔ N2(X,v=54) + CO(X,v=8) | nitrogenCO2VV | 36.8502, -109.049, -194.728 |
| N2(X,v=56) + CO(X,v=7) ↔ N2(X,v=55) + CO(X,v=8) | nitrogenCO2VV | 51.2108, -362.507, 903.89 |
| N2(X,v=57) + CO(X,v=7) ↔ N2(X,v=56) + CO(X,v=8) | nitrogenCO2VV | 23.6017, 152.696, -1504.7 |
| N2(X,v=58) + CO(X,v=7) ↔ N2(X,v=57) + CO(X,v=8) | nitrogenCO2VV | 29.7725, 28.5361, -897.909 |
| N2(X,v=59) + CO(X,v=7) ↔ N2(X,v=58) + CO(X,v=8) | nitrogenCO2VV | 35.7783, -88.7268, -343.463 |
| N2(X,v=1) + CO(X,v=8) ↔ N2(X,v=0) + CO(X,v=9) | nitrogenCO2VV | 32.3618, -88.3454, 140.387 |

| Reaction | Rate |
|---|---|
| N2(X,v=2) + CO(X,v=8) ↔ N2(X,v=1) + CO(X,v=9) | \| nitrogenCO2VV\| -3.73436, 533.706, -2459.22\| |
| N2(X,v=3) + CO(X,v=8) ↔ N2(X,v=2) + CO(X,v=9) | \| nitrogenCO2VV\| 15.912, 191.598, -932.546\| |
| N2(X,v=4) + CO(X,v=8) ↔ N2(X,v=3) + CO(X,v=9) | \| nitrogenCO2VV\| 25.2813, 37.4707, -270.39\| |
| N2(X,v=5) + CO(X,v=8) ↔ N2(X,v=4) + CO(X,v=9) | \| nitrogenCO2VV\| 29.6215, -28.2553, 1.66105\| |
| N2(X,v=6) + CO(X,v=8) ↔ N2(X,v=5) + CO(X,v=9) | \| nitrogenCO2VV\| 32.032, -61.9251, 139.847\| |
| N2(X,v=7) + CO(X,v=8) ↔ N2(X,v=6) + CO(X,v=9) | \| nitrogenCO2VV\| 33.1967, -76.1716, 202.899\| |
| N2(X,v=8) + CO(X,v=8) ↔ N2(X,v=7) + CO(X,v=9) | \| nitrogenCO2VV\| 33.4787, -77.0806, 216.061\| |
| N2(X,v=9) + CO(X,v=8) ↔ N2(X,v=8) + CO(X,v=9) | \| nitrogenCO2VV\| 31.9344, -48.1254, 107.766\| |
| N2(X,v=10) + CO(X,v=8) ↔ N2(X,v=9) + CO(X,v=9) | \| nitrogenCO2VV\| 31.0228, -32.201, 61.8396\| |
| N2(X,v=11) + CO(X,v=8) ↔ N2(X,v=10) + CO(X,v=9) | \| nitrogenCO2VV\| 29.3031, -7.14863, -7.77146\| |
| N2(X,v=12) + CO(X,v=8) ↔ N2(X,v=11) + CO(X,v=9) | \| nitrogenCO2VV\| 29.5279, -10.9223, 23.6776\| |
| N2(X,v=13) + CO(X,v=8) ↔ N2(X,v=12) + CO(X,v=9) | \| nitrogenCO2VV\| 30.1699, -20.215, 69.4873\| |
| N2(X,v=14) + CO(X,v=8) ↔ N2(X,v=13) + CO(X,v=9) | \| nitrogenCO2VV\| 31.0208, -31.9396, 118.412\| |
| N2(X,v=15) + CO(X,v=8) ↔ N2(X,v=14) + CO(X,v=9) | \| nitrogenCO2VV\| 32.0295, -45.3682, 167.699\| |
| N2(X,v=16) + CO(X,v=8) ↔ N2(X,v=15) + CO(X,v=9) | \| nitrogenCO2VV\| 32.8678, -55.2454, 196.072\| |
| N2(X,v=17) + CO(X,v=8) ↔ N2(X,v=16) + CO(X,v=9) | \| nitrogenCO2VV\| 32.8027, -53.8685, 188.364\| |
| N2(X,v=18) + CO(X,v=8) ↔ N2(X,v=17) + CO(X,v=9) | \| nitrogenCO2VV\| 33.9073, -69.9873, 251.113\| |
| N2(X,v=19) + CO(X,v=8) ↔ N2(X,v=18) + CO(X,v=9) | \| nitrogenCO2VV\| 32.4666, -44.9351, 140.834\| |
| N2(X,v=20) + CO(X,v=8) ↔ N2(X,v=19) + CO(X,v=9) | \| nitrogenCO2VV\| 33.1698, -53.2823, 155.337\| |
| N2(X,v=21) + CO(X,v=8) ↔ N2(X,v=20) + CO(X,v=9) | \| nitrogenCO2VV\| 33.4986, -55.5803, 143.452\| |
| N2(X,v=22) + CO(X,v=8) ↔ N2(X,v=21) + CO(X,v=9) | \| nitrogenCO2VV\| 33.2677, -48.7401, 93.0705\| |
| N2(X,v=23) + CO(X,v=8) ↔ N2(X,v=22) + CO(X,v=9) | \| nitrogenCO2VV\| 32.0049, -24.8499, -27.8647\| |
| N2(X,v=24) + CO(X,v=8) ↔ N2(X,v=23) + CO(X,v=9) | \| nitrogenCO2VV\| 30.042, 10.3659, -195.065\| |
| N2(X,v=25) + CO(X,v=8) ↔ N2(X,v=24) + CO(X,v=9) | \| nitrogenCO2VV\| 29.8201, 14.6561, -228.041\| |
| N2(X,v=26) + CO(X,v=8) ↔ N2(X,v=25) + CO(X,v=9) | \| nitrogenCO2VV\| 31.6916, -17.5779, -105.225\| |
| N2(X,v=27) + CO(X,v=8) ↔ N2(X,v=26) + CO(X,v=9) | \| nitrogenCO2VV\| 32.517, -29.8576, -78.6974\| |
| N2(X,v=28) + CO(X,v=8) ↔ N2(X,v=27) + CO(X,v=9) | \| nitrogenCO2VV\| 32.0567, -19.5722, -151.69\| |
| N2(X,v=29) + CO(X,v=8) ↔ N2(X,v=28) + CO(X,v=9) | \| nitrogenCO2VV\| 38.6269, -125.262, 248.641\| |
| N2(X,v=30) + CO(X,v=8) ↔ N2(X,v=29) + CO(X,v=9) | \| nitrogenCO2VV\| 34.8377, -54.0826, -98.4217\| |
| N2(X,v=31) + CO(X,v=8) ↔ N2(X,v=30) + CO(X,v=9) | \| nitrogenCO2VV\| 26.8297, 90.8501, -765.404\| |
| N2(X,v=32) + CO(X,v=8) ↔ N2(X,v=31) + CO(X,v=9) | \| nitrogenCO2VV\| 27.8804, 68.4718, -665.245\| |
| N2(X,v=33) + CO(X,v=8) ↔ N2(X,v=32) + CO(X,v=9) | \| nitrogenCO2VV\| 31.7351, -4.18328, -342.516\| |
| N2(X,v=34) + CO(X,v=8) ↔ N2(X,v=33) + CO(X,v=9) | \| nitrogenCO2VV\| 34.2136, -49.9261, -151.515\| |
| N2(X,v=35) + CO(X,v=8) ↔ N2(X,v=34) + CO(X,v=9) | \| nitrogenCO2VV\| 35.2671, -69.1465, -83.5518\| |
| N2(X,v=36) + CO(X,v=8) ↔ N2(X,v=35) + CO(X,v=9) | \| nitrogenCO2VV\| 35.972, -82.2682, -41.7865\| |
| N2(X,v=37) + CO(X,v=8) ↔ N2(X,v=36) + CO(X,v=9) | \| nitrogenCO2VV\| 36.7554, -96.9259, 7.83565\| |

| Reaction | Rate |
|---|---|
| N2(X,v=38) + CO(X,v=8) ↔ N2(X,v=37) + CO(X,v=9) | \| nitrogenCO2VV\| 46.5894, -264.667, 701.364\| |
| N2(X,v=39) + CO(X,v=8) ↔ N2(X,v=38) + CO(X,v=9) | \| nitrogenCO2VV\| 32.8274, -12.9261, -457.74\| |
| N2(X,v=40) + CO(X,v=8) ↔ N2(X,v=39) + CO(X,v=9) | \| nitrogenCO2VV\| 30.1723, 30.1307, -641.936\| |
| N2(X,v=41) + CO(X,v=8) ↔ N2(X,v=40) + CO(X,v=9) | \| nitrogenCO2VV\| 33.3654, -33.8782, -339.516\| |
| N2(X,v=42) + CO(X,v=8) ↔ N2(X,v=41) + CO(X,v=9) | \| nitrogenCO2VV\| 36.2567, -90.28, -82.5366\| |
| N2(X,v=43) + CO(X,v=8) ↔ N2(X,v=42) + CO(X,v=9) | \| nitrogenCO2VV\| 37.2602, -110.071, -2.68413\| |
| N2(X,v=44) + CO(X,v=8) ↔ N2(X,v=43) + CO(X,v=9) | \| nitrogenCO2VV\| 37.2675, -110.81, -13.6114\| |
| N2(X,v=45) + CO(X,v=8) ↔ N2(X,v=44) + CO(X,v=9) | \| nitrogenCO2VV\| 37.1955, -109.964, -32.3401\| |
| N2(X,v=46) + CO(X,v=8) ↔ N2(X,v=45) + CO(X,v=9) | \| nitrogenCO2VV\| 36.8942, -104.931, -70.0985\| |
| N2(X,v=47) + CO(X,v=8) ↔ N2(X,v=46) + CO(X,v=9) | \| nitrogenCO2VV\| 48.5121, -306.762, 786.572\| |
| N2(X,v=48) + CO(X,v=8) ↔ N2(X,v=47) + CO(X,v=9) | \| nitrogenCO2VV\| 28.0918, 71.5555, -971.601\| |
| N2(X,v=49) + CO(X,v=8) ↔ N2(X,v=48) + CO(X,v=9) | \| nitrogenCO2VV\| 29.3963, 40.5812, -809.933\| |
| N2(X,v=50) + CO(X,v=8) ↔ N2(X,v=49) + CO(X,v=9) | \| nitrogenCO2VV\| 34.2713, -55.0634, -358.291\| |
| N2(X,v=51) + CO(X,v=8) ↔ N2(X,v=50) + CO(X,v=9) | \| nitrogenCO2VV\| 36.7601, -103.258, -142.477\| |
| N2(X,v=52) + CO(X,v=8) ↔ N2(X,v=51) + CO(X,v=9) | \| nitrogenCO2VV\| 37.0864, -109.721, -127.82\| |
| N2(X,v=53) + CO(X,v=8) ↔ N2(X,v=52) + CO(X,v=9) | \| nitrogenCO2VV\| 37.0995, -110.175, -141.985\| |
| N2(X,v=54) + CO(X,v=8) ↔ N2(X,v=53) + CO(X,v=9) | \| nitrogenCO2VV\| 37.1256, -110.875, -155.04\| |
| N2(X,v=55) + CO(X,v=8) ↔ N2(X,v=54) + CO(X,v=9) | \| nitrogenCO2VV\| 36.9683, -108.236, -183.307\| |
| N2(X,v=56) + CO(X,v=8) ↔ N2(X,v=55) + CO(X,v=9) | \| nitrogenCO2VV\| 51.0817, -357.009, 893.191\| |
| N2(X,v=57) + CO(X,v=8) ↔ N2(X,v=56) + CO(X,v=9) | \| nitrogenCO2VV\| 24.0612, 146.943, -1461.72\| |
| N2(X,v=58) + CO(X,v=8) ↔ N2(X,v=57) + CO(X,v=9) | \| nitrogenCO2VV\| 29.7843, 31.4039, -896.377\| |
| N2(X,v=59) + CO(X,v=8) ↔ N2(X,v=58) + CO(X,v=9) | \| nitrogenCO2VV\| 35.7668, -85.4168, -344.006\| |
| N2(X,v=1) + CO(X,v=9) ↔ N2(X,v=0) + CO(X,v=10) | \| nitrogenCO2VV\| 33.2204, -102.447, 189.316\| |
| N2(X,v=2) + CO(X,v=9) ↔ N2(X,v=1) + CO(X,v=10) | \| nitrogenCO2VV\| -5.6317, 569.782, -2637.94\| |
| N2(X,v=3) + CO(X,v=9) ↔ N2(X,v=2) + CO(X,v=10) | \| nitrogenCO2VV\| 16.3071, 186.471, -927.03\| |
| N2(X,v=4) + CO(X,v=9) ↔ N2(X,v=3) + CO(X,v=10) | \| nitrogenCO2VV\| 26.4668, 19.1097, -210.435\| |
| N2(X,v=5) + CO(X,v=9) ↔ N2(X,v=4) + CO(X,v=10) | \| nitrogenCO2VV\| 30.9276, -47.9501, 63.8296\| |
| N2(X,v=6) + CO(X,v=9) ↔ N2(X,v=5) + CO(X,v=10) | \| nitrogenCO2VV\| 33.0599, -76.5064, 179.239\| |
| N2(X,v=7) + CO(X,v=9) ↔ N2(X,v=6) + CO(X,v=10) | \| nitrogenCO2VV\| 33.7306, -82.2853, 207.325\| |
| N2(X,v=8) + CO(X,v=9) ↔ N2(X,v=7) + CO(X,v=10) | \| nitrogenCO2VV\| 33.523, -74.9006, 186.907\| |
| N2(X,v=9) + CO(X,v=9) ↔ N2(X,v=8) + CO(X,v=10) | \| nitrogenCO2VV\| 31.8635, -43.8771, 70.5171\| |
| N2(X,v=10) + CO(X,v=9) ↔ N2(X,v=9) + CO(X,v=10) | \| nitrogenCO2VV\| 31.018, -29.1196, 31.0993\| |
| N2(X,v=11) + CO(X,v=9) ↔ N2(X,v=10) + CO(X,v=10) | \| nitrogenCO2VV\| 28.8321, 2.99391, -62.8544\| |
| N2(X,v=12) + CO(X,v=9) ↔ N2(X,v=11) + CO(X,v=10) | \| nitrogenCO2VV\| 29.1205, -2.32015, -21.3736\| |

| N2(X,v=13) + CO(X,v=9) ↔ N2(X,v=12) + CO(X,v=10) | nitrogenCO2VV | 29.8514, -13.5554, 36.1305 |
| N2(X,v=14) + CO(X,v=9) ↔ N2(X,v=13) + CO(X,v=10) | nitrogenCO2VV | 30.7857, -27.1095, 96.3588 |
| N2(X,v=15) + CO(X,v=9) ↔ N2(X,v=14) + CO(X,v=10) | nitrogenCO2VV | 31.8973, -42.6205, 157.965 |
| N2(X,v=16) + CO(X,v=9) ↔ N2(X,v=15) + CO(X,v=10) | nitrogenCO2VV | 32.9819, -56.9686, 209.272 |
| N2(X,v=17) + CO(X,v=9) ↔ N2(X,v=16) + CO(X,v=10) | nitrogenCO2VV | 33.1225, -56.1964, 198.935 |
| N2(X,v=18) + CO(X,v=9) ↔ N2(X,v=17) + CO(X,v=10) | nitrogenCO2VV | 34.1266, -72.9717, 269.1 |
| N2(X,v=19) + CO(X,v=9) ↔ N2(X,v=18) + CO(X,v=10) | nitrogenCO2VV | 32.7486, -48.0928, 162.049 |
| N2(X,v=20) + CO(X,v=9) ↔ N2(X,v=19) + CO(X,v=10) | nitrogenCO2VV | 33.1203, -51.818, 164.014 |
| N2(X,v=21) + CO(X,v=9) ↔ N2(X,v=20) + CO(X,v=10) | nitrogenCO2VV | 33.5025, -55.2404, 159.984 |
| N2(X,v=22) + CO(X,v=9) ↔ N2(X,v=21) + CO(X,v=10) | nitrogenCO2VV | 33.4518, -51.6074, 125.475 |
| N2(X,v=23) + CO(X,v=9) ↔ N2(X,v=22) + CO(X,v=10) | nitrogenCO2VV | 32.6541, -35.7329, 40.0799 |
| N2(X,v=24) + CO(X,v=9) ↔ N2(X,v=23) + CO(X,v=10) | nitrogenCO2VV | 30.9868, -5.67572, -103.473 |
| N2(X,v=25) + CO(X,v=9) ↔ N2(X,v=24) + CO(X,v=10) | nitrogenCO2VV | 30.0975, 10.3894, -186.584 |
| N2(X,v=26) + CO(X,v=9) ↔ N2(X,v=25) + CO(X,v=10) | nitrogenCO2VV | 31.4012, -11.8884, -105.723 |
| N2(X,v=27) + CO(X,v=9) ↔ N2(X,v=26) + CO(X,v=10) | nitrogenCO2VV | 32.7705, -33.9485, -34.4103 |
| N2(X,v=28) + CO(X,v=9) ↔ N2(X,v=27) + CO(X,v=10) | nitrogenCO2VV | 32.0509, -19.1746, -125.744 |
| N2(X,v=29) + CO(X,v=9) ↔ N2(X,v=28) + CO(X,v=10) | nitrogenCO2VV | 37.7578, -109.427, 206.721 |
| N2(X,v=30) + CO(X,v=9) ↔ N2(X,v=29) + CO(X,v=10) | nitrogenCO2VV | 35.3059, -62.2261, -33.2488 |
| N2(X,v=31) + CO(X,v=9) ↔ N2(X,v=30) + CO(X,v=10) | nitrogenCO2VV | 27.6315, 76.9357, -675.108 |
| N2(X,v=32) + CO(X,v=9) ↔ N2(X,v=31) + CO(X,v=10) | nitrogenCO2VV | 27.4622, 77.5115, -682.173 |
| N2(X,v=33) + CO(X,v=9) ↔ N2(X,v=32) + CO(X,v=10) | nitrogenCO2VV | 31.0404, 10.261, -385.5 |
| N2(X,v=34) + CO(X,v=9) ↔ N2(X,v=33) + CO(X,v=10) | nitrogenCO2VV | 33.7322, -39.0703, -179.677 |
| N2(X,v=35) + CO(X,v=9) ↔ N2(X,v=34) + CO(X,v=10) | nitrogenCO2VV | 34.7918, -57.9553, -115.578 |
| N2(X,v=36) + CO(X,v=9) ↔ N2(X,v=35) + CO(X,v=10) | nitrogenCO2VV | 35.5545, -71.7656, -72.7516 |
| N2(X,v=37) + CO(X,v=9) ↔ N2(X,v=36) + CO(X,v=10) | nitrogenCO2VV | 36.4783, -88.6724, -14.7636 |
| N2(X,v=38) + CO(X,v=9) ↔ N2(X,v=37) + CO(X,v=10) | nitrogenCO2VV | 46.2905, -255.678, 673.509 |
| N2(X,v=39) + CO(X,v=9) ↔ N2(X,v=38) + CO(X,v=10) | nitrogenCO2VV | 33.1156, -14.8287, -435.521 |
| N2(X,v=40) + CO(X,v=9) ↔ N2(X,v=39) + CO(X,v=10) | nitrogenCO2VV | 30.0684, 35.4771, -653.336 |
| N2(X,v=41) + CO(X,v=9) ↔ N2(X,v=40) + CO(X,v=10) | nitrogenCO2VV | 33.2786, -28.8421, -349.83 |
| N2(X,v=42) + CO(X,v=9) ↔ N2(X,v=41) + CO(X,v=10) | nitrogenCO2VV | 36.228, -86.3459, -87.9188 |
| N2(X,v=43) + CO(X,v=9) ↔ N2(X,v=42) + CO(X,v=10) | nitrogenCO2VV | 37.3816, -109.034, 5.63488 |
| N2(X,v=44) + CO(X,v=9) ↔ N2(X,v=43) + CO(X,v=10) | nitrogenCO2VV | 37.4509, -111.026, 0.820871 |
| N2(X,v=45) + CO(X,v=9) ↔ N2(X,v=44) + CO(X,v=10) | nitrogenCO2VV | 37.3873, -110.424, -16.465 |
| N2(X,v=46) + CO(X,v=9) ↔ N2(X,v=45) + CO(X,v=10) | nitrogenCO2VV | 37.0877, -105.513, -53.3103 |
| N2(X,v=47) + CO(X,v=9) ↔ N2(X,v=46) + CO(X,v=10) | nitrogenCO2VV | 48.4428, -302.366, 779.847 |
| N2(X,v=48) + CO(X,v=9) ↔ N2(X,v=47) + CO(X,v=10) | nitrogenCO2VV | 28.8408, 60.34, -904.086 |

| N2(X,v=49) + CO(X,v=9) ↔ N2(X,v=48) + CO(X,v=10) | nitrogenCO2VV | 29.6015, 39.645, -790.914 |
| N2(X,v=50) + CO(X,v=9) ↔ N2(X,v=49) + CO(X,v=10) | nitrogenCO2VV | 34.2623, -51.9532, -358.373 |
| N2(X,v=51) + CO(X,v=9) ↔ N2(X,v=50) + CO(X,v=10) | nitrogenCO2VV | 36.8435, -101.954, -133.755 |
| N2(X,v=52) + CO(X,v=9) ↔ N2(X,v=51) + CO(X,v=10) | nitrogenCO2VV | 37.2134, -109.272, -114.909 |
| N2(X,v=53) + CO(X,v=9) ↔ N2(X,v=52) + CO(X,v=10) | nitrogenCO2VV | 37.2188, -109.595, -129.61 |
| N2(X,v=54) + CO(X,v=9) ↔ N2(X,v=53) + CO(X,v=10) | nitrogenCO2VV | 37.2385, -110.184, -143.131 |
| N2(X,v=55) + CO(X,v=9) ↔ N2(X,v=54) + CO(X,v=10) | nitrogenCO2VV | 37.0752, -107.442, -171.815 |
| N2(X,v=56) + CO(X,v=9) ↔ N2(X,v=55) + CO(X,v=10) | nitrogenCO2VV | 50.9359, -351.43, 882.095 |
| N2(X,v=57) + CO(X,v=9) ↔ N2(X,v=56) + CO(X,v=10) | nitrogenCO2VV | 24.5304, 140.774, -1416.81 |
| N2(X,v=58) + CO(X,v=9) ↔ N2(X,v=57) + CO(X,v=10) | nitrogenCO2VV | 29.7967, 34.0224, -893.654 |
| N2(X,v=59) + CO(X,v=9) ↔ N2(X,v=58) + CO(X,v=10) | nitrogenCO2VV | 35.7364, -81.9823, -345.153 |

ii. $N_2$-$CO_2$ V-V:

The process:

$CO_2(00^0v_i1) + N_2(w_i) \leftrightarrow CO_2(00^0v_f1) + N_2(w_f)$   | KustovaScaling_CO2v3_N2 | $v_i$*$w_f$ | has a rate coefficient of $k = v_i \cdot w_f \cdot 1.66 \cdot 10^{-30} \cdot \exp(27.221 + 10.8178 \cdot T^{-1/3} - 224.158 \cdot T^{-2/3})$ in m³s⁻¹.

| CO2(X,v=00011) + N2(X,v=0) ↔ CO2(X,v=00001) + N2(X,v=1) | nitrogenCO2VV | 43.8, -306, 1288 |
%k=k0*(v3+1)(w+1)
| CO2(X,v=00011) + N2(X,v=1) ↔ CO2(X,v=00021) + N2(X,v=0) | nitrogenCO2VV | 46.953, -349.851, 1482 |
| CO2(X,v=00021) + N2(X,v=1) ↔ CO2(X,v=00031) + N2(X,v=0) | nitrogenCO2VV | 41.6815, -239.359, 944.929 |
| CO2(X,v=00031) + N2(X,v=1) ↔ CO2(X,v=00041) + N2(X,v=0) | KustovaScaling_CO2v3_N2 | 4 |
| CO2(X,v=00041) + N2(X,v=1) ↔ CO2(X,v=00051) + N2(X,v=0) | KustovaScaling_CO2v3_N2 | 5 |

| CO2(X,v=00011) + N2(X,v=1) ↔ CO2(X,v=00001) + N2(X,v=2) | nitrogenCO2VV | 35.8752, -138.147, 471.585 |
| CO2(X,v=00021) + N2(X,v=1) ↔ CO2(X,v=00011) + N2(X,v=2) | nitrogenCO2VV | 44.8346, -294.813, 1213.08 |
| CO2(X,v=00021) + N2(X,v=2) ↔ CO2(X,v=00031) + N2(X,v=1) | nitrogenCO2VV | 48.0886, -350.201, 1481.91 |
| CO2(X,v=00031) + N2(X,v=2) ↔ CO2(X,v=00041) + N2(X,v=1) | nitrogenCO2VV | 44.0283, -265.507, 1069.99 |
| CO2(X,v=00041) + N2(X,v=2) ↔ CO2(X,v=00051) + N2(X,v=1) | KustovaScaling_CO2v3_N2 | 10 |

| CO2(X,v=00011) + N2(X,v=2) ↔ CO2(X,v=00001) + N2(X,v=3) | KustovaScaling_CO2v3_N2 | 3 |
| CO2(X,v=00021) + N2(X,v=2) ↔ CO2(X,v=00011) + N2(X,v=3) | KustovaScaling_CO2v3_N2 | 6 |
| CO2(X,v=00031) + N2(X,v=2) ↔ CO2(X,v=00021) + N2(X,v=3) | nitrogenCO2VV | 45.2075, -286.167, 1170.46 |
| CO2(X,v=00041) + N2(X,v=2) ↔ CO2(X,v=00031) + N2(X,v=3) | nitrogenCO2VV | 48.6965, -348.305, 1471.57 |
| CO2(X,v=00041) + N2(X,v=3) ↔ CO2(X,v=00051) + N2(X,v=2) | nitrogenCO2VV | 45.8051, -287.873, 1178.84 |

CO2(X,v=00011) + N2(X,v=3) ↔ CO2(X,v=00001) + N2(X,v=4)  |  KustovaScaling_CO2v3_N2 | 4 |
CO2(X,v=00021) + N2(X,v=3) ↔ CO2(X,v=00011) + N2(X,v=4)  |  KustovaScaling_CO2v3_N2 | 8 |
CO2(X,v=00031) + N2(X,v=3) ↔ CO2(X,v=00021) + N2(X,v=4)  |  KustovaScaling_CO2v3_N2 | 12 |
CO2(X,v=00041) + N2(X,v=3) ↔ CO2(X,v=00031) + N2(X,v=4)  |  nitrogenCO2VV           | 44.6275, -263.695, 1061.25 |
CO2(X,v=00051) + N2(X,v=3) ↔ CO2(X,v=00041) + N2(X,v=4)  |  nitrogenCO2VV           | 49.2916, -350.245, 1482.44 |

CO2(X,v=00011) + N2(X,v=4) ↔ CO2(X,v=00001) + N2(X,v=5)  |  KustovaScaling_CO2v3_N2 | 5 |
CO2(X,v=00021) + N2(X,v=4) ↔ CO2(X,v=00011) + N2(X,v=5)  |  KustovaScaling_CO2v3_N2 | 10 |
CO2(X,v=00031) + N2(X,v=4) ↔ CO2(X,v=00021) + N2(X,v=5)  |  KustovaScaling_CO2v3_N2 | 15 |
CO2(X,v=00041) + N2(X,v=4) ↔ CO2(X,v=00031) + N2(X,v=5)  |  KustovaScaling_CO2v3_N2 | 20 |
CO2(X,v=00051) + N2(X,v=4) ↔ CO2(X,v=00041) + N2(X,v=5)  |  nitrogenCO2VV           | 43.6765, -236.98, 933.642 |

CO2(X,v=00011) + N2(X,v=5) ↔ CO2(X,v=00001) + N2(X,v=6)  |  KustovaScaling_CO2v3_N2 | 6 |
CO2(X,v=00021) + N2(X,v=5) ↔ CO2(X,v=00011) + N2(X,v=6)  |  KustovaScaling_CO2v3_N2 | 12 |
CO2(X,v=00031) + N2(X,v=5) ↔ CO2(X,v=00021) + N2(X,v=6)  |  KustovaScaling_CO2v3_N2 | 18 |
CO2(X,v=00041) + N2(X,v=5) ↔ CO2(X,v=00031) + N2(X,v=6)  |  KustovaScaling_CO2v3_N2 | 24 |
CO2(X,v=00051) + N2(X,v=5) ↔ CO2(X,v=00041) + N2(X,v=6)  |  KustovaScaling_CO2v3_N2 | 30 |

CO2(X,v=00011) + N2(X,v=6) ↔ CO2(X,v=00001) + N2(X,v=7)  |  KustovaScaling_CO2v3_N2 | 7 |
CO2(X,v=00021) + N2(X,v=6) ↔ CO2(X,v=00011) + N2(X,v=7)  |  KustovaScaling_CO2v3_N2 | 14 |
CO2(X,v=00031) + N2(X,v=6) ↔ CO2(X,v=00021) + N2(X,v=7)  |  KustovaScaling_CO2v3_N2 | 21 |
CO2(X,v=00041) + N2(X,v=6) ↔ CO2(X,v=00031) + N2(X,v=7)  |  KustovaScaling_CO2v3_N2 | 28 |
CO2(X,v=00051) + N2(X,v=6) ↔ CO2(X,v=00041) + N2(X,v=7)  |  KustovaScaling_CO2v3_N2 | 35 |

CO2(X,v=00011) + N2(X,v=7) ↔ CO2(X,v=00001) + N2(X,v=8)  |  KustovaScaling_CO2v3_N2 | 8 |
CO2(X,v=00021) + N2(X,v=7) ↔ CO2(X,v=00011) + N2(X,v=8)  |  KustovaScaling_CO2v3_N2 | 16 |
CO2(X,v=00031) + N2(X,v=7) ↔ CO2(X,v=00021) + N2(X,v=8)  |  KustovaScaling_CO2v3_N2 | 24 |
CO2(X,v=00041) + N2(X,v=7) ↔ CO2(X,v=00031) + N2(X,v=8)  |  KustovaScaling_CO2v3_N2 | 32 |
CO2(X,v=00051) + N2(X,v=7) ↔ CO2(X,v=00041) + N2(X,v=8)  |  KustovaScaling_CO2v3_N2 | 40 |

CO2(X,v=00011) + N2(X,v=8) ↔ CO2(X,v=00001) + N2(X,v=9)  |  KustovaScaling_CO2v3_N2 | 9 |
CO2(X,v=00021) + N2(X,v=8) ↔ CO2(X,v=00011) + N2(X,v=9)  |  KustovaScaling_CO2v3_N2 | 18 |
CO2(X,v=00031) + N2(X,v=8) ↔ CO2(X,v=00021) + N2(X,v=9)  |  KustovaScaling_CO2v3_N2 | 27 |
CO2(X,v=00041) + N2(X,v=8) ↔ CO2(X,v=00031) + N2(X,v=9)  |  KustovaScaling_CO2v3_N2 | 36 |
CO2(X,v=00051) + N2(X,v=8) ↔ CO2(X,v=00041) + N2(X,v=9)  |  KustovaScaling_CO2v3_N2 | 45 |

CO2(X,v=00011) + N2(X,v=9) ↔ CO2(X,v=00001) + N2(X,v=10)  |  KustovaScaling_CO2v3_N2 | 10 |
CO2(X,v=00021) + N2(X,v=9) ↔ CO2(X,v=00011) + N2(X,v=10)  |  KustovaScaling_CO2v3_N2 | 20 |
CO2(X,v=00031) + N2(X,v=9) ↔ CO2(X,v=00021) + N2(X,v=10)  |  KustovaScaling_CO2v3_N2 | 30 |
CO2(X,v=00041) + N2(X,v=9) ↔ CO2(X,v=00031) + N2(X,v=10)  |  KustovaScaling_CO2v3_N2 | 40 |
CO2(X,v=00051) + N2(X,v=9) ↔ CO2(X,v=00041) + N2(X,v=10)  |  KustovaScaling_CO2v3_N2 | 50 |

CO2(X,v=00011) + N2(X,v=10) ↔ CO2(X,v=00001) + N2(X,v=11)  |  KustovaScaling_CO2v3_N2 | 11 |
CO2(X,v=00021) + N2(X,v=10) ↔ CO2(X,v=00011) + N2(X,v=11)  |  KustovaScaling_CO2v3_N2 | 22 |
CO2(X,v=00031) + N2(X,v=10) ↔ CO2(X,v=00021) + N2(X,v=11)  |  KustovaScaling_CO2v3_N2 | 33 |
CO2(X,v=00041) + N2(X,v=10) ↔ CO2(X,v=00031) + N2(X,v=11)  |  KustovaScaling_CO2v3_N2 | 44 |
CO2(X,v=00051) + N2(X,v=10) ↔ CO2(X,v=00041) + N2(X,v=11)  |  KustovaScaling_CO2v3_N2 | 55 |

CO2(X,v=00011) + N2(X,v=11) ↔ CO2(X,v=00001) + N2(X,v=12)  |  KustovaScaling_CO2v3_N2 | 12 |
CO2(X,v=00021) + N2(X,v=11) ↔ CO2(X,v=00011) + N2(X,v=12)  |  KustovaScaling_CO2v3_N2 | 24 |
CO2(X,v=00031) + N2(X,v=11) ↔ CO2(X,v=00021) + N2(X,v=12)  |  KustovaScaling_CO2v3_N2 | 36 |
CO2(X,v=00041) + N2(X,v=11) ↔ CO2(X,v=00031) + N2(X,v=12)  |  KustovaScaling_CO2v3_N2 | 48 |
CO2(X,v=00051) + N2(X,v=11) ↔ CO2(X,v=00041) + N2(X,v=12)  |  KustovaScaling_CO2v3_N2 | 60 |

CO2(X,v=00011) + N2(X,v=12) ↔ CO2(X,v=00001) + N2(X,v=13)  |  KustovaScaling_CO2v3_N2 | 13 |
CO2(X,v=00021) + N2(X,v=12) ↔ CO2(X,v=00011) + N2(X,v=13)  |  KustovaScaling_CO2v3_N2 | 26 |
CO2(X,v=00031) + N2(X,v=12) ↔ CO2(X,v=00021) + N2(X,v=13)  |  KustovaScaling_CO2v3_N2 | 39 |
CO2(X,v=00041) + N2(X,v=12) ↔ CO2(X,v=00031) + N2(X,v=13)  |  KustovaScaling_CO2v3_N2 | 52 |
CO2(X,v=00051) + N2(X,v=12) ↔ CO2(X,v=00041) + N2(X,v=13)  |  KustovaScaling_CO2v3_N2 | 65 |

CO2(X,v=00011) + N2(X,v=13) ↔ CO2(X,v=00001) + N2(X,v=14)  |  KustovaScaling_CO2v3_N2 | 14 |
CO2(X,v=00021) + N2(X,v=13) ↔ CO2(X,v=00011) + N2(X,v=14)  |  KustovaScaling_CO2v3_N2 | 28 |
CO2(X,v=00031) + N2(X,v=13) ↔ CO2(X,v=00021) + N2(X,v=14)  |  KustovaScaling_CO2v3_N2 | 42 |
CO2(X,v=00041) + N2(X,v=13) ↔ CO2(X,v=00031) + N2(X,v=14)  |  KustovaScaling_CO2v3_N2 | 56 |
CO2(X,v=00051) + N2(X,v=13) ↔ CO2(X,v=00041) + N2(X,v=14)  |  KustovaScaling_CO2v3_N2 | 70 |

CO2(X,v=00011) + N2(X,v=14) ↔ CO2(X,v=00001) + N2(X,v=15)  |  KustovaScaling_CO2v3_N2 | 15 |
CO2(X,v=00021) + N2(X,v=14) ↔ CO2(X,v=00011) + N2(X,v=15)  |  KustovaScaling_CO2v3_N2 | 30 |
CO2(X,v=00031) + N2(X,v=14) ↔ CO2(X,v=00021) + N2(X,v=15)  |  KustovaScaling_CO2v3_N2 | 45 |
CO2(X,v=00041) + N2(X,v=14) ↔ CO2(X,v=00031) + N2(X,v=15)  |  KustovaScaling_CO2v3_N2 | 60 |
CO2(X,v=00051) + N2(X,v=14) ↔ CO2(X,v=00041) + N2(X,v=15)  |  KustovaScaling_CO2v3_N2 | 75 |

CO2(X,v=00011) + N2(X,v=15) ↔ CO2(X,v=00001) + N2(X,v=16) | KustovaScaling_CO2v3_N2 | 16 |

CO2(X,v=00021) + N2(X,v=15) ↔ CO2(X,v=00011) + N2(X,v=16) | KustovaScaling_CO2v3_N2 | 32 |

CO2(X,v=00031) + N2(X,v=15) ↔ CO2(X,v=00021) + N2(X,v=16) | KustovaScaling_CO2v3_N2 | 48 |

CO2(X,v=00041) + N2(X,v=15) ↔ CO2(X,v=00031) + N2(X,v=16) | KustovaScaling_CO2v3_N2 | 64 |

CO2(X,v=00051) + N2(X,v=15) ↔ CO2(X,v=00041) + N2(X,v=16) | KustovaScaling_CO2v3_N2 | 80 |

CO2(X,v=00011) + N2(X,v=16) ↔ CO2(X,v=00001) + N2(X,v=17) | KustovaScaling_CO2v3_N2 | 17 |

CO2(X,v=00021) + N2(X,v=16) ↔ CO2(X,v=00011) + N2(X,v=17) | KustovaScaling_CO2v3_N2 | 34 |

CO2(X,v=00031) + N2(X,v=16) ↔ CO2(X,v=00021) + N2(X,v=17) | KustovaScaling_CO2v3_N2 | 51 |

CO2(X,v=00041) + N2(X,v=16) ↔ CO2(X,v=00031) + N2(X,v=17) | KustovaScaling_CO2v3_N2 | 68 |

CO2(X,v=00051) + N2(X,v=16) ↔ CO2(X,v=00041) + N2(X,v=17) | KustovaScaling_CO2v3_N2 | 85 |

CO2(X,v=00011) + N2(X,v=17) ↔ CO2(X,v=00001) + N2(X,v=18) | KustovaScaling_CO2v3_N2 | 18 |

CO2(X,v=00021) + N2(X,v=17) ↔ CO2(X,v=00011) + N2(X,v=18) | KustovaScaling_CO2v3_N2 | 36 |

CO2(X,v=00031) + N2(X,v=17) ↔ CO2(X,v=00021) + N2(X,v=18) | KustovaScaling_CO2v3_N2 | 54 |

CO2(X,v=00041) + N2(X,v=17) ↔ CO2(X,v=00031) + N2(X,v=18) | KustovaScaling_CO2v3_N2 | 72 |

CO2(X,v=00051) + N2(X,v=17) ↔ CO2(X,v=00041) + N2(X,v=18) | KustovaScaling_CO2v3_N2 | 90 |

CO2(X,v=00011) + N2(X,v=18) ↔ CO2(X,v=00001) + N2(X,v=19) | KustovaScaling_CO2v3_N2 | 19 |

CO2(X,v=00021) + N2(X,v=18) ↔ CO2(X,v=00011) + N2(X,v=19) | KustovaScaling_CO2v3_N2 | 38 |

CO2(X,v=00031) + N2(X,v=18) ↔ CO2(X,v=00021) + N2(X,v=19) | KustovaScaling_CO2v3_N2 | 57 |

CO2(X,v=00041) + N2(X,v=18) ↔ CO2(X,v=00031) + N2(X,v=19) | KustovaScaling_CO2v3_N2 | 76 |

CO2(X,v=00051) + N2(X,v=18) ↔ CO2(X,v=00041) + N2(X,v=19) | KustovaScaling_CO2v3_N2 | 95 |

CO2(X,v=00011) + N2(X,v=19) ↔ CO2(X,v=00001) + N2(X,v=20) | KustovaScaling_CO2v3_N2 | 20 |

CO2(X,v=00021) + N2(X,v=19) ↔ CO2(X,v=00011) + N2(X,v=20) | KustovaScaling_CO2v3_N2 | 40 |

CO2(X,v=00031) + N2(X,v=19) ↔ CO2(X,v=00021) + N2(X,v=20) | KustovaScaling_CO2v3_N2 | 60 |

CO2(X,v=00041) + N2(X,v=19) ↔ CO2(X,v=00031) + N2(X,v=20) | KustovaScaling_CO2v3_N2 | 80 |

CO2(X,v=00051) + N2(X,v=19) ↔ CO2(X,v=00041) + N2(X,v=20) | KustovaScaling_CO2v3_N2 | 100 |

CO2(X,v=00011) + N2(X,v=20) ↔ CO2(X,v=00001) + N2(X,v=21) | KustovaScaling_CO2v3_N2 | 21 |

CO2(X,v=00021) + N2(X,v=20) ↔ CO2(X,v=00011) + N2(X,v=21) | KustovaScaling_CO2v3_N2 | 42 |

CO2(X,v=00031) + N2(X,v=20) ↔ CO2(X,v=00021) + N2(X,v=21) | KustovaScaling_CO2v3_N2 | 63 |

CO2(X,v=00041) + N2(X,v=20) ↔ CO2(X,v=00031) + N2(X,v=21) | KustovaScaling_CO2v3_N2 | 84 |

CO2(X,v=00051) + N2(X,v=20) ↔ CO2(X,v=00041) + N2(X,v=21) | KustovaScaling_CO2v3_N2 | 105 |

CO2(X,v=00011) + N2(X,v=21) ↔ CO2(X,v=00001) + N2(X,v=22) | KustovaScaling_CO2v3_N2 | 22 |
CO2(X,v=00021) + N2(X,v=21) ↔ CO2(X,v=00011) + N2(X,v=22) | KustovaScaling_CO2v3_N2 | 44 |
CO2(X,v=00031) + N2(X,v=21) ↔ CO2(X,v=00021) + N2(X,v=22) | KustovaScaling_CO2v3_N2 | 66 |
CO2(X,v=00041) + N2(X,v=21) ↔ CO2(X,v=00031) + N2(X,v=22) | KustovaScaling_CO2v3_N2 | 88 |
CO2(X,v=00051) + N2(X,v=21) ↔ CO2(X,v=00041) + N2(X,v=22) | KustovaScaling_CO2v3_N2 | 110 |

CO2(X,v=00011) + N2(X,v=22) ↔ CO2(X,v=00001) + N2(X,v=23) | KustovaScaling_CO2v3_N2 | 43 |
CO2(X,v=00021) + N2(X,v=22) ↔ CO2(X,v=00021) + N2(X,v=23) | KustovaScaling_CO2v3_N2 | 46 |
CO2(X,v=00031) + N2(X,v=22) ↔ CO2(X,v=00021) + N2(X,v=23) | KustovaScaling_CO2v3_N2 | 69 |
CO2(X,v=00041) + N2(X,v=22) ↔ CO2(X,v=00031) + N2(X,v=23) | KustovaScaling_CO2v3_N2 | 92 |
CO2(X,v=00051) + N2(X,v=22) ↔ CO2(X,v=00041) + N2(X,v=23) | KustovaScaling_CO2v3_N2 | 115 |

CO2(X,v=00011) + N2(X,v=23) ↔ CO2(X,v=00001) + N2(X,v=24) | KustovaScaling_CO2v3_N2 | 24 |
CO2(X,v=00021) + N2(X,v=23) ↔ CO2(X,v=00011) + N2(X,v=24) | KustovaScaling_CO2v3_N2 | 48 |
CO2(X,v=00031) + N2(X,v=23) ↔ CO2(X,v=00021) + N2(X,v=24) | KustovaScaling_CO2v3_N2 | 72 |
CO2(X,v=00041) + N2(X,v=23) ↔ CO2(X,v=00031) + N2(X,v=24) | KustovaScaling_CO2v3_N2 | 96 |
CO2(X,v=00051) + N2(X,v=23) ↔ CO2(X,v=00041) + N2(X,v=24) | KustovaScaling_CO2v3_N2 | 120 |

CO2(X,v=00011) + N2(X,v=24) ↔ CO2(X,v=00001) + N2(X,v=25) | KustovaScaling_CO2v3_N2 | 25 |
CO2(X,v=00021) + N2(X,v=24) ↔ CO2(X,v=00011) + N2(X,v=25) | KustovaScaling_CO2v3_N2 | 50 |
CO2(X,v=00031) + N2(X,v=24) ↔ CO2(X,v=00021) + N2(X,v=25) | KustovaScaling_CO2v3_N2 | 75 |
CO2(X,v=00041) + N2(X,v=24) ↔ CO2(X,v=00031) + N2(X,v=25) | KustovaScaling_CO2v3_N2 | 100 |
CO2(X,v=00051) + N2(X,v=24) ↔ CO2(X,v=00041) + N2(X,v=25) | KustovaScaling_CO2v3_N2 | 125 |

CO2(X,v=00011) + N2(X,v=25) ↔ CO2(X,v=00001) + N2(X,v=26) | KustovaScaling_CO2v3_N2 | 26 |
CO2(X,v=00021) + N2(X,v=25) ↔ CO2(X,v=00011) + N2(X,v=26) | KustovaScaling_CO2v3_N2 | 52 |
CO2(X,v=00031) + N2(X,v=25) ↔ CO2(X,v=00021) + N2(X,v=26) | KustovaScaling_CO2v3_N2 | 78 |
CO2(X,v=00041) + N2(X,v=25) ↔ CO2(X,v=00031) + N2(X,v=26) | KustovaScaling_CO2v3_N2 | 104 |
CO2(X,v=00051) + N2(X,v=25) ↔ CO2(X,v=00041) + N2(X,v=26) | KustovaScaling_CO2v3_N2 | 130 |

CO2(X,v=00011) + N2(X,v=26) ↔ CO2(X,v=00001) + N2(X,v=27) | KustovaScaling_CO2v3_N2 | 27 |
CO2(X,v=00021) + N2(X,v=26) ↔ CO2(X,v=00011) + N2(X,v=27) | KustovaScaling_CO2v3_N2 | 54 |
CO2(X,v=00031) + N2(X,v=26) ↔ CO2(X,v=00021) + N2(X,v=27) | KustovaScaling_CO2v3_N2 | 81 |
CO2(X,v=00041) + N2(X,v=26) ↔ CO2(X,v=00031) + N2(X,v=27) | KustovaScaling_CO2v3_N2 | 108 |
CO2(X,v=00051) + N2(X,v=26) ↔ CO2(X,v=00041) + N2(X,v=27) | KustovaScaling_CO2v3_N2 | 135 |

CO2(X,v=00011) + N2(X,v=27) ↔ CO2(X,v=00001) + N2(X,v=28) | KustovaScaling_CO2v3_N2 | 28 |
CO2(X,v=00021) + N2(X,v=27) ↔ CO2(X,v=00011) + N2(X,v=28) | KustovaScaling_CO2v3_N2 | 56 |
CO2(X,v=00031) + N2(X,v=27) ↔ CO2(X,v=00021) + N2(X,v=28) | KustovaScaling_CO2v3_N2 | 84 |
CO2(X,v=00041) + N2(X,v=27) ↔ CO2(X,v=00031) + N2(X,v=28) | KustovaScaling_CO2v3_N2 | 112 |
CO2(X,v=00051) + N2(X,v=27) ↔ CO2(X,v=00041) + N2(X,v=28) | KustovaScaling_CO2v3_N2 | 140 |

CO2(X,v=00011) + N2(X,v=28) ↔ CO2(X,v=00001) + N2(X,v=29) | KustovaScaling_CO2v3_N2 | 29 |
CO2(X,v=00021) + N2(X,v=28) ↔ CO2(X,v=00011) + N2(X,v=29) | KustovaScaling_CO2v3_N2 | 58 |
CO2(X,v=00031) + N2(X,v=28) ↔ CO2(X,v=00021) + N2(X,v=29) | KustovaScaling_CO2v3_N2 | 87 |
CO2(X,v=00041) + N2(X,v=28) ↔ CO2(X,v=00031) + N2(X,v=29) | KustovaScaling_CO2v3_N2 | 116 |
CO2(X,v=00051) + N2(X,v=28) ↔ CO2(X,v=00041) + N2(X,v=29) | KustovaScaling_CO2v3_N2 | 145 |

CO2(X,v=00011) + N2(X,v=29) ↔ CO2(X,v=00001) + N2(X,v=30) | KustovaScaling_CO2v3_N2 | 30 |
CO2(X,v=00021) + N2(X,v=29) ↔ CO2(X,v=00011) + N2(X,v=30) | KustovaScaling_CO2v3_N2 | 60 |
CO2(X,v=00031) + N2(X,v=29) ↔ CO2(X,v=00021) + N2(X,v=30) | KustovaScaling_CO2v3_N2 | 90 |
CO2(X,v=00041) + N2(X,v=29) ↔ CO2(X,v=00031) + N2(X,v=30) | KustovaScaling_CO2v3_N2 | 120 |
CO2(X,v=00051) + N2(X,v=29) ↔ CO2(X,v=00041) + N2(X,v=30) | KustovaScaling_CO2v3_N2 | 150 |

CO2(X,v=00011) + N2(X,v=30) ↔ CO2(X,v=00001) + N2(X,v=31) | KustovaScaling_CO2v3_N2 | 31 |
CO2(X,v=00021) + N2(X,v=30) ↔ CO2(X,v=00011) + N2(X,v=31) | KustovaScaling_CO2v3_N2 | 62 |
CO2(X,v=00031) + N2(X,v=30) ↔ CO2(X,v=00021) + N2(X,v=31) | KustovaScaling_CO2v3_N2 | 93 |
CO2(X,v=00041) + N2(X,v=30) ↔ CO2(X,v=00031) + N2(X,v=31) | KustovaScaling_CO2v3_N2 | 124 |
CO2(X,v=00051) + N2(X,v=30) ↔ CO2(X,v=00041) + N2(X,v=31) | KustovaScaling_CO2v3_N2 | 155 |

CO2(X,v=00011) + N2(X,v=31) ↔ CO2(X,v=00001) + N2(X,v=32) | KustovaScaling_CO2v3_N2 | 32 |
CO2(X,v=00021) + N2(X,v=31) ↔ CO2(X,v=00011) + N2(X,v=32) | KustovaScaling_CO2v3_N2 | 64 |
CO2(X,v=00031) + N2(X,v=31) ↔ CO2(X,v=00021) + N2(X,v=32) | KustovaScaling_CO2v3_N2 | 96 |
CO2(X,v=00041) + N2(X,v=31) ↔ CO2(X,v=00031) + N2(X,v=32) | KustovaScaling_CO2v3_N2 | 128 |
CO2(X,v=00051) + N2(X,v=31) ↔ CO2(X,v=00041) + N2(X,v=32) | KustovaScaling_CO2v3_N2 | 160 |

CO2(X,v=00011) + N2(X,v=32) ↔ CO2(X,v=00001) + N2(X,v=33) | KustovaScaling_CO2v3_N2 | 33 |
CO2(X,v=00021) + N2(X,v=32) ↔ CO2(X,v=00001) + N2(X,v=33) | KustovaScaling_CO2v3_N2 | 66 |
CO2(X,v=00031) + N2(X,v=32) ↔ CO2(X,v=00021) + N2(X,v=33) | KustovaScaling_CO2v3_N2 | 99 |
CO2(X,v=00041) + N2(X,v=32) ↔ CO2(X,v=00031) + N2(X,v=33) | KustovaScaling_CO2v3_N2 | 132 |
CO2(X,v=00051) + N2(X,v=32) ↔ CO2(X,v=00041) + N2(X,v=33) | KustovaScaling_CO2v3_N2 | 165 |

CO2(X,v=00011) + N2(X,v=33) ↔ CO2(X,v=00001) + N2(X,v=34) | KustovaScaling_CO2v3_N2 | 34 |
CO2(X,v=00021) + N2(X,v=33) ↔ CO2(X,v=00011) + N2(X,v=34) | KustovaScaling_CO2v3_N2 | 68 |
CO2(X,v=00031) + N2(X,v=33) ↔ CO2(X,v=00021) + N2(X,v=34) | KustovaScaling_CO2v3_N2 | 102 |
CO2(X,v=00041) + N2(X,v=33) ↔ CO2(X,v=00031) + N2(X,v=34) | KustovaScaling_CO2v3_N2 | 136 |
CO2(X,v=00051) + N2(X,v=33) ↔ CO2(X,v=00041) + N2(X,v=34) | KustovaScaling_CO2v3_N2 | 170 |

CO2(X,v=00011) + N2(X,v=34) ↔ CO2(X,v=00001) + N2(X,v=35) | KustovaScaling_CO2v3_N2 | 35 |
CO2(X,v=00021) + N2(X,v=34) ↔ CO2(X,v=00011) + N2(X,v=35) | KustovaScaling_CO2v3_N2 | 70 |
CO2(X,v=00031) + N2(X,v=34) ↔ CO2(X,v=00021) + N2(X,v=35) | KustovaScaling_CO2v3_N2 | 105 |
CO2(X,v=00041) + N2(X,v=34) ↔ CO2(X,v=00031) + N2(X,v=35) | KustovaScaling_CO2v3_N2 | 140 |
CO2(X,v=00051) + N2(X,v=34) ↔ CO2(X,v=00041) + N2(X,v=35) | KustovaScaling_CO2v3_N2 | 175 |

CO2(X,v=00011) + N2(X,v=35) ↔ CO2(X,v=00001) + N2(X,v=36) | KustovaScaling_CO2v3_N2 | 36 |
CO2(X,v=00021) + N2(X,v=35) ↔ CO2(X,v=00011) + N2(X,v=36) | KustovaScaling_CO2v3_N2 | 72 |
CO2(X,v=00031) + N2(X,v=35) ↔ CO2(X,v=00021) + N2(X,v=36) | KustovaScaling_CO2v3_N2 | 108 |
CO2(X,v=00041) + N2(X,v=35) ↔ CO2(X,v=00031) + N2(X,v=36) | KustovaScaling_CO2v3_N2 | 144 |
CO2(X,v=00051) + N2(X,v=35) ↔ CO2(X,v=00041) + N2(X,v=36) | KustovaScaling_CO2v3_N2 | 180 |

CO2(X,v=00011) + N2(X,v=36) ↔ CO2(X,v=00001) + N2(X,v=37) | KustovaScaling_CO2v3_N2 | 37 |
CO2(X,v=00021) + N2(X,v=36) ↔ CO2(X,v=00011) + N2(X,v=37) | KustovaScaling_CO2v3_N2 | 74 |
CO2(X,v=00031) + N2(X,v=36) ↔ CO2(X,v=00021) + N2(X,v=37) | KustovaScaling_CO2v3_N2 | 111 |
CO2(X,v=00041) + N2(X,v=36) ↔ CO2(X,v=00031) + N2(X,v=37) | KustovaScaling_CO2v3_N2 | 148 |
CO2(X,v=00051) + N2(X,v=36) ↔ CO2(X,v=00041) + N2(X,v=37) | KustovaScaling_CO2v3_N2 | 185 |

CO2(X,v=00011) + N2(X,v=37) ↔ CO2(X,v=00001) + N2(X,v=38) | KustovaScaling_CO2v3_N2 | 38 |
CO2(X,v=00021) + N2(X,v=37) ↔ CO2(X,v=00011) + N2(X,v=38) | KustovaScaling_CO2v3_N2 | 76 |
CO2(X,v=00031) + N2(X,v=37) ↔ CO2(X,v=00021) + N2(X,v=38) | KustovaScaling_CO2v3_N2 | 114 |
CO2(X,v=00041) + N2(X,v=37) ↔ CO2(X,v=00031) + N2(X,v=38) | KustovaScaling_CO2v3_N2 | 152 |
CO2(X,v=00051) + N2(X,v=37) ↔ CO2(X,v=00041) + N2(X,v=38) | KustovaScaling_CO2v3_N2 | 190 |

CO2(X,v=00011) + N2(X,v=38) ↔ CO2(X,v=00001) + N2(X,v=39) | KustovaScaling_CO2v3_N2 | 39 |
CO2(X,v=00021) + N2(X,v=38) ↔ CO2(X,v=00011) + N2(X,v=39) | KustovaScaling_CO2v3_N2 | 78 |
CO2(X,v=00031) + N2(X,v=38) ↔ CO2(X,v=00021) + N2(X,v=39) | KustovaScaling_CO2v3_N2 | 117 |
CO2(X,v=00041) + N2(X,v=38) ↔ CO2(X,v=00031) + N2(X,v=39) | KustovaScaling_CO2v3_N2 | 156 |
CO2(X,v=00051) + N2(X,v=38) ↔ CO2(X,v=00041) + N2(X,v=39) | KustovaScaling_CO2v3_N2 | 195 |

CO2(X,v=00011) + N2(X,v=39) ↔ CO2(X,v=00001) + N2(X,v=40) | KustovaScaling_CO2v3_N2 | 40 |
CO2(X,v=00021) + N2(X,v=39) ↔ CO2(X,v=00011) + N2(X,v=40) | KustovaScaling_CO2v3_N2 | 80 |
CO2(X,v=00031) + N2(X,v=39) ↔ CO2(X,v=00021) + N2(X,v=40) | KustovaScaling_CO2v3_N2 | 120 |
CO2(X,v=00041) + N2(X,v=39) ↔ CO2(X,v=00031) + N2(X,v=40) | KustovaScaling_CO2v3_N2 | 160 |
CO2(X,v=00051) + N2(X,v=39) ↔ CO2(X,v=00041) + N2(X,v=40) | KustovaScaling_CO2v3_N2 | 200 |

CO2(X,v=00011) + N2(X,v=40) ↔ CO2(X,v=00001) + N2(X,v=41) | KustovaScaling_CO2v3_N2 | 41 |
CO2(X,v=00021) + N2(X,v=40) ↔ CO2(X,v=00011) + N2(X,v=41) | KustovaScaling_CO2v3_N2 | 82 |
CO2(X,v=00031) + N2(X,v=40) ↔ CO2(X,v=00021) + N2(X,v=41) | KustovaScaling_CO2v3_N2 | 123 |
CO2(X,v=00041) + N2(X,v=40) ↔ CO2(X,v=00031) + N2(X,v=41) | KustovaScaling_CO2v3_N2 | 164 |
CO2(X,v=00051) + N2(X,v=40) ↔ CO2(X,v=00041) + N2(X,v=41) | KustovaScaling_CO2v3_N2 | 205 |

CO2(X,v=00011) + N2(X,v=41) ↔ CO2(X,v=00001) + N2(X,v=42) | KustovaScaling_CO2v3_N2 | 42 |
CO2(X,v=00021) + N2(X,v=41) ↔ CO2(X,v=00011) + N2(X,v=42) | KustovaScaling_CO2v3_N2 | 84 |
CO2(X,v=00031) + N2(X,v=41) ↔ CO2(X,v=00021) + N2(X,v=42) | KustovaScaling_CO2v3_N2 | 126 |
CO2(X,v=00041) + N2(X,v=41) ↔ CO2(X,v=00031) + N2(X,v=42) | KustovaScaling_CO2v3_N2 | 168 |
CO2(X,v=00051) + N2(X,v=41) ↔ CO2(X,v=00041) + N2(X,v=42) | KustovaScaling_CO2v3_N2 | 210 |

CO2(X,v=00011) + N2(X,v=42) ↔ CO2(X,v=00001) + N2(X,v=43) | KustovaScaling_CO2v3_N2 | 43 |
CO2(X,v=00021) + N2(X,v=42) ↔ CO2(X,v=00011) + N2(X,v=43) | KustovaScaling_CO2v3_N2 | 86 |
CO2(X,v=00031) + N2(X,v=42) ↔ CO2(X,v=00021) + N2(X,v=43) | KustovaScaling_CO2v3_N2 | 129 |
CO2(X,v=00041) + N2(X,v=42) ↔ CO2(X,v=00031) + N2(X,v=43) | KustovaScaling_CO2v3_N2 | 172 |
CO2(X,v=00051) + N2(X,v=42) ↔ CO2(X,v=00041) + N2(X,v=43) | KustovaScaling_CO2v3_N2 | 215 |

CO2(X,v=00011) + N2(X,v=43) ↔ CO2(X,v=00001) + N2(X,v=44) | KustovaScaling_CO2v3_N2 | 44 |
CO2(X,v=00021) + N2(X,v=43) ↔ CO2(X,v=00011) + N2(X,v=44) | KustovaScaling_CO2v3_N2 | 88 |
CO2(X,v=00031) + N2(X,v=43) ↔ CO2(X,v=00021) + N2(X,v=44) | KustovaScaling_CO2v3_N2 | 132 |
CO2(X,v=00041) + N2(X,v=43) ↔ CO2(X,v=00031) + N2(X,v=44) | KustovaScaling_CO2v3_N2 | 176 |
CO2(X,v=00051) + N2(X,v=43) ↔ CO2(X,v=00041) + N2(X,v=44) | KustovaScaling_CO2v3_N2 | 220 |

CO2(X,v=00011) + N2(X,v=44) ↔ CO2(X,v=00001) + N2(X,v=45) | KustovaScaling_CO2v3_N2 | 45 |
CO2(X,v=00021) + N2(X,v=44) ↔ CO2(X,v=00011) + N2(X,v=45) | KustovaScaling_CO2v3_N2 | 90 |
CO2(X,v=00031) + N2(X,v=44) ↔ CO2(X,v=00021) + N2(X,v=45) | KustovaScaling_CO2v3_N2 | 135 |
CO2(X,v=00041) + N2(X,v=44) ↔ CO2(X,v=00031) + N2(X,v=45) | KustovaScaling_CO2v3_N2 | 180 |
CO2(X,v=00051) + N2(X,v=44) ↔ CO2(X,v=00041) + N2(X,v=45) | KustovaScaling_CO2v3_N2 | 225 |

| | | | |
|---|---|---|---|
| CO2(X,v=00011) + N2(X,v=45) ↔ CO2(X,v=00001) + N2(X,v=46) | KustovaScaling_CO2v3_N2 | 46 |
| CO2(X,v=00021) + N2(X,v=45) ↔ CO2(X,v=00011) + N2(X,v=46) | KustovaScaling_CO2v3_N2 | 92 |
| CO2(X,v=00031) + N2(X,v=45) ↔ CO2(X,v=00021) + N2(X,v=46) | KustovaScaling_CO2v3_N2 | 138 |
| CO2(X,v=00041) + N2(X,v=45) ↔ CO2(X,v=00031) + N2(X,v=46) | KustovaScaling_CO2v3_N2 | 184 |
| CO2(X,v=00051) + N2(X,v=45) ↔ CO2(X,v=00041) + N2(X,v=46) | KustovaScaling_CO2v3_N2 | 230 |
| | | | |
| CO2(X,v=00011) + N2(X,v=46) ↔ CO2(X,v=00001) + N2(X,v=47) | KustovaScaling_CO2v3_N2 | 47 |
| CO2(X,v=00021) + N2(X,v=46) ↔ CO2(X,v=00011) + N2(X,v=47) | KustovaScaling_CO2v3_N2 | 94 |
| CO2(X,v=00031) + N2(X,v=46) ↔ CO2(X,v=00021) + N2(X,v=47) | KustovaScaling_CO2v3_N2 | 141 |
| CO2(X,v=00041) + N2(X,v=46) ↔ CO2(X,v=00031) + N2(X,v=47) | KustovaScaling_CO2v3_N2 | 188 |
| CO2(X,v=00051) + N2(X,v=46) ↔ CO2(X,v=00041) + N2(X,v=47) | KustovaScaling_CO2v3_N2 | 235 |
| | | | |
| CO2(X,v=00011) + N2(X,v=47) ↔ CO2(X,v=00001) + N2(X,v=48) | KustovaScaling_CO2v3_N2 | 48 |
| CO2(X,v=00021) + N2(X,v=47) ↔ CO2(X,v=00011) + N2(X,v=48) | KustovaScaling_CO2v3_N2 | 96 |
| CO2(X,v=00031) + N2(X,v=47) ↔ CO2(X,v=00021) + N2(X,v=48) | KustovaScaling_CO2v3_N2 | 144 |
| CO2(X,v=00041) + N2(X,v=47) ↔ CO2(X,v=00031) + N2(X,v=48) | KustovaScaling_CO2v3_N2 | 192 |
| CO2(X,v=00051) + N2(X,v=47) ↔ CO2(X,v=00041) + N2(X,v=48) | KustovaScaling_CO2v3_N2 | 240 |
| | | | |
| CO2(X,v=00011) + N2(X,v=48) ↔ CO2(X,v=00001) + N2(X,v=49) | KustovaScaling_CO2v3_N2 | 49 |
| CO2(X,v=00021) + N2(X,v=48) ↔ CO2(X,v=00011) + N2(X,v=49) | KustovaScaling_CO2v3_N2 | 98 |
| CO2(X,v=00031) + N2(X,v=48) ↔ CO2(X,v=00021) + N2(X,v=49) | KustovaScaling_CO2v3_N2 | 147 |
| CO2(X,v=00041) + N2(X,v=48) ↔ CO2(X,v=00031) + N2(X,v=49) | KustovaScaling_CO2v3_N2 | 196 |
| CO2(X,v=00051) + N2(X,v=48) ↔ CO2(X,v=00041) + N2(X,v=49) | KustovaScaling_CO2v3_N2 | 245 |
| | | | |
| CO2(X,v=00011) + N2(X,v=49) ↔ CO2(X,v=00001) + N2(X,v=50) | KustovaScaling_CO2v3_N2 | 50 |
| CO2(X,v=00021) + N2(X,v=49) ↔ CO2(X,v=00011) + N2(X,v=50) | KustovaScaling_CO2v3_N2 | 100 |
| CO2(X,v=00031) + N2(X,v=49) ↔ CO2(X,v=00021) + N2(X,v=50) | KustovaScaling_CO2v3_N2 | 150 |
| CO2(X,v=00041) + N2(X,v=49) ↔ CO2(X,v=00031) + N2(X,v=50) | KustovaScaling_CO2v3_N2 | 200 |
| CO2(X,v=00051) + N2(X,v=49) ↔ CO2(X,v=00041) + N2(X,v=50) | KustovaScaling_CO2v3_N2 | 250 |
| | | | |
| CO2(X,v=00011) + N2(X,v=50) ↔ CO2(X,v=00001) + N2(X,v=51) | KustovaScaling_CO2v3_N2 | 51 |
| CO2(X,v=00021) + N2(X,v=50) ↔ CO2(X,v=00011) + N2(X,v=51) | KustovaScaling_CO2v3_N2 | 102 |
| CO2(X,v=00031) + N2(X,v=50) ↔ CO2(X,v=00021) + N2(X,v=51) | KustovaScaling_CO2v3_N2 | 153 |
| CO2(X,v=00041) + N2(X,v=50) ↔ CO2(X,v=00031) + N2(X,v=51) | KustovaScaling_CO2v3_N2 | 204 |
| CO2(X,v=00051) + N2(X,v=50) ↔ CO2(X,v=00041) + N2(X,v=51) | KustovaScaling_CO2v3_N2 | 255 |

CO2(X,v=00011) + N2(X,v=51) ↔ CO2(X,v=00001) + N2(X,v=52)  |  KustovaScaling_CO2v3_N2 |  52 |
CO2(X,v=00021) + N2(X,v=51) ↔ CO2(X,v=00011) + N2(X,v=52)  |  KustovaScaling_CO2v3_N2 | 104 |
CO2(X,v=00031) + N2(X,v=51) ↔ CO2(X,v=00021) + N2(X,v=52)  |  KustovaScaling_CO2v3_N2 | 156 |
CO2(X,v=00041) + N2(X,v=51) ↔ CO2(X,v=00031) + N2(X,v=52)  |  KustovaScaling_CO2v3_N2 | 208 |
CO2(X,v=00051) + N2(X,v=51) ↔ CO2(X,v=00041) + N2(X,v=52)  |  KustovaScaling_CO2v3_N2 | 260 |

CO2(X,v=00011) + N2(X,v=52) ↔ CO2(X,v=00001) + N2(X,v=53)  |  KustovaScaling_CO2v3_N2 |  53 |
CO2(X,v=00021) + N2(X,v=52) ↔ CO2(X,v=00011) + N2(X,v=53)  |  KustovaScaling_CO2v3_N2 | 106 |
CO2(X,v=00031) + N2(X,v=52) ↔ CO2(X,v=00021) + N2(X,v=53)  |  KustovaScaling_CO2v3_N2 | 159 |
CO2(X,v=00041) + N2(X,v=52) ↔ CO2(X,v=00031) + N2(X,v=53)  |  KustovaScaling_CO2v3_N2 | 212 |
CO2(X,v=00051) + N2(X,v=52) ↔ CO2(X,v=00041) + N2(X,v=53)  |  KustovaScaling_CO2v3_N2 | 265 |

CO2(X,v=00011) + N2(X,v=53) ↔ CO2(X,v=00001) + N2(X,v=54)  |  KustovaScaling_CO2v3_N2 |  54 |
CO2(X,v=00021) + N2(X,v=53) ↔ CO2(X,v=00011) + N2(X,v=54)  |  KustovaScaling_CO2v3_N2 | 108 |
CO2(X,v=00031) + N2(X,v=53) ↔ CO2(X,v=00021) + N2(X,v=54)  |  KustovaScaling_CO2v3_N2 | 162 |
CO2(X,v=00041) + N2(X,v=53) ↔ CO2(X,v=00031) + N2(X,v=54)  |  KustovaScaling_CO2v3_N2 | 216 |
CO2(X,v=00051) + N2(X,v=53) ↔ CO2(X,v=00041) + N2(X,v=54)  |  KustovaScaling_CO2v3_N2 | 270 |

CO2(X,v=00011) + N2(X,v=54) ↔ CO2(X,v=00001) + N2(X,v=55)  |  KustovaScaling_CO2v3_N2 |  55 |
CO2(X,v=00021) + N2(X,v=54) ↔ CO2(X,v=00011) + N2(X,v=55)  |  KustovaScaling_CO2v3_N2 | 110 |
CO2(X,v=00031) + N2(X,v=54) ↔ CO2(X,v=00021) + N2(X,v=55)  |  KustovaScaling_CO2v3_N2 | 165 |
CO2(X,v=00041) + N2(X,v=54) ↔ CO2(X,v=00031) + N2(X,v=55)  |  KustovaScaling_CO2v3_N2 | 220 |
CO2(X,v=00051) + N2(X,v=54) ↔ CO2(X,v=00041) + N2(X,v=55)  |  KustovaScaling_CO2v3_N2 | 275 |

CO2(X,v=00011) + N2(X,v=55) ↔ CO2(X,v=00001) + N2(X,v=56)  |  KustovaScaling_CO2v3_N2 |  56 |
CO2(X,v=00021) + N2(X,v=55) ↔ CO2(X,v=00011) + N2(X,v=56)  |  KustovaScaling_CO2v3_N2 | 112 |
CO2(X,v=00031) + N2(X,v=55) ↔ CO2(X,v=00021) + N2(X,v=56)  |  KustovaScaling_CO2v3_N2 | 168 |
CO2(X,v=00041) + N2(X,v=55) ↔ CO2(X,v=00031) + N2(X,v=56)  |  KustovaScaling_CO2v3_N2 | 224 |
CO2(X,v=00051) + N2(X,v=55) ↔ CO2(X,v=00041) + N2(X,v=56)  |  KustovaScaling_CO2v3_N2 | 280 |

CO2(X,v=00011) + N2(X,v=56) ↔ CO2(X,v=00001) + N2(X,v=57)  |  KustovaScaling_CO2v3_N2 |  57 |
CO2(X,v=00021) + N2(X,v=56) ↔ CO2(X,v=00011) + N2(X,v=57)  |  KustovaScaling_CO2v3_N2 | 114 |
CO2(X,v=00031) + N2(X,v=56) ↔ CO2(X,v=00021) + N2(X,v=57)  |  KustovaScaling_CO2v3_N2 | 171 |
CO2(X,v=00041) + N2(X,v=56) ↔ CO2(X,v=00031) + N2(X,v=57)  |  KustovaScaling_CO2v3_N2 | 228 |
CO2(X,v=00051) + N2(X,v=56) ↔ CO2(X,v=00041) + N2(X,v=57)  |  KustovaScaling_CO2v3_N2 | 285 |

CO2(X,v=00011) + N2(X,v=57) ↔ CO2(X,v=00001) + N2(X,v=58)  |  KustovaScaling_CO2v3_N2 |  58 |
CO2(X,v=00021) + N2(X,v=57) ↔ CO2(X,v=00011) + N2(X,v=58)  |  KustovaScaling_CO2v3_N2 | 116 |
CO2(X,v=00031) + N2(X,v=57) ↔ CO2(X,v=00021) + N2(X,v=58)  |  KustovaScaling_CO2v3_N2 | 174 |
CO2(X,v=00041) + N2(X,v=57) ↔ CO2(X,v=00031) + N2(X,v=58)  |  KustovaScaling_CO2v3_N2 | 232 |
CO2(X,v=00051) + N2(X,v=57) ↔ CO2(X,v=00041) + N2(X,v=58)  |  KustovaScaling_CO2v3_N2 | 290 |

CO2(X,v=00011) + N2(X,v=58) ↔ CO2(X,v=00001) + N2(X,v=59)  |  KustovaScaling_CO2v3_N2 |  59 |
CO2(X,v=00021) + N2(X,v=58) ↔ CO2(X,v=00011) + N2(X,v=59)  |  KustovaScaling_CO2v3_N2 | 118 |
CO2(X,v=00031) + N2(X,v=58) ↔ CO2(X,v=00021) + N2(X,v=59)  |  KustovaScaling_CO2v3_N2 | 177 |
CO2(X,v=00041) + N2(X,v=58) ↔ CO2(X,v=00031) + N2(X,v=59)  |  KustovaScaling_CO2v3_N2 | 236 |
CO2(X,v=00051) + N2(X,v=58) ↔ CO2(X,v=00041) + N2(X,v=59)  |  KustovaScaling_CO2v3_N2 | 295 |

# B. Calculated concentrations of all the species considered in the model

Table 4: Densities in $m^{-3}$ of all the species included in the model, at 2 Torr and 50 mA, for different $CO_2$-$N_2$ mixtures for the short tube (length 0.23 cm) where the $CO_2$ dissociation fraction and NO densities are measured.

| | 2 Torr | | | | | |
|---|---|---|---|---|---|---|
| | short tube 0.23 cm | | | | | |
| $CO_2$ fraction | 0.1 | 0.25 | 0.5 | 0.75 | 0.9 | 1 |
| species density ($m^{-3}$) | | | | | | |
| $O_2(X^3\Sigma_g^-)$ | 1.59E+20 | 5.81E+20 | 1.37E+21 | 2.14E+21 | 2.59E+21 | 2.95E+21 |
| $O_2(a^1\Delta_g)$ | 1.98E+19 | 5.22E+19 | 1.05E+20 | 1.67E+20 | 2.10E+20 | 2.46E+20 |
| $O_2(b^1\Sigma_g^+)$ | 5.12E+17 | 6.52E+17 | 7.00E+17 | 8.11E+17 | 9.15E+17 | 1.04E+18 |
| $O_2(A'^3\Delta_u, A^3\Sigma_u^+, c^1\Sigma_u^-)$ | 2.32E+16 | 4.29E+16 | 8.00E+16 | 1.22E+17 | 1.49E+17 | 1.66E+17 |
| $O_2^+$ | 2.45E+15 | 2.17E+15 | 1.54E+15 | 1.56E+15 | 2.23E+15 | 9.23E+15 |
| $O(^3P)$ | 1.34E+21 | 2.00E+21 | 2.24E+21 | 2.44E+21 | 2.61E+21 | 2.85E+21 |
| $O(^1D)$ | 1.65E+17 | 7.42E+16 | 4.17E+16 | 3.55E+16 | 3.56E+16 | 3.80E+16 |
| $O^+$ | 6.16E+13 | 1.74E+13 | 6.47E+12 | 3.89E+12 | 3.31E+12 | 3.40E+12 |
| $O^-$ | 3.87E+13 | 5.39E+13 | 8.22E+13 | 1.13E+14 | 1.34E+14 | 1.47E+14 |
| $O_3$ | 9.12E+14 | 3.25E+15 | 6.29E+15 | 8.80E+15 | 1.01E+16 | 1.18E+16 |
| $O_3^*$ | 6.58E+11 | 8.55E+12 | 4.15E+13 | 9.22E+13 | 1.28E+14 | 1.64E+14 |
| $CO_2(X^1\Sigma_g^+)$ | 1.20E+21 | 3.77E+21 | 8.31E+21 | 1.26E+22 | 1.49E+22 | 1.63E+22 |
| $CO_2^+$ | 2.86E+14 | 2.11E+14 | 1.81E+14 | 1.61E+14 | 1.54E+14 | 1.56E+14 |
| $C(^3P)$ | 4.25E+16 | 1.73E+16 | 9.45E+15 | 7.56E+15 | 7.22E+15 | 7.38E+15 |
| $CO(a^3\Pi_r)$ | 4.07E+16 | 3.63E+16 | 3.54E+16 | 3.74E+16 | 3.95E+16 | 4.18E+16 |
| $CO(X^1\Sigma^+)$ | 1.78E+21 | 3.39E+21 | 5.37E+21 | 7.24E+21 | 8.35E+21 | 9.24E+21 |
| $CO^+$ | 3.21E+13 | 9.01E+12 | 3.43E+12 | 2.12E+12 | 1.83E+12 | 1.91E+12 |
| $N_2(X^1\Sigma_g^+)$ | 2.68E+22 | 2.14E+22 | 1.36E+22 | 6.53E+21 | 2.53E+21 | |
| $N_2(A^3\Sigma_u^+)$ | 1.59E+18 | 5.62E+17 | 1.62E+17 | 4.68E+16 | 1.45E+16 | |
| $N_2(B^3\Pi_g)$ | 6.89E+16 | 2.08E+16 | 9.23E+15 | 4.00E+15 | 1.55E+15 | |
| $N_2(C^3\Pi_u)$ | 2.17E+14 | 1.16E+14 | 5.32E+13 | 2.10E+13 | 7.63E+12 | |
| $N_2(w^1\Delta_u)$ | 1.15E+16 | 7.55E+15 | 4.01E+15 | 1.73E+15 | 6.47E+14 | |
| $N_2(a^1\Pi_g)$ | 1.37E+16 | 7.25E+15 | 2.70E+15 | 8.63E+14 | 2.76E+14 | |
| $N_2(a'^1\Sigma_u^-)$ | 2.49E+17 | 7.33E+16 | 1.46E+16 | 3.81E+15 | 1.35E+15 | |
| $N_2^+$ | 4.01E+14 | 1.23E+14 | 3.63E+13 | 1.08E+13 | 3.59E+12 | |
| $N_2^+(B^2\Sigma_u^+)$ | 8.11E+10 | 2.17E+10 | 6.37E+09 | 1.80E+09 | 5.77E+08 | |
| $N(^4S)$ | 6.21E+18 | 1.26E+18 | 2.66E+17 | 7.85E+16 | 3.11E+16 | |
| $N(^2D)$ | 1.38E+18 | 4.57E+17 | 9.19E+16 | 2.37E+16 | 7.89E+15 | |
| $N(^2P)$ | 5.14E+16 | 5.30E+15 | 4.44E+14 | 9.03E+13 | 3.49E+13 | |
| $N^+$ | 1.64E+12 | 1.26E+11 | 7.22E+09 | 9.67E+08 | 2.61E+08 | |
| $N_4^+$ | 1.53E+13 | 1.56E+12 | 1.27E+11 | 7.61E+09 | 4.10E+08 | |
| $N_3^+$ | 1.06E+13 | 6.19E+11 | 2.98E+10 | 1.95E+09 | 1.87E+08 | |
| $NO(X^2\Pi_r)$ | 7.91E+19 | 1.18E+20 | 1.87E+20 | 1.88E+20 | 1.25E+20 | |
| $NO(A^2\Sigma^+)$ | 1.83E+15 | 9.45E+14 | 4.22E+14 | 1.20E+14 | 2.43E+13 | |
| $NO(B^2\Pi_r)$ | 7.56E+10 | 1.81E+10 | 2.73E+09 | 5.01E+08 | 1.28E+08 | |
| $NO^+$ | 7.14E+15 | 7.71E+15 | 8.09E+15 | 7.84E+15 | 7.08E+15 | |
| NO$_2$(X$^2$A$_1$) | 8.92E+15 | 1.10E+16 | 1.28E+16 | 8.55E+15 | 4.09E+15 | |
| NO$_2$(a$^4$A$_2$) | 6.44E+13 | 1.42E+14 | 2.52E+14 | 2.71E+14 | 1.87E+14 | |
| CN(X$^2\Sigma^+$) | 1.61E+11 | 5.15E+10 | 1.66E+10 | 5.25E+09 | 1.39E+09 | |
| CN(B$^2\Sigma^+$) | 6.98E+08 | 2.51E+07 | 2.65E+06 | 5.49E+05 | 1.63E+05 | |
| n$_e$ | 1.04E+16 | 1.02E+16 | 9.78E+15 | 9.46E+15 | 9.34E+15 | 9.24E+15 |

*Table 5: Densities in m$^{-3}$ of all the species included in the model, at 2 Torr and 50 mA, for different CO$_2$-N$_2$ mixtures for the long tube (length 0.67 cm) where the O density is measured.*

| | 2 Torr | | | | | |
|---|---|---|---|---|---|---|
| | long tube 0.67 cm | | | | | |
| CO$_2$ fraction | 0.1 | 0.25 | 0.5 | 0.75 | 0.9 | 1 |
| species density (m$^{-3}$) | | | | | | |
| O$_2$(X$^3\Sigma_g^-$) | 2.38E+20 | 9.65E+20 | 2.18E+21 | 3.26E+21 | 3.83E+21 | 4.25E+21 |
| O$_2$(a$^1\Delta_g$) | 3.10E+19 | 8.72E+19 | 1.69E+20 | 2.53E+20 | 3.06E+20 | 3.47E+20 |
| O$_2$(b$^1\Sigma_g^+$) | 1.29E+18 | 1.70E+18 | 1.68E+18 | 1.77E+18 | 1.88E+18 | 2.05E+18 |
| O$_2$(A'$^3\Delta_u$, A$^3\Sigma_u^+$, c$^1\Sigma_u^-$) | 2.79E+16 | 5.79E+16 | 1.03E+17 | 1.44E+17 | 1.67E+17 | 1.80E+17 |
| O$_2^+$ | 2.20E+15 | 1.69E+15 | 1.21E+15 | 1.22E+15 | 1.67E+15 | 9.44E+15 |
| O($^3$P) | 1.70E+21 | 2.46E+21 | 2.71E+21 | 2.95E+21 | 3.15E+21 | 3.43E+21 |
| O($^1$D) | 3.19E+17 | 1.25E+17 | 6.38E+16 | 5.08E+16 | 4.95E+16 | 5.18E+16 |
| O$^+$ | 1.27E+14 | 2.93E+13 | 9.59E+12 | 5.43E+12 | 4.49E+12 | 4.54E+12 |
| O$^-$ | 3.23E+13 | 5.01E+13 | 7.41E+13 | 9.57E+13 | 1.09E+14 | 1.19E+14 |
| O$_3$ | 1.31E+15 | 5.38E+15 | 1.00E+16 | 1.34E+16 | 1.49E+16 | 1.73E+16 |
| O$_3$* | 1.49E+12 | 2.26E+13 | 9.83E+13 | 1.98E+14 | 2.60E+14 | 3.21E+14 |
| CO$_2$(X$^1\Sigma^+_g$) | 5.95E+20 | 2.20E+21 | 5.30E+21 | 8.35E+21 | 9.98E+21 | 1.09E+22 |
| CO$_2^+$ | 1.96E+14 | 1.43E+14 | 1.24E+14 | 1.09E+14 | 1.03E+14 | 1.04E+14 |
| C($^3$P) | 3.89E+16 | 1.48E+16 | 8.44E+15 | 6.86E+15 | 6.54E+15 | 6.62E+15 |
| CO(a$^3\Pi_r$) | 4.34E+16 | 3.96E+16 | 3.83E+16 | 3.90E+16 | 4.03E+16 | 4.17E+16 |
| CO(X$^1\Sigma^+$) | 2.34E+21 | 4.74E+21 | 7.68E+21 | 1.02E+22 | 1.16E+22 | 1.26E+22 |
| CO$^+$ | 7.94E+13 | 2.05E+13 | 7.57E+12 | 4.50E+12 | 3.76E+12 | 3.79E+12 |
| N$_2$(X $^1\Sigma_g^+$) | 2.64E+22 | 2.07E+22 | 1.29E+22 | 6.06E+21 | 2.31E+21 | |
| N$_2$(A $^3\Sigma_u^+$) | 1.19E+18 | 3.37E+17 | 8.76E+16 | 2.46E+16 | 7.58E+15 | |
| N$_2$(B$^3\Pi_g$) | 5.51E+16 | 1.59E+16 | 6.42E+15 | 2.54E+15 | 9.32E+14 | |
| N$_2$(C $^3\Pi_u$) | 2.01E+14 | 1.02E+14 | 4.54E+13 | 1.74E+13 | 6.15E+12 | |
| N$_2$(w $^1\Delta_u$) | 1.09E+16 | 6.09E+15 | 2.70E+15 | 1.03E+15 | 3.65E+14 | |
| N$_2$(a $^1\Pi_g$) | 1.25E+16 | 5.14E+15 | 1.58E+15 | 4.77E+14 | 1.51E+14 | |
| N$_2$(a' $^1\Sigma_u^-$) | 1.88E+17 | 3.85E+16 | 7.27E+15 | 2.03E+15 | 7.28E+14 | |
| N$_2^+$ | 3.14E+14 | 9.22E+13 | 2.81E+13 | 8.51E+12 | 2.78E+12 | |
| N$_2^+$(B $^2\Sigma_u^+$) | 7.17E+10 | 2.24E+10 | 7.44E+09 | 2.27E+09 | 7.25E+08 | |
| N($^4$S) | 5.23E+18 | 7.42E+17 | 1.48E+17 | 4.57E+16 | 1.82E+16 | |
| N($^2$D) | 1.11E+18 | 2.44E+17 | 4.29E+16 | 1.09E+16 | 3.58E+15 | |
| N($^2$P) | 3.27E+16 | 1.79E+15 | 1.40E+14 | 3.02E+13 | 1.16E+13 | |
| N$^+$ | 1.07E+12 | 4.25E+10 | 2.15E+09 | 2.93E+08 | 7.94E+07 | |
| N$_4^+$ | 7.87E+12 | 6.74E+11 | 5.88E+10 | 3.50E+09 | 1.68E+08 | |
| N$_3^+$ | 5.28E+12 | 1.93E+11 | 8.81E+09 | 5.80E+08 | 5.45E+07 | |
| NO(X $^2\Pi_r$) | 9.61E+19 | 1.68E+20 | 2.58E+20 | 2.55E+20 | 1.80E+20 | |
| NO(A $^2\Sigma^+$) | 1.66E+15 | 7.99E+14 | 3.05E+14 | 8.22E+13 | 1.76E+13 | |

| | | | | | |
|---|---|---|---|---|---|
| NO(B²Π_r) | 7.88E+10 | 1.24E+10 | 1.72E+09 | 3.51E+08 | 1.00E+08 |
| NO⁺ | 7.47E+15 | 8.36E+15 | 8.68E+15 | 8.47E+15 | 7.90E+15 |
| NO₂(X²A₁) | 6.90E+15 | 1.14E+16 | 1.39E+16 | 8.94E+15 | 4.23E+15 |
| NO₂(a⁴A₂) | 1.89E+14 | 3.73E+14 | 5.14E+14 | 4.92E+14 | 3.43E+14 |
| CN(X²Σ⁺) | 1.45E+11 | 4.29E+10 | 1.45E+10 | 4.87E+09 | 1.39E+09 |
| CN(B²Σ⁺) | 5.14E+08 | 1.46E+07 | 1.43E+06 | 3.07E+05 | 9.25E+04 |
| $n_e$ | 1.04E+16 | 1.03E+16 | 9.99E+15 | 9.73E+15 | 9.58E+15 | 9.44E+15 |

*Table 6: Densities in $m^{-3}$ of all the species included in the model, at 2 Torr and 50mA, for different $CO_2$-$N_2$ mixtures and including the reactions with $N_2(A,B)$ from section III.1.1. for the short tube (length 0.23 cm) where the $CO_2$ dissociation fraction and NO densities are measured.*

| | 2 Torr | | | | |
|---|---|---|---|---|---|
| | short tube 0.23 cm | | | | |
| CO₂ fraction | 0.1 | 0.25 | 0.5 | 0.75 | 0.9 |
| species density (m⁻³) | | | | | |
| O₂(X³Σ_g⁻) | 2.49E+20 | 8.16E+20 | 1.65E+21 | 2.31E+21 | 2.66E+21 |
| O₂(a¹Δ_g) | 3.01E+19 | 7.03E+19 | 1.25E+20 | 1.80E+20 | 2.16E+20 |
| O₂(b¹Σ_g⁺) | 1.15E+18 | 1.07E+18 | 9.23E+17 | 9.01E+17 | 9.48E+17 |
| O₂(A'³Δ_u, A³Σ_u⁺, c¹Σ_u⁻) | 3.17E+16 | 5.69E+16 | 9.51E+16 | 1.31E+17 | 1.52E+17 |
| O₂⁺ | 2.33E+15 | 2.12E+15 | 1.62E+15 | 1.61E+15 | 2.17E+15 |
| O(³P) | 1.56E+21 | 2.11E+21 | 2.26E+21 | 2.42E+21 | 2.60E+21 |
| O(¹D) | 2.63E+17 | 8.93E+16 | 4.54E+16 | 3.61E+16 | 3.58E+16 |
| O⁺ | 1.01E+14 | 2.06E+13 | 6.87E+12 | 3.91E+12 | 3.31E+12 |
| O⁻ | 3.71E+13 | 5.61E+13 | 8.43E+13 | 1.14E+14 | 1.33E+14 |
| O₃ | 1.39E+15 | 4.53E+15 | 7.59E+15 | 9.50E+15 | 1.04E+16 |
| O₃* | 1.61E+12 | 1.62E+13 | 5.78E+13 | 1.05E+14 | 1.34E+14 |
| CO₂(X¹Σ⁺_g) | 7.33E+20 | 3.06E+21 | 7.53E+21 | 1.22E+22 | 1.47E+22 |
| CO₂⁺ | 2.21E+14 | 1.83E+14 | 1.69E+14 | 1.57E+14 | 1.52E+14 |
| C(³P) | 3.49E+16 | 1.47E+16 | 8.79E+15 | 7.32E+15 | 7.15E+15 |
| CO(a³Π_r) | 4.38E+16 | 3.88E+16 | 3.72E+16 | 3.80E+16 | 3.97E+16 |
| CO(X¹Σ⁺) | 2.21E+21 | 4.00E+21 | 5.99E+21 | 7.59E+21 | 8.49E+21 |
| CO⁺ | 6.27E+13 | 1.28E+13 | 4.21E+12 | 2.29E+12 | 1.88E+12 |
| N₂(X ¹Σ_g⁺) | 2.65E+22 | 2.11E+22 | 1.34E+22 | 6.50E+21 | 2.52E+21 |
| N₂(A ³Σ_u⁺) | 1.10E+18 | 3.78E+17 | 1.23E+17 | 4.20E+16 | 1.48E+16 |
| N₂(B³Π_g) | 3.69E+16 | 9.00E+15 | 3.04E+15 | 1.06E+15 | 3.71E+14 |
| N₂(C ³Π_u) | 2.01E+14 | 1.08E+14 | 5.10E+13 | 2.05E+13 | 7.54E+12 |
| N₂(w ¹Δ_u) | 1.09E+16 | 6.68E+15 | 3.52E+15 | 1.59E+15 | 6.25E+14 |
| N₂(a ¹Π_g) | 1.25E+16 | 5.91E+15 | 2.24E+15 | 7.83E+14 | 2.66E+14 |
| N₂(a' ¹Σ_u⁻) | 1.99E+17 | 5.52E+16 | 1.25E+16 | 3.60E+15 | 1.30E+15 |
| N₂⁺ | 3.39E+14 | 1.11E+14 | 3.52E+13 | 1.07E+13 | 3.56E+12 |
| N₂⁺(B ²Σ_u⁺) | 7.54E+10 | 2.20E+10 | 6.65E+09 | 1.84E+09 | 5.81E+08 |
| N(⁴S) | 5.41E+18 | 9.63E+17 | 2.32E+17 | 7.47E+16 | 3.01E+16 |
| N(²D) | 9.82E+17 | 2.87E+17 | 6.64E+16 | 2.06E+16 | 7.86E+15 |
| N(²P) | 3.14E+16 | 2.82E+15 | 3.15E+14 | 7.89E+13 | 3.34E+13 |
| N⁺ | 9.71E+11 | 6.14E+10 | 4.56E+09 | 7.87E+08 | 2.51E+08 |
| N₄⁺ | 8.94E+12 | 1.07E+12 | 1.08E+11 | 7.12E+09 | 3.96E+08 |
| $N_3^+$ | 5.33E+12 | 3.12E+11 | 1.98E+10 | 1.66E+09 | 1.85E+08 |
| $NO(X\ ^2\Pi_r)$ | 8.89E+19 | 1.24E+20 | 1.79E+20 | 1.81E+20 | 1.30E+20 |
| $NO(A\ ^2\Sigma^+)$ | 1.42E+15 | 6.64E+14 | 3.03E+14 | 1.03E+14 | 2.58E+13 |
| $NO(B\ ^2\Pi_r)$ | 7.55E+10 | 1.44E+10 | 2.40E+09 | 4.80E+08 | 1.25E+08 |
| $NO^+$ | 7.33E+15 | 7.80E+15 | 8.10E+15 | 7.79E+15 | 7.14E+15 |
| $NO_2(X^2A_1)$ | 9.88E+15 | 1.13E+16 | 1.13E+16 | 6.59E+15 | 2.75E+15 |
| $NO_2(a^4A_2)$ | 8.43E+13 | 1.58E+14 | 2.43E+14 | 2.59E+14 | 1.92E+14 |
| $CN(X^2\Sigma^+)$ | 1.30E+11 | 4.28E+10 | 1.51E+10 | 4.98E+09 | 1.35E+09 |
| $CN(B^2\Sigma^+)$ | 5.01E+08 | 1.69E+07 | 2.05E+06 | 4.92E+05 | 1.58E+05 |
| $n_e$ | 1.04E+16 | 1.02E+16 | 9.86E+15 | 9.46E+15 | 9.34E+15 |

*Table 7: Densities in $m^{-3}$ of all the species included in the model, for a 50 % $CO_2$ 50 % $N_2$ initial gas mixture, for different pressures, at 50mA and including the reactions with $N_2(A,B)$ from section III.1.1. for the short tube (length 0.23 cm) where the $CO_2$ dissociation fraction and NO densities are measured.*

| | 50 % CO2 50 % N2 | | | |
|---|---|---|---|---|
| | short tube 0.23 cm | | | |
| Pressure (Torr) | 0.6 | 1 | 2 | 4 |
| species density ($m^{-3}$) | | | | |
| $O_2(X^3\Sigma_g^-)$ | 4.98E+20 | 6.34E+20 | 1.36E+21 | 3.11E+21 |
| $O_2(a^1\Delta_g)$ | 4.37E+19 | 5.42E+19 | 1.05E+20 | 2.03E+20 |
| $O_2(b^1\Sigma_g^+)$ | 6.30E+17 | 6.20E+17 | 6.97E+17 | 6.50E+17 |
| $O_2(A'^3\Delta_u, A^3\Sigma_u^+, c^1\Sigma_u^-)$ | 8.50E+16 | 6.63E+16 | 7.93E+16 | 1.26E+17 |
| $O_2^+$ | 3.74E+15 | 2.97E+15 | 1.54E+15 | 7.86E+14 |
| $O(^3P)$ | 1.04E+21 | 1.60E+21 | 2.25E+21 | 2.29E+21 |
| $O(^1D)$ | 6.21E+16 | 5.90E+16 | 4.18E+16 | 2.35E+16 |
| $O^+$ | 3.59E+13 | 2.09E+13 | 6.49E+12 | 1.62E+12 |
| $O^-$ | 1.14E+14 | 8.73E+13 | 8.17E+13 | 9.04E+13 |
| $O_3$ | 9.48E+15 | 5.48E+15 | 6.27E+15 | 7.67E+15 |
| $O_3^*$ | 1.21E+13 | 1.35E+13 | 4.13E+13 | 1.19E+14 |
| $CO_2(X^1\Sigma_g^+)$ | 3.77E+21 | 5.03E+21 | 8.31E+21 | 1.41E+22 |
| $CO_2^+$ | 9.17E+14 | 4.60E+14 | 1.81E+14 | 8.53E+13 |
| $C(^3P)$ | 2.49E+16 | 2.03E+16 | 9.48E+15 | 4.25E+15 |
| $CO(a^3\Pi_r)$ | 5.69E+16 | 4.65E+16 | 3.54E+16 | 2.76E+16 |
| $CO(X^1\Sigma^+)$ | 2.22E+21 | 3.10E+21 | 5.37E+21 | 9.16E+21 |
| $CO^+$ | 2.01E+13 | 1.05E+13 | 3.43E+12 | 1.05E+12 |
| $N_2(X\ ^1\Sigma_g^+)$ | 5.95E+21 | 8.07E+21 | 1.36E+22 | 2.31E+22 |
| $N_2(A\ ^3\Sigma_u^+)$ | 3.60E+17 | 2.80E+17 | 1.62E+17 | 8.71E+16 |
| $N_2(B^3\Pi_g)$ | 1.86E+16 | 1.49E+16 | 9.25E+15 | 5.24E+15 |
| $N_2(C\ ^3\Pi_u)$ | 7.46E+13 | 6.66E+13 | 5.32E+13 | 4.06E+13 |
| $N_2(w\ ^1\Delta_u)$ | 9.32E+15 | 7.34E+15 | 4.02E+15 | 1.99E+15 |
| $N_2(a\ ^1\Pi_g)$ | 6.59E+15 | 5.28E+15 | 2.71E+15 | 1.25E+15 |
| $N_2(a'\ ^1\Sigma_u^-)$ | 3.15E+16 | 2.60E+16 | 1.46E+16 | 8.05E+15 |
| $N_2^+$ | 1.78E+14 | 9.74E+13 | 3.62E+13 | 1.48E+13 |
| $N_2^+(B\ ^2\Sigma_u^+)$ | 3.78E+10 | 1.95E+10 | 6.37E+09 | 2.04E+09 |
| $N(^4S)$ | 5.58E+17 | 4.52E+17 | 2.67E+17 | 1.37E+17 |
| $N(^2D)$ | 1.99E+17 | 1.85E+17 | 9.25E+16 | 2.94E+16 |
| N($^2$P) | 2.20E+15 | 1.47E+15 | 4.46E+14 | 9.14E+13 |
| N$^+$ | 1.59E+11 | 7.00E+10 | 7.29E+09 | 4.45E+08 |
| N$_4^+$ | 5.09E+11 | 2.79E+11 | 1.27E+11 | 7.56E+10 |
| N$_3^+$ | 2.92E+11 | 1.53E+11 | 2.98E+10 | 4.06E+09 |
| NO(X $^2\Pi_r$) | 9.12E+19 | 1.20E+20 | 1.87E+20 | 2.40E+20 |
| NO(A $^2\Sigma^+$) | 4.71E+14 | 4.82E+14 | 4.23E+14 | 2.75E+14 |
| NO(B$^2\Pi_r$) | 2.16E+09 | 2.67E+09 | 2.75E+09 | 1.75E+09 |
| NO$^+$ | 3.51E+15 | 5.38E+15 | 8.09E+15 | 9.79E+15 |
| NO$_2$(X$^2$A$_1$) | 9.96E+14 | 2.67E+15 | 1.01E+16 | 3.11E+16 |
| NO$_2$(a$^4$A$_2$) | 7.55E+13 | 1.49E+14 | 3.09E+14 | 3.80E+14 |
| CN(X$^2\Sigma^+$) | 1.90E+10 | 2.12E+10 | 1.67E+10 | 1.26E+10 |
| CN(B$^2\Sigma^+$) | 2.49E+07 | 1.11E+07 | 2.67E+06 | 5.87E+05 |
| n$_e$ | 8.29E+15 | 8.86E+15 | 9.78E+15 | 1.06E+16 |

*Table 8: Densities in m$^{-3}$ of all the species included in the model, for a 50 % CO$_2$ 50 % N$_2$ initial gas mixture, for different pressures, at 50mA and including the reactions with N$_2$(A,B) from section III.1.1. for the long tube (length 0.67 cm) where the O density is measured.*

| | 50 % CO2 50 % N2 | | | |
|---|---|---|---|---|
| | long tube 0.67 cm | | | |
| Pressure (Torr) | 0.6 | 1 | 2 | 4 |
| species density (m$^{-3}$) | | | | |
| O$_2$(X$^3\Sigma_g^-$) | 8.46E+20 | 1.08E+21 | 2.18E+21 | 4.63E+21 |
| O$_2$(a$^1\Delta_g$) | 7.70E+19 | 9.45E+19 | 1.69E+20 | 2.97E+20 |
| O$_2$(b$^1\Sigma_g^+$) | 1.65E+18 | 1.64E+18 | 1.68E+18 | 1.35E+18 |
| O$_2$(A'$^3\Delta_u$, A$^3\Sigma_u^+$, c$^1\Sigma_u^-$) | 1.16E+17 | 8.89E+16 | 1.03E+17 | 1.57E+17 |
| O$_2^+$ | 3.26E+15 | 2.41E+15 | 1.21E+15 | 6.56E+14 |
| O($^3$P) | 1.34E+21 | 2.03E+21 | 2.71E+21 | 2.63E+21 |
| O($^1$D) | 1.02E+17 | 9.67E+16 | 6.38E+16 | 3.31E+16 |
| O$^+$ | 5.78E+13 | 3.35E+13 | 9.59E+12 | 2.14E+12 |
| O$^-$ | 1.01E+14 | 7.66E+13 | 7.41E+13 | 8.26E+13 |
| O$_3$ | 1.46E+16 | 8.89E+15 | 1.00E+16 | 1.13E+16 |
| O$_3$* | 3.51E+13 | 3.77E+13 | 9.83E+13 | 2.34E+14 |
| CO$_2$(X$^1\Sigma^+_g$) | 2.33E+21 | 3.10E+21 | 5.30E+21 | 9.45E+21 |
| CO$_2^+$ | 6.14E+14 | 3.07E+14 | 1.24E+14 | 5.95E+13 |
| C($^3$P) | 2.17E+16 | 1.75E+16 | 8.44E+15 | 3.98E+15 |
| CO(a$^3\Pi_r$) | 6.28E+16 | 5.09E+16 | 3.83E+16 | 2.94E+16 |
| CO(X$^1\Sigma^+$) | 3.32E+21 | 4.55E+21 | 7.68E+21 | 1.28E+22 |
| CO$^+$ | 4.80E+13 | 2.47E+13 | 7.57E+12 | 2.16E+12 |
| N$_2$(X $^1\Sigma_g^+$) | 5.58E+21 | 7.57E+21 | 1.29E+22 | 2.21E+22 |
| N$_2$(A $^3\Sigma_u^+$) | 1.91E+17 | 1.48E+17 | 8.76E+16 | 4.91E+16 |
| N$_2$(B$^3\Pi_g$) | 1.24E+16 | 1.02E+16 | 6.42E+15 | 3.71E+15 |
| N$_2$(C $^3\Pi_u$) | 6.20E+13 | 5.60E+13 | 4.54E+13 | 3.50E+13 |
| N$_2$(w $^1\Delta_u$) | 6.13E+15 | 4.89E+15 | 2.70E+15 | 1.37E+15 |
| N$_2$(a $^1\Pi_g$) | 3.71E+15 | 3.00E+15 | 1.58E+15 | 7.66E+14 |
| N$_2$(a' $^1\Sigma_u^-$) | 1.46E+16 | 1.20E+16 | 7.27E+15 | 4.49E+15 |
| N$_2^+$ | 1.31E+14 | 7.23E+13 | 2.81E+13 | 1.20E+13 |
| $N_2^+(B\ ^2\Sigma_u^+)$ | 4.02E+10 | 2.16E+10 | 7.44E+09 | 2.47E+09 |
| $N(^4S)$ | 3.50E+17 | 2.56E+17 | 1.48E+17 | 8.21E+16 |
| $N(^2D)$ | 1.00E+17 | 8.90E+16 | 4.29E+16 | 1.41E+16 |
| $N(^2P)$ | 7.82E+14 | 4.74E+14 | 1.40E+14 | 3.23E+13 |
| $N^+$ | 5.29E+10 | 2.12E+10 | 2.15E+09 | 1.45E+08 |
| $N_4^+$ | 1.83E+11 | 1.04E+11 | 5.88E+10 | 4.01E+10 |
| $N_3^+$ | 9.66E+10 | 4.68E+10 | 8.81E+09 | 1.31E+09 |
| $NO(X\ ^2\Pi_r)$ | 1.32E+20 | 1.75E+20 | 2.58E+20 | 3.02E+20 |
| $NO(A\ ^2\Sigma^+)$ | 3.59E+14 | 3.64E+14 | 3.05E+14 | 1.86E+14 |
| $NO(B^2\Pi_r)$ | 1.67E+09 | 1.81E+09 | 1.72E+09 | 1.13E+09 |
| $NO^+$ | 4.29E+15 | 6.19E+15 | 8.68E+15 | 1.03E+16 |
| $NO_2(X^2A_1)$ | 1.38E+15 | 3.76E+15 | 1.39E+16 | 3.94E+16 |
| $NO_2(a^4A_2)$ | 1.43E+14 | 2.78E+14 | 5.14E+14 | 5.49E+14 |
| $CN(X^2\Sigma^+)$ | 1.62E+10 | 1.77E+10 | 1.45E+10 | 1.17E+10 |
| $CN(B^2\Sigma^+)$ | 1.61E+07 | 6.22E+06 | 1.43E+06 | 3.42E+05 |
| $n_e$ | 8.30E+15 | 8.95E+15 | 9.99E+15 | 1.09E+16 |